\pdfoutput=1
\documentclass[twocolumn,prd,floatfix,nofootinbib,longbibliography,notitlepage,superscriptaddress,aps]{revtex4-2}

\usepackage{amsmath}
\usepackage{amssymb}
\usepackage{bm}
\usepackage{subfigure}
\usepackage[usenames,svgnames,dvipsnames]{xcolor}
\usepackage{array}
\usepackage{bbm}
\usepackage{booktabs}
\usepackage{multirow}
\usepackage{xcolor}

\newcommand{\Figref}[1]{Figure~\ref{#1}}

\newcommand{\secref}[1]{section~\ref{#1}}
\newcommand{\subsecref}[1]{subsection~\ref{#1}}
\newcommand{\Appref}[1]{Appendix~\ref{#1}}
\newcommand{\eg}{e.g.,\,}
\newcommand{\ie}{i.e.,\,}

\usepackage{graphicx}

\graphicspath{{./Figures/}}

\usepackage[T1]{fontenc}
\usepackage[utf8]{inputenc}
\usepackage{lmodern}

\usepackage[unicode, colorlinks, allcolors=blue!70!black, linktocpage, pdfusetitle]{hyperref}
\hypersetup{
    colorlinks=true,
    urlcolor=SteelBlue,
    linkcolor=red,
    citecolor=blue,
}
\usepackage[all]{hypcap}
\usepackage{orcidlink}
\usepackage{microtype}

\newcommand{\infmapsf}{%
\begin{figure}[t!]
\centering
\includegraphics[width=0.5\textwidth]{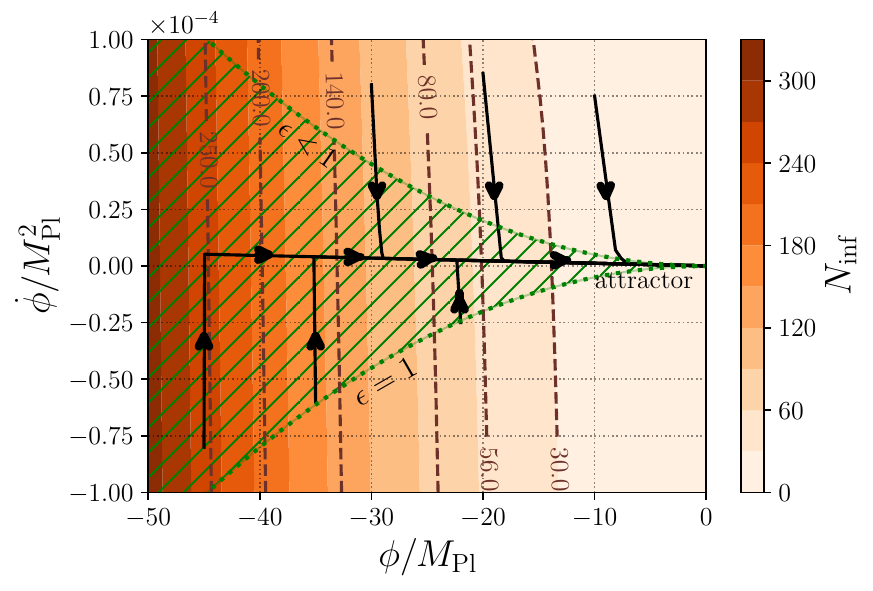}
\caption{\label{fig:inf_map_lp4} Map of initial conditions 
for the potential $V(\phi)=\lambda\phi^4/4$ with $\lambda=10^{-14}$.
In this case, background trajectories converge to the attractor 
exponentially fast. The green dashed curves mark the boundaries 
where $\epsilon=1$. Under the slow-roll approximation, the allowed 
region of background field configurations is restricted to the wedge 
shaded with green lines.}
\end{figure}
}

\newcommand{\ellevolsf}{%
\begin{figure}[t!]
\centering
\includegraphics[width=0.45\textwidth]{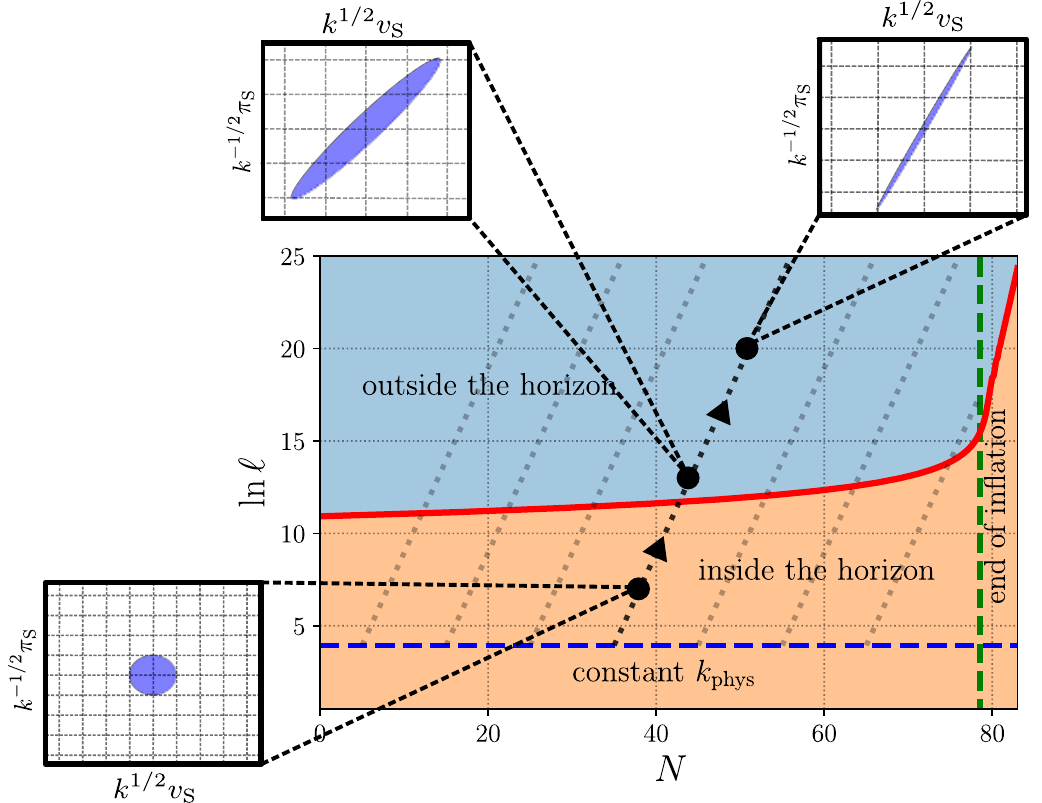}
\caption{\label{fig:ell_evol_sf} Evolution scheme for perturbation 
modes during cosmic inflation. The x-axis shows the number of 
e-folds, while the y-axis tracks the evolution of physical 
length scales $(\ell)$, including the horizon $1/H$ (red curve). 
Approximate time-translational symmetry permits choosing initial 
conditions on a surface of large constant physical wavelength 
$k_{\rm phys}$. Lateral insets illustrate the deformation of the 
Wigner ellipse at different times, shown in dimensionless 
units. The evolution reflects the squeezing of the ellipse as modes 
cross the horizon.}
\end{figure}
}

\newcommand{\sfaccidents}{%
\begin{figure}[ht!]
\centering
\includegraphics[width=0.49\textwidth]{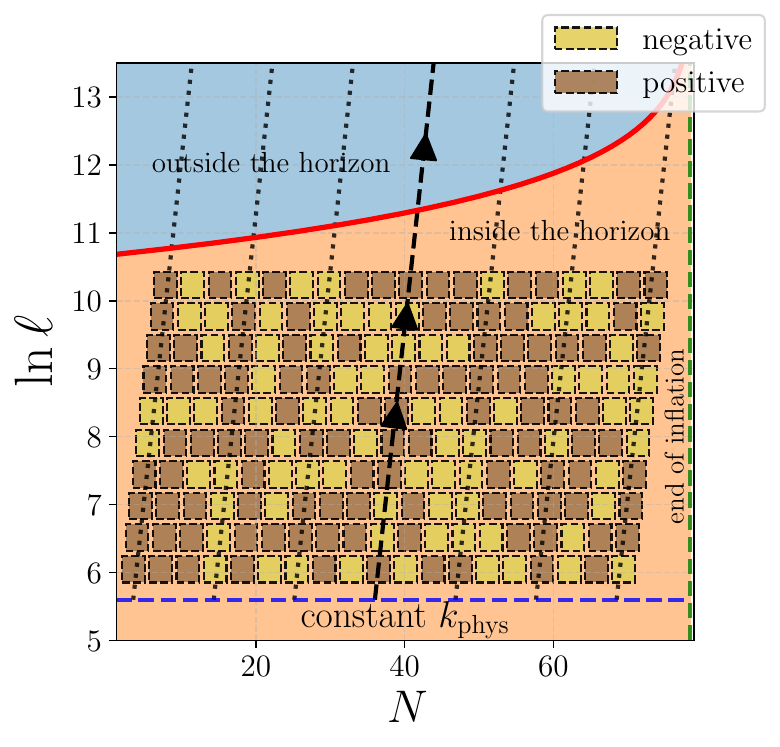}
\caption{\label{fig:scheme_acc_sf} Example of an arbitrary 
configuration of decoherence events (``accidents'') in a portion of 
the $(N,\ln\ell)$ plane. These accidents are arranged as ``tiles'' 
within a 
region bounded above by the horizon and below by a surface 
of constant $k_{\rm phys}$. The source term can either increase or reduce 
the determinant of the covariance matrix. This setup enables the 
evaluation of an arbitrary time-dependent power spectrum of the 
environment.
}
\end{figure}
}

\newcommand{\sfrndm}{%
\begin{figure*}[t!]
\centering
\includegraphics[width=0.9\textwidth]{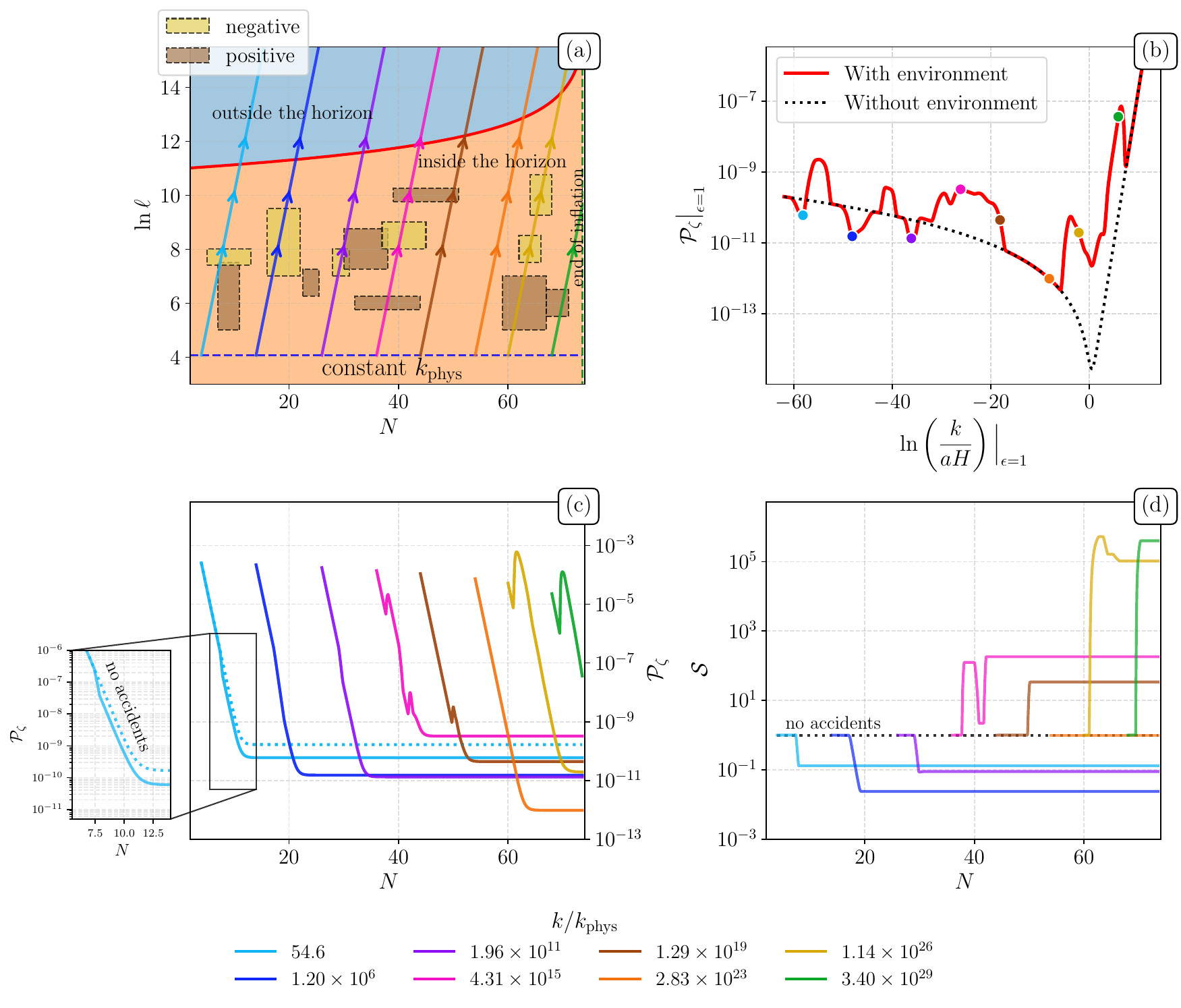}
\caption{\label{fig:imp_rndm_acc_sf} Implementation 
of a random distribution of accidents spanning 
different wavenumber ranges and durations. We use the same color convention as in \Figref{fig:scheme_acc_sf} to denote accidents with positive $\alpha_\Gamma$
in different tones of brown, and yellow in the cases 
with $\alpha_{\Gamma}<0$. Panel 
(a): Distribution of accidents together with the 
mode injection scheme. Several modes are selected 
to evaluate their evolution, their contribution to 
the power spectrum, and the resulting gain or loss 
in state availability. Panel (b): Power spectrum of
primordial curvature fluctuations. The red curve 
corresponds to the spectrum in the presence of 
accidents, while the black dotted curve shows the 
spectrum free of decoherence events.  Panel (c): 
Evolution of the curvature spectrum at fixed 
wavenumber. Dotted lines depict the evolution 
with no accidents highlighting how 
accidents alter the mode dynamics. The inset 
on the left depicts the reduction of amplitude 
caused by negative changes in the determinant.
Panel (d): Changes in the area of the Wigner 
ellipse for several modes. The label ``no accidents'' refers to the conservation of the 
determinant in the accident-free scenarios. In contrast, decoherence events with finite 
duration modify the determinant during their action but leave it unchanged once they switch off.}
\end{figure*}
}

\newcommand{\sfcmpnstd}{%
\begin{figure*}[t!]
\centering
\includegraphics[width=.9\textwidth]{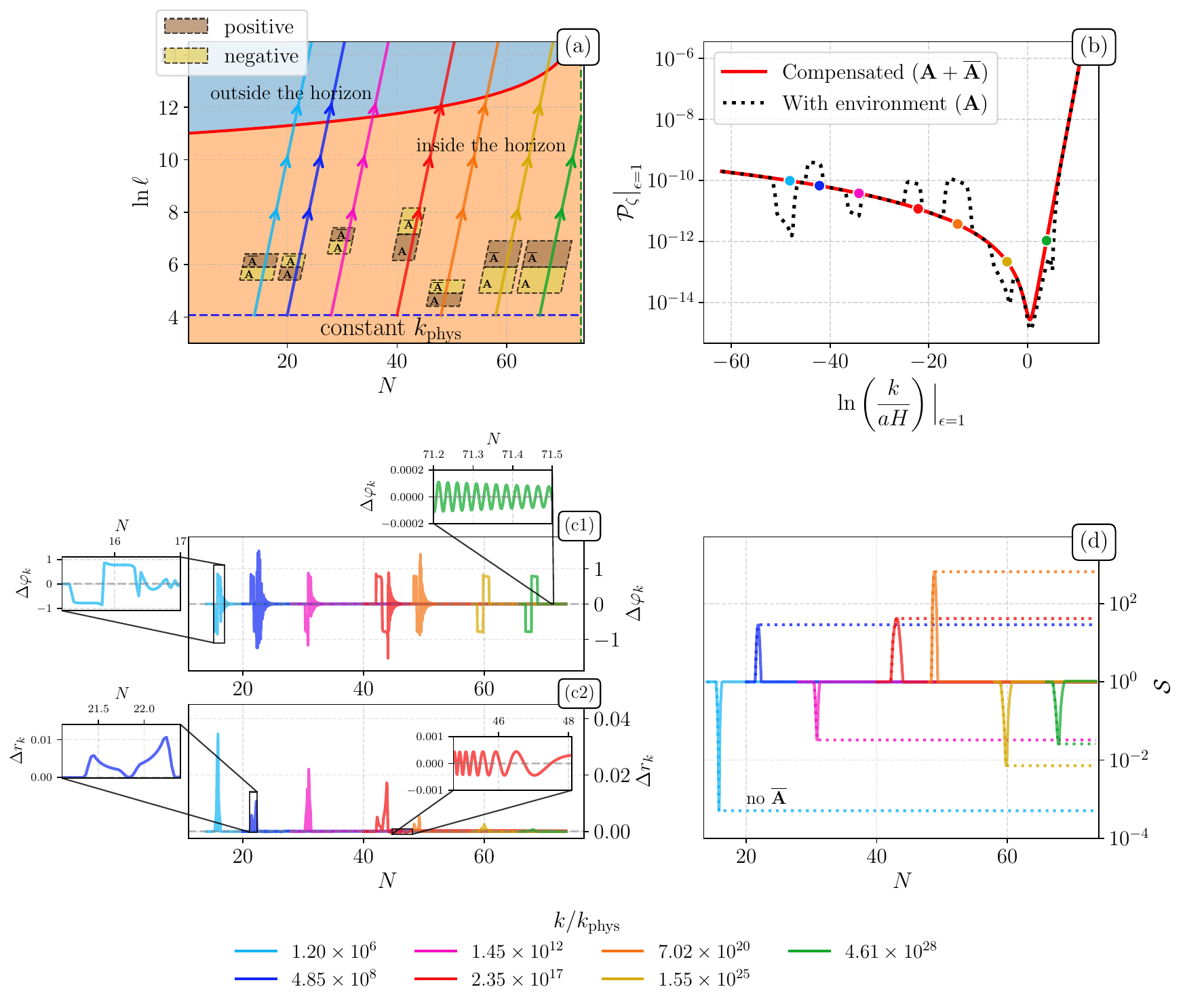}
\caption{\label{fig:cmpnstd_acc_sf} Implementation 
of a distribution of decoherence events that leave the primordial power spectrum unaltered. The same color convention as in Figures \ref{fig:scheme_acc_sf} and \ref{fig:imp_rndm_acc_sf} is adopted: accidents 
with positive $\alpha_\Gamma$ are shown in different tones of brown, while those
with $\alpha_{\Gamma}<0$ are shown in yellow. Panel 
(a): Mode injection scheme and distribution of accidents $(\rm A)$, together with their corresponding 
anti-accidents $(\bar{\rm A})$ that cancel effects in mode evolution. We use the source term 
parameterization proposed in Eq.~\eqref{eq:tiling_source_mod} to confine the effects to specific bands 
of wavelengths. Panel (b): Power spectrum of primordial curvature fluctuations in the presence of accidents 
(black dotted line) and after including accident/anti-accident pairs. The compensation mechanism restores the
spectrum, leaving it unchanged. Panels (c1) and (c2): 
Evolution of the deformations in the squeezing parameters $\Delta r_k\equiv r_k-r^{(0)}_k$ and 
$\Delta \varphi_k\equiv \varphi_k-\varphi^{(0)}_k$, measured relative to the accident-free scenario. 
The insets show that the final state is represented by a Wigner ellipse with unit area (approx.) and 
almost negligible eccentricity, which rotates until crossing the horizon.
Panel (d): Evolution of the Wigner ellipse area for several modes. After passing through an 
accident/anti-accident pair, the area relaxes back to unity. The label ``no $(\bar{\mathrm{A}})$'' 
indicates cases without anti-accidents, where the enlarged (or reduced) area persists without reversal.}
\end{figure*}
}

\newcommand{\flowersf}{%
\begin{figure*}
\centering
\subfigure{
\includegraphics[width=.95\textwidth]{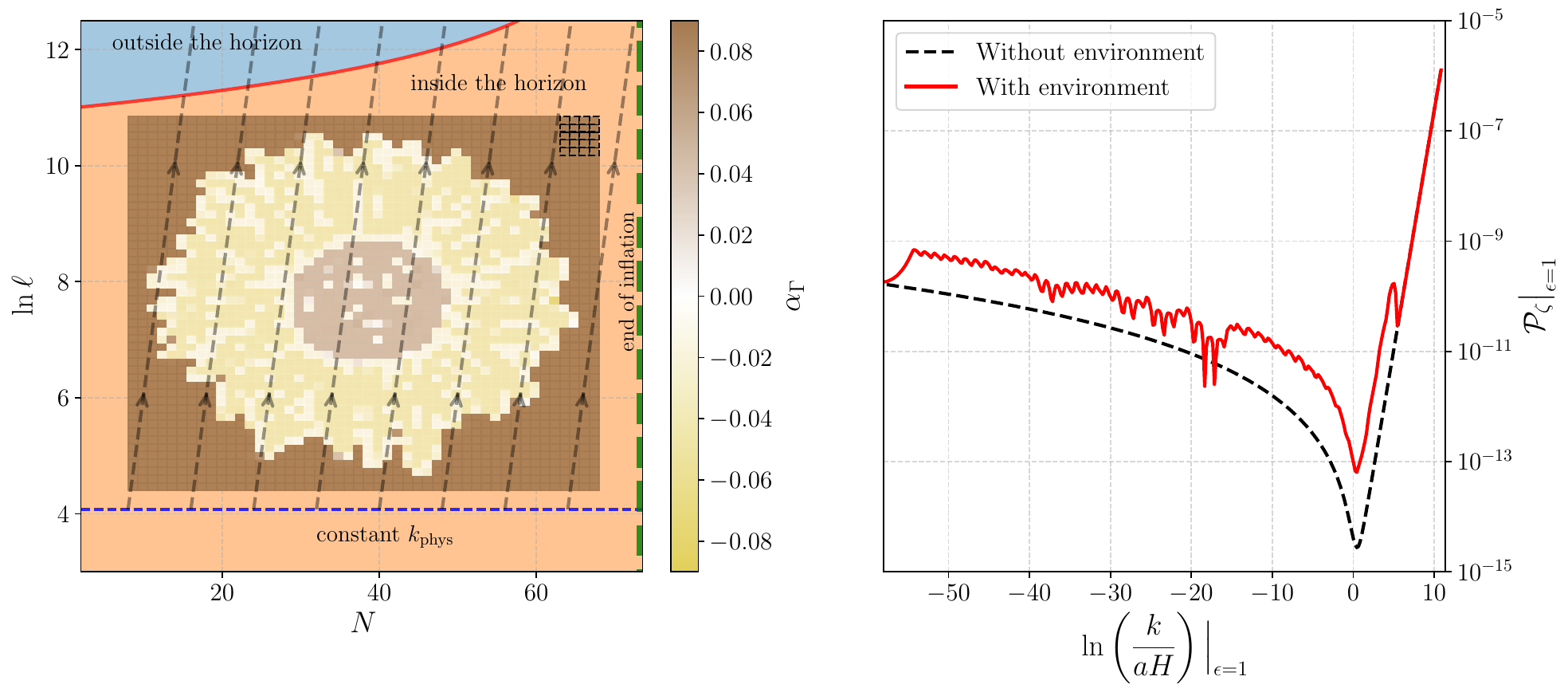}} \,
\caption{\label{fig:flwr_pwr} Deformations in the curvature power spectrum for a large number of decoherence events. Left panel: mode injection scheme with a 
$48\times48$ array of decoherence and recoherence events forming a sunflower pattern. The small $5\times4$ array in the upper right corner shows the duration and the $k$-span for each accident in real magnitude. Using a single processor, the code resolves the dynamics of 365 modes within a few minutes, without showing 
signs of numerical instabilities. Right panel: resulting power spectrum of curvature fluctuations compared with the environment-free scenario. Small oscillations appear at all scales due to the presence of accidents.}
\end{figure*} 
}

\newcommand{\mfreprod}{%
\begin{figure*}
\centering
\subfigure{
\includegraphics[width=\textwidth]{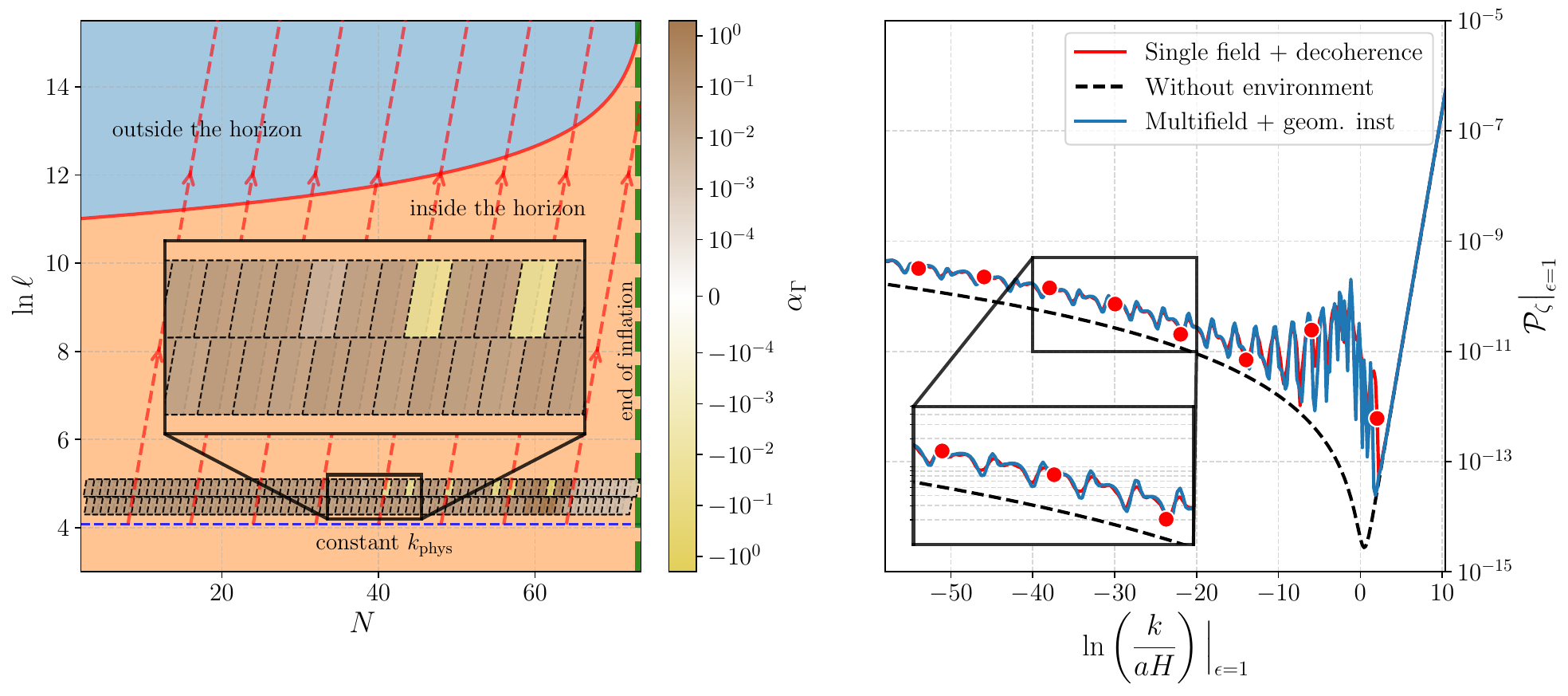}} \,
\caption{\label{fig:mf_inst_reprod} Attempt to reproduce the primordial curvature power 
spectrum generated by a multifield model with geometrical instabilities. Left panel: Two 
layers of decoherence events are applied to progressively deform each $k$-mode. 
Following the same procedure as in the case of reversible decoherence accidents, the 
values of $\alpha_\Gamma$ are adjusted through a bisection method to match the  
power of the multifield spectrum. Right panel: Comparison of the resulting power 
spectra, illustrating the deformation of the single-field spectrum into the multifield 
one, which exhibits characteristic features induced by geometrical instabilities. The 
inset highlights that the discrepancies remain small but non-negligible. This 
reproduction strategy could be further refined by incorporating machine learning 
algorithms, an idea that will be explored in future work.
}
\end{figure*} 
}

\newcommand{\backmultinlin}{%
\begin{figure}
\centering
\subfigure{
\includegraphics[width=.49\textwidth]{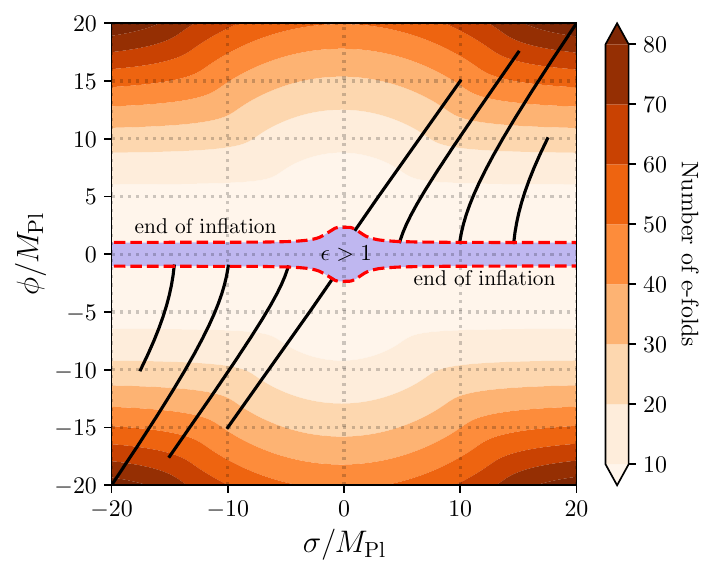}} \,
\caption{\label{fig:back_multi_lp4} Map of initial conditions that produce $\mathcal{N}$ e-folds of 
inflation in the multifield potential $V(\phi,\sigma)=\frac{\lambda}{4}\phi^4+\frac{g}{2}\phi^2\sigma^2$. We selected a few background trajectories (in solid black lines) to show how these converge toward the dotted black lines that denote the end of inflation. The blue region indicates the region where the first slow-roll parameter $\epsilon$ exceeds unity, signaling the termination of the phase of exponential expansion. 
}
\end{figure} 
}

\newcommand{\trnglrschm}{%
\begin{figure*}
\centering
\subfigure{
\includegraphics[width=1.0\textwidth]{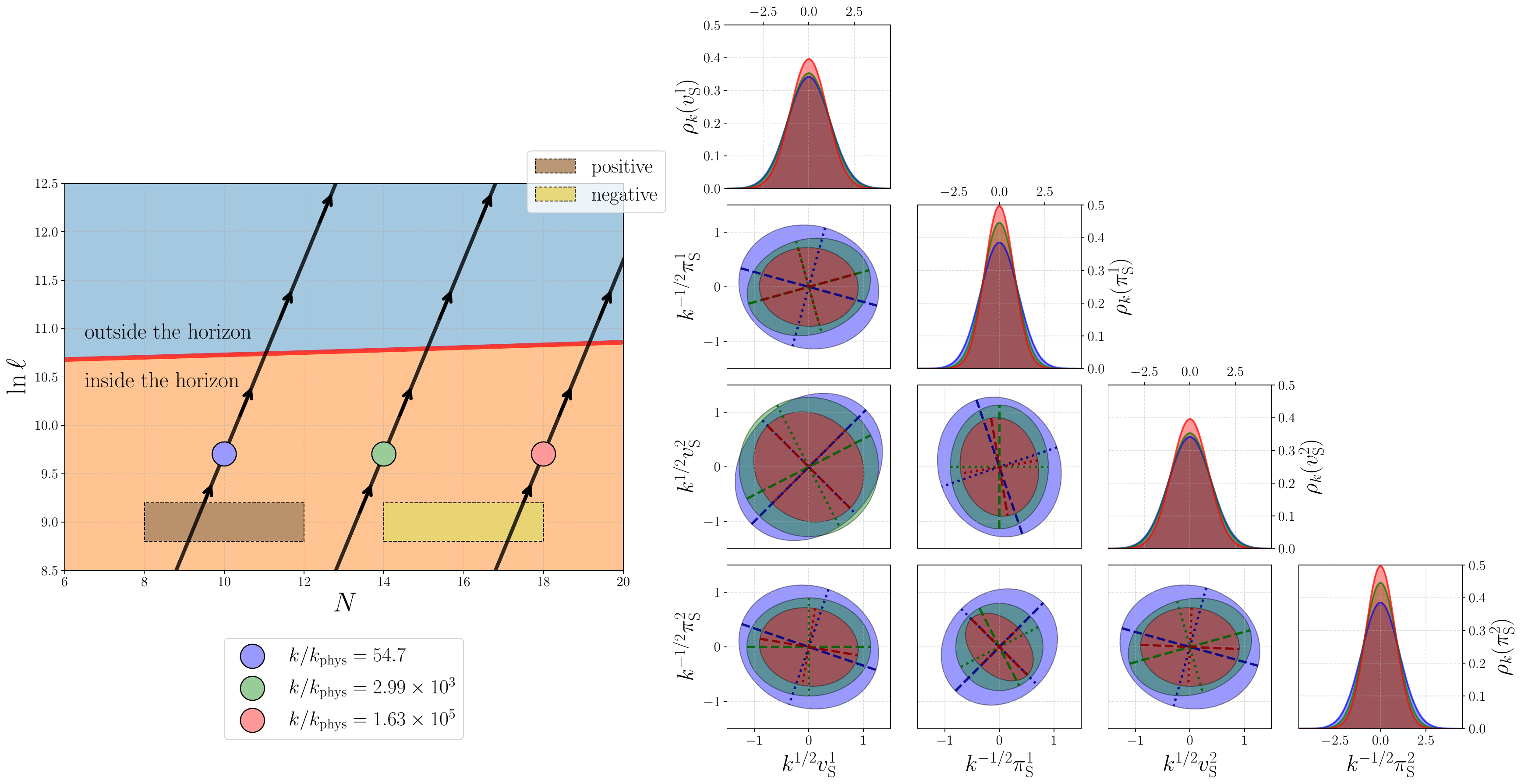}} \,
\caption{\label{fig:Wigner_mf} Wigner ellipse evolution through decoherence and 
recoherence events. Left panel: Injection scheme and configuration of decoherence events with 
$\alpha_\Gamma>0$ in brown and 
$\alpha_\Gamma<0$ in yellow. Even in the 
accident-free scenario, different modes 
exhibit slight dephasing at subhorizon 
scales; however, the main source of Wigner 
ellipse deformations -- and hence of state 
modifications -- is the occurrence of 
decoherence or recoherence events, which 
produce distinguishable signatures. Right 
panel: Triangular plot showing each 
projection of the joint Wigner 
function after each decoherence event, 
highlighting the change in area with 
respect to the ellipse in the environment-free case (in green). We observe 
that the states deformed by decoherence (or 
decoherence) are slightly more squeezed 
than in the case free of accidents. On the 
rightmost side, we illustrate 
the deformations of the marginalized probability distributions $\rho_k(v_{\rm S}^A,\pi_{\rm S}^B)$. In addition to this plot, we generated an animation representing the evolution of the joint distribution, which is included as an ancillary file.  
}
\end{figure*} 
}

\newcommand{\sevolmf}{%
\begin{figure*}[t!]
\centering
\includegraphics[width=.8\textwidth]{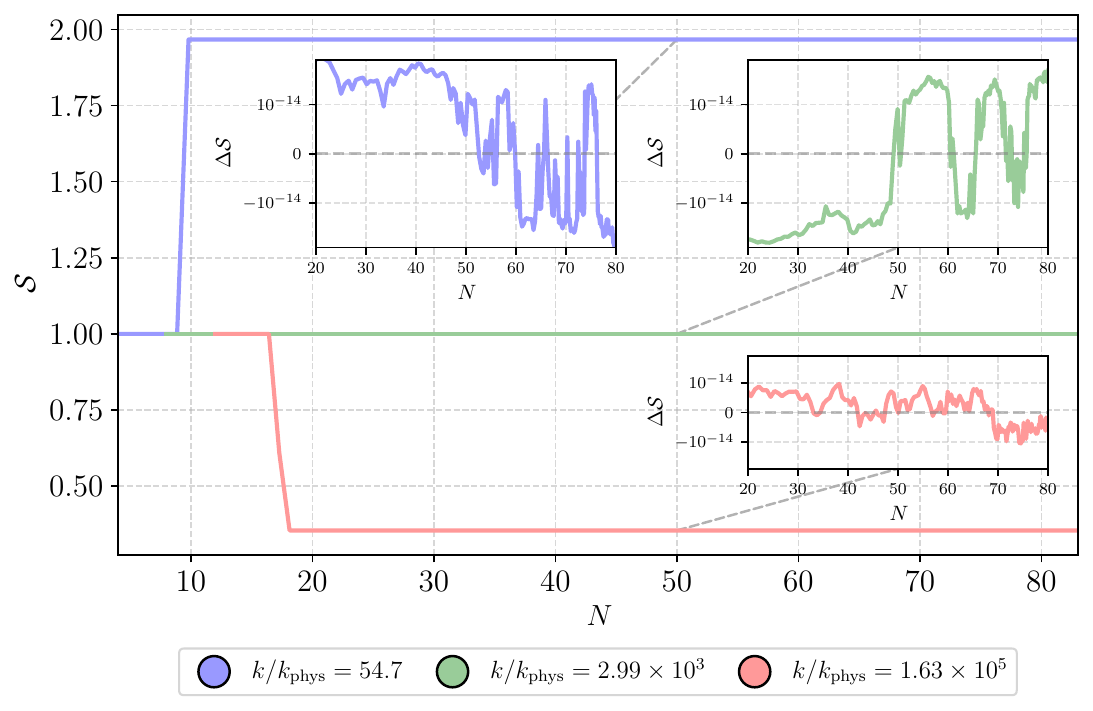}
\caption{\label{fig:s_evol_mf} Evolution of the determinant for the modes 
considered in \Figref{fig:Wigner_mf}. As expected, changes in the determinant 
only occur while the decoherence events are active. The insets depict the deviation of the determinant from its mean value, computed from the period following the decoherence event up to the end of inflation. These deviations confirm that the conservation of $\mathcal{S}$ is preserved to machine precision, and also that errors do not propagate after 
traversing the state-deformation tiles. 
}
\end{figure*}
}

\newcommand{\deccrrmf}{%
\begin{figure*}[t!]
\centering
\includegraphics[width=.9\textwidth]{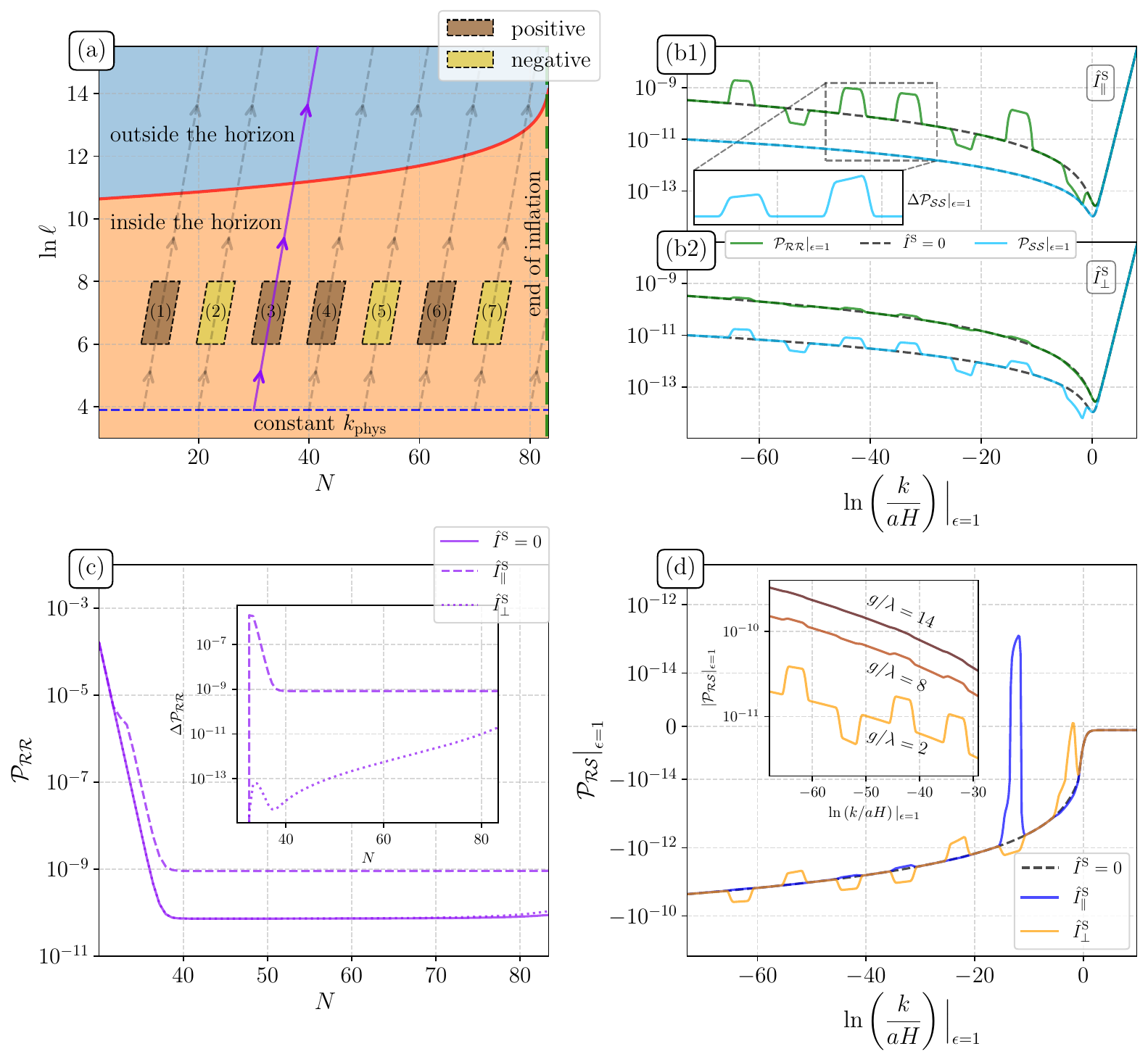}
\caption{\label{fig:dec_crr_mf} Mode correlators resulting from the implementation 
of the two decoherence channels built from the environmental operators $\hat{I}^{\rm S}_{\parallel}$ and 
$\hat{I}^{\rm S}_{\perp}$ defined in Eqns.~(\ref{eq:oper_par}--\ref{eq:oper_perp}). Panel (a):
Mode-injection scheme with an array of decoherence events, using the same tiling in the $(N,\ln \ell)$ plane for both operators. The highlighted mode (purple dashed line) shows the evolution of mode correlators across each decoherence region. Panels (b1) and (b2): Parallel $(\mathcal{P}_{\mathcal{RR}})$ and perpendicular $(\mathcal{P}_{\mathcal{SS}})$ spectra for the $\hat{I}^{\rm S}_{\parallel}$ and $\hat{I}^{\rm S}_{\perp}$ operators. Each operator imprints features primarily on its corresponding spectrum, with residual structures in $\mathcal{P}_{\mathcal{RR}}$ arising from small correlations between isocurvature and adiabatic modes.
Panel (c): Evolution of the $\mathcal{P}_{\cal RR}$ $k$-mode 
highlighted in Panel (a), as it crosses parallel and perpendicular decoherence events. The inset shows the 
variation $\Delta \mathcal{P}_{\cal RR}$ 
relative to the accident-free scenario 
$(\hat{I}^{\rm S}=0)$, indicating that 
$\mathcal{P}_{\cal RR}$ modes grow when crossing the 
$\hat{I}_{\perp}^{\rm S}$ configuration, which is consistent with the presence of the 
small-amplitude features visible (in green) in panel (b2). Panel (d): Cross correlators $(\mathcal{P}_{\mathcal{RS}})$ for the same setup. The large spike (in blue) marks a sign flip 
rather than a change in correlation magnitude. The 
inset illustrates how increasing the field coupling 
enhances the amplitude of cross correlations. Features 
are present in all of the cases.
}
\end{figure*}
}

\newcommand{\tablemf}{
\begin{table}[t!]
\centering
\renewcommand{\arraystretch}{1.05}
\setlength{\tabcolsep}{3pt} % slightly tighter spacing
\begin{tabular}{@{\hspace{0pt}}|c|c|c|c|@{\hspace{0pt}}}
\hline
\textbf{Event} &
\multicolumn{2}{c|}{\textbf{$\alpha_\Gamma$}} &
\multirow{2}{*}{\textbf{\makebox[1.4cm][c]{$\ln\!\left(\frac{k_j}{k_{\mathrm{phys}}}\right)$}}}\\
\cline{2-3}
 & $\hat{I}^{\rm S}_{\parallel}$ & $\hat{I}^{\rm S}_{\perp}$ &  \\
\hline
(1) & 0.25 & 0.46 & 10.0   \\
(2) & -0.075 & -0.280 & 20.0  \\
(3) & 0.25 & 0.37 & 30.0 \\
(4) & 0.25 & 0.34 & 40.0 \\
(5) & -0.058 & -0.2 & 50.0  \\
(6) & 0.25 & 0.2 & 60.0 \\
(7) & -0.04 & -0.1 & 70.0 \\
\hline
\end{tabular}
\caption{Tile parameters for the configuration of decoherence events sourced by parallel 
$(\hat{I}^{\rm S}_{\parallel})$ and perpendicular $(\hat{I}^{\rm S}_{\perp})$ operators 
in \Figref{fig:dec_crr_mf}, panel (a). The remaining tile parameters (\ie the widths, vertical locations along the $\ln\ell$ axis and heights) are constant.}
\label{tab:events}
\end{table}
}

\newcommand{\landscapemf}{%
\begin{figure*}
\centering
\subfigure{
\includegraphics[width=\textwidth]{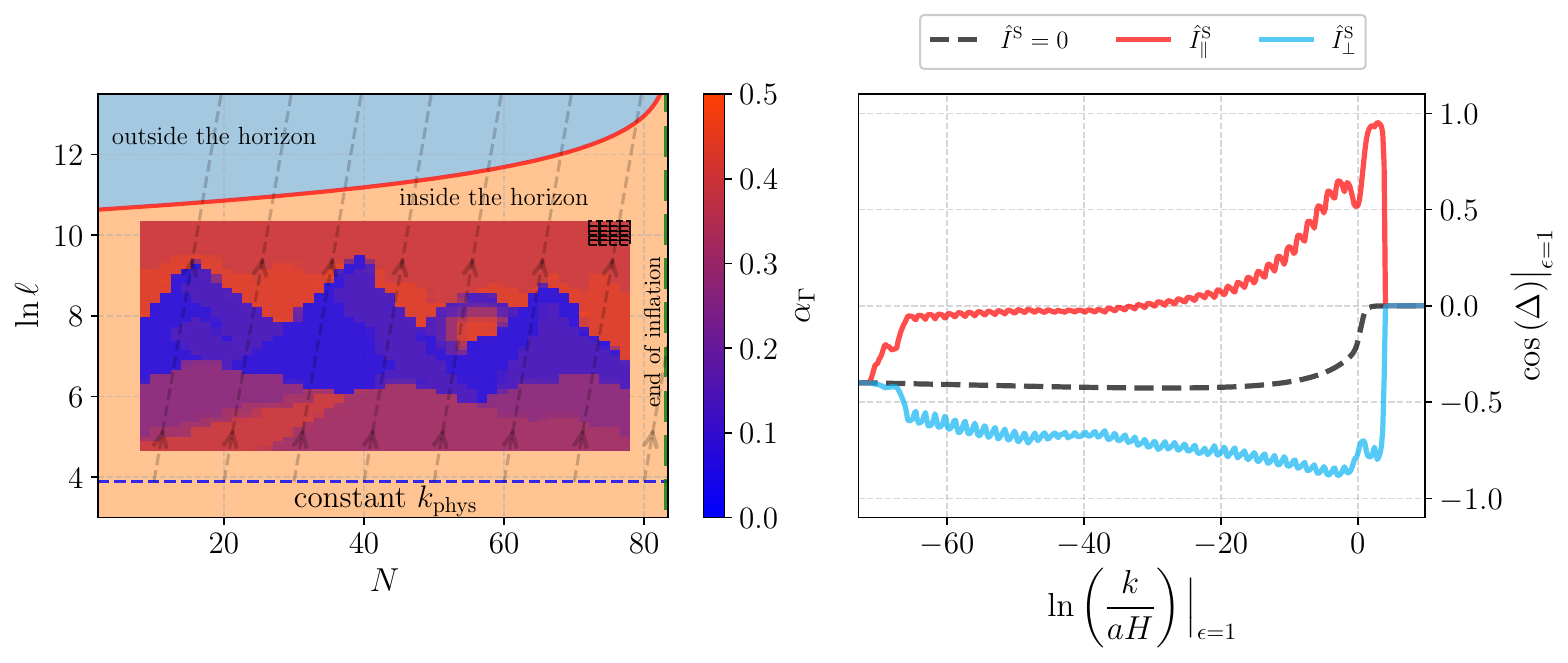}} \,
\caption{\label{fig:lndscp_crr_mf} Deformations in the mode cross-correlations for a 
large number of decoherence tiles generated by parallel and perpendicular environmental 
operators. Left panel: mode injection scheme with a $48\times48$ array of decoherence 
events (with $\alpha_\Gamma>0$) forming a landscape pattern. The small $5\times4$ array 
in the upper right corner shows the duration and the $k$-span for each accident in real 
magnitude. As in the single-field scenario depicted in \Figref{fig:flwr_pwr}, the code 
resolves the dynamics of 365 modes within a few minutes, without showing 
signs of numerical instabilities. Right panel: resulting normalized mode cross-correlations, $\cos(\Delta)$, at the end of inflation in the cases $\hat{I}^{\rm S}=0$, $\hat{I}^{\rm S}_\parallel$ and 
$\hat{I}^{\rm S}_\perp$ using the same tile pattern displayed in the left panel. 
Apart from the presence of features, it is evident that the choice of 
parallel/perpendicular operators flips the 
signs of the cross-correlations. The sharp decrease in $\cos(\Delta)$ at $k\gtrsim aH$ originates from a region free of events in the $(N,\ln\ell)$ plane. Wider arrays of tiles, incorporating mixed contributions from parallel and perpendicular operators, will enhance the cross-correlation power over that wavelength range. }
\end{figure*}
}

\newcommand{\boxes}{%
\begin{figure*}[ht!]
\centering
\subfigure{
\includegraphics[width=\textwidth]{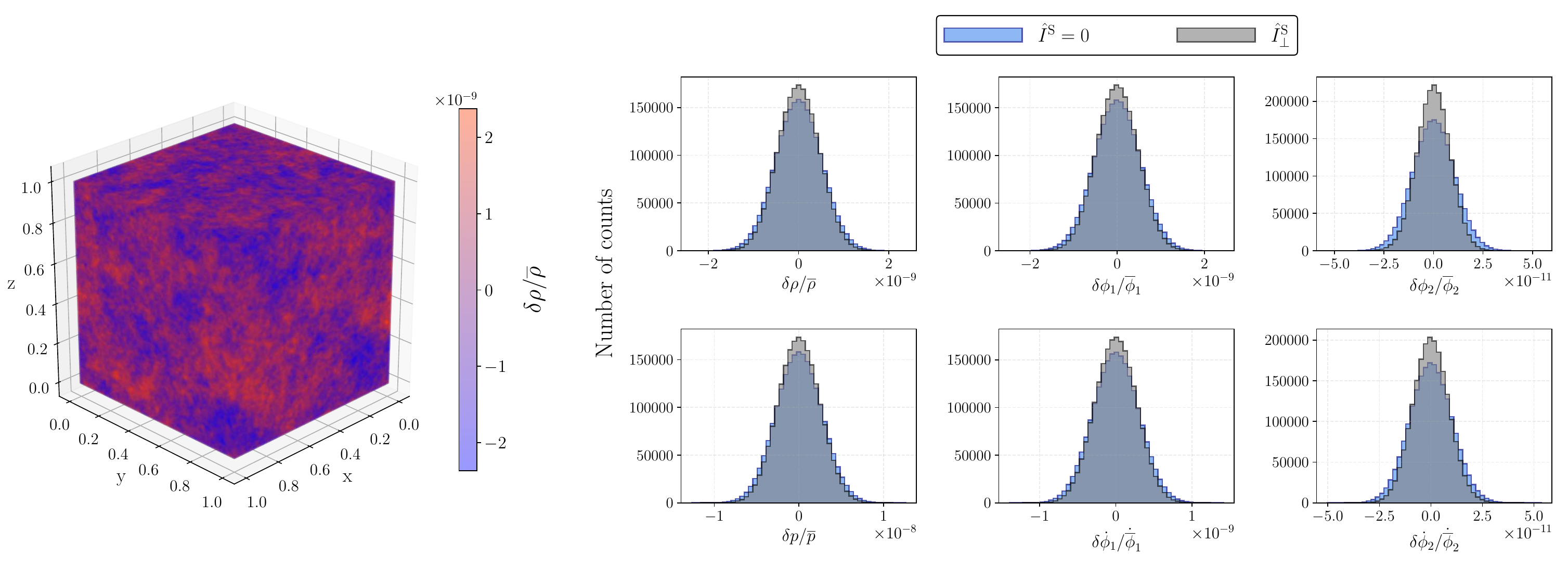}} \,
\caption{\label{fig:3d_real} Three-dimensional realization of the field and energy 
density fluctuations at the end of inflation obtained through dynamical coloring 
transformations. Left panel: spatial distribution of energy density fluctuations in a 
$128^3$ box, computed with the nonlinear potential in Eq.~\eqref{eq:double_lp4_sfs} and 
Fourier-based spatial derivatives for the gradient energy. Right panel: 
histograms of the field, momentum, density and pressure fluctuations after the system
crosses an array of decoherence tiles generated by perpendicular 
environmental operators $\hat{I}^{\rm S}_\perp$. The resulting Gaussian distributions 
remain close to their decoherence-free counterparts, indicating that decoherence acting 
on isocurvature modes does not necessarily induce large changes in the initial 
conditions for reheating.}
\end{figure*}
}

\begin{document}

\title{Numerical tiling-based simulations of decoherence\\ in multifield models of inflation}

\author{Johor D.\ \surname{Pe\~nalba Quispitupa}\,\orcidlink{0009-0004-2594-139X}}
\email{jpenalbaq@uni.pe}
\affiliation{Facultad de Ciencias - Universidad Nacional de Ingenier\'ia; Av. Tupac Amaru 210 - Rimac, Lima, Per\'u}

\author{Guillermo F.\ \surname{Quispe Pe\~na}\,\orcidlink{0009-0000-3775-6520}}
\email{gfq@sfu.ca}
\affiliation{Simon Fraser University, 8888 University Dr. W, Burnaby, BC V5A 1S6}

\author{Jos\'e T.\ \surname{G\'alvez Ghersi}\,\orcidlink{0000-0001-7289-3846}}
\email{jgalvezg@utec.edu.pe}
\affiliation{Universidad de Ingenieria y Tecnologia; 
Jr. Medrano Silva 165 - Barranco 15063, Lima, Per\'u}

\date{\today}

% Because hyperref only gets the *last* author, we need to be explicit.
\hypersetup{pdfauthor={Pe\~nalba Quispitupa, Quispe Pe\~na and G\'alvez Ghersi}}

\begin{abstract}
In previous work, we developed a method for computing two-point 
correlators by decomposing the mode degrees of freedom into fast 
and slow components. Building on this framework, we present a numerical 
implementation to study the evolution of primordial scalar perturbations under 
controlled state deformations induced by the simplest environment 
corrections from the Lindblad equation. The primary contribution of this work is the development of a numerically stable and flexible framework for implementing open-system effects in inflationary perturbations, rather than the extraction of concrete predictions for the primordial spectrum. Our approach generalizes to an arbitrary 
number of degrees of freedom and does not rely on the 
slow-roll approximation. The computational routine is numerically efficient and allows users to configure arbitrary sequences of decoherence events, with full control over their duration, shape, amplitude and effective wavelength range. The resulting outputs are compatible with nonlinear numerical codes, enabling studies of how decoherence effects propagate during reheating. 
\end{abstract}
\maketitle

\section{Introduction}
\setcounter{equation}{0}

At the perturbative level, the inflationary model 
has been remarkably successful since its inception 
\cite{Guth:1980zm, Bardeen:1983qw, Mukhanov:1990me, 
Starobinsky:1992ts}. It provides an effective mechanism by which 
quantum-mechanical fluctuations produce the
primordial inhomogeneities consistent with the nearly isotropic and 
homogeneous state of the early universe shown in the Cosmic Microwave Background (CMB) 
\cite{Smoot:1998jt, Bennett_2013, Planck:2018jri}. In the years to 
come, the enhanced resolution of the next--generation CMB 
observations \cite{SimonsObservatory:2018koc, CMB-S4:2020lpa, Belkner:2023duz, 
Hertig:2024adq} will allow for more stringent constraints on 
inflation, as well as in the emergence of primordial non-Gaussianities
and potential gravitational wave backgrounds from polarization maps 
free from the effects of intergalactic dust. 

Modifications to the dynamics of the inflationary 
adiabatic or the isocurvature fields -- through 
potential 
\cite{Kallosh:2016gqp,Kallosh:2019jnl,Kallosh:2022ggf, Kallosh:2025rni}, 
field geometry \cite{Renaux-Petel:2015mga} and kinetic 
 deformations \cite{Armendariz-Picon:1999hyi} -- have 
motivated investigations into possible connections 
with dark matter, oscillon production, formation of primordial black 
holes and non-Gaussianities, while remaining consistent 
with observations. In addition to these approaches, the 
open-quantum systems framework has been proposed to 
incorporate the effect of environmental 
operators via non-unitary distortions of the vacuum state 
\cite{Martin:2016qta,Martin:2018lin,Martin:2021qkg,
Martin:2021znx, Burgess:2022nwu, Kranas:2025jgm}. We 
conjecture that this formalism is sufficiently 
general to span a substantial portion of the 
configuration space of initial Gaussian states, 
thereby enabling controlled state manipulations 
akin to the implementation of a pumping protocol in quantum optics \cite{Session:2023hoq, Medina:2025bto}.
Even though only a subset of these initial conditions leads to modifications 
of the primordial spectrum consistent with current observations, 
the entire set remains essential. Together, they offer a comprehensive 
set of scenarios that (a) enable efficient forecasting of primordial 
non-Gaussianities during inflation with minimal computational cost, 
and (b) supply well-defined initial conditions to study the nonlinear evolution of
overdensities during reheating, and posteriorly during large structure formation 
\cite{Bond:1995yt, Stahl:2025qru}. 

The primary objective of this paper is to demonstrate how 
the first modifications 
introduced by the open quantum systems framework -- which naturally 
induces initial state deformations through decoherence and 
recoherence -- can be consistently combined with the 
nonlinear background evolution driven by an arbitrary 
number of fields during inflation, avoiding the 
use of the 
so-called slow-roll approximations. Obtaining exact solutions 
from a nonlinear system of background equations becomes increasingly
challenging as the number of fields grows. In addition to this, 
one needs to resolve fast oscillations -- which are computationally 
expensive -- to find the exact evolution of perturbation mode 
equations. Therefore, numerical solutions are necessary to compute 
dynamical observables consistent with the background expansion, 
such as the evolving two-point correlation functions of scalar 
fluctuations. To this end, we have developed a technique based 
on the separation of fast and slow degrees of freedom 
\cite{GalvezGhersi:2016wbu,GalvezGhersi:2018haa}. This 
approach reduces the computational cost of solving exactly both the 
evolution of the nonlinear background equations and the 
perturbation field correlations deep inside the horizon -- precisely 
where decoherence events modify the behavior of the mode solutions. 
Hence, this procedure offers a scalable framework for introducing 
features into two-point correlators compatible with any sufficiently 
smooth inflationary background evolution, while remaining feasible on 
consumer-grade computers. 

Building on this separation method, we modify the evolution 
scheme to solve the simplest realization of the Lindblad equation presented in 
\cite{Martin:2016qta,Martin:2018lin,Martin:2021qkg,
Martin:2021znx, Burgess:2022nwu}, and extend it to the multifield case. 
The numerical routines are designed to accommodate an arbitrary sequence of 
decoherence events (or ``accidents'') acting on different $k$-modes at different times, 
without impacting on computational performance. This sequence defines the 
time-dependent power spectrum of the 
environmental operators and determines the resulting deformed state of each mode. 
The duration, amplitude and range of wavelengths of these events can be tuned to generate
distinctive features in the evolution of Gaussian states and in the 
resulting power spectrum of primordial curvature fluctuations. This framework is agnostic about the origins of decoherence, and the physical interpretation of specific 
arrays of decoherence events is deferred for future work.

The fast and slow decomposition scheme employed here can also be interpreted as a dynamical coloring 
transformation based on Cholesky decomposition. Such transformations can be applied to any set of 
unit-variance random variables to generate initial conditions for the inflationary fields, which 
can be subsequently evolved nonlinearly using reheating codes \cite{Frolov:2008hy}. 
The computational routines developed for this study are publicly available in 
\href{https://github.com/JohorD}{\texttt{https://github.com/JohorD}}. 

The remainder of this paper is organized as follows. In 
Section \ref{sec:OQSF}, we review the open quantum system 
formalism in the context of inflationary models driven by 
a single scalar field. We also introduce the foundations of 
our fast--slow scale separation scheme and compute the curvature 
power spectrum under the influence of multiple accidents introducing 
decoherence. In Section \ref{sec:Multi_deco}, we extend our 
component separation method to incorporate decoherence effects in 
multifield inflation models. Our results demonstrate that it is 
possible to generate features in the power spectrum of primordial 
curvature fluctuations across any range of wavelengths. Using 
cross-correlations, we illustrate how decoherence propagates 
between isocurvature and isentropic degrees of freedom. Following Eq.~\eqref{eq:antisym_A_mod}, this section also includes a detailed discussion of the limitations and regime of validity of the present implementation. In 
Appendix \ref{app:sym_sep}, we show that  
our decomposition scheme can also be applied to other  
forms of coloring transformations beyond the Cholesky decomposition. 
In Appendix \ref{app:Gaussian}, 
we construct three-dimensional field realizations from the outputs of our 
numerical routine. These field realizations can serve as initial 
conditions for simulations of the reheating phase. Finally, in 
Section \ref{sec:conc}, we discuss potential future applications of 
these routines and present our conclusions.

\vspace{-.5em}
\section{Review of the open quantum system framework}
\label{sec:OQSF}

In this section, we present an overview of the background 
and the perturbative mode dynamics when inflation is driven by a 
single scalar field. Building on the open quantum system 
framework described in 
\cite{Martin:2016qta,Martin:2018lin,Martin:2021qkg,
Martin:2021znx}, the following subsections develop a 
numerical formalism to address the general case of an arbitrary 
configuration of state accidents -- arising from decoherence and 
recoherence events -- occurring at arbitrary instants 
of time and wavelengths. By tuning these configurations of decoherence events, we postulate that 
one can deform the initial states of a simple inflationary model to reproduce features in the primordial curvature spectrum typically associated to more intricate models.

\subsection{Single field inflation: dynamical setup for 
closed systems}
\label{subsec:Inf_back_pert}

We begin with a quick revision of the characteristic exponential 
expansion of the universe during cosmic inflation, which is driven 
by the dynamics of the inflaton field minimally coupled to gravity. 
Both geometry and field dynamics are governed by the following 
effective action

\begin{align}
S=\int d^4x\sqrt{-g}\left[\frac{M^2_{\mathrm{Pl}}}{2} R-\frac{1}{2}g^{\mu\nu}\nabla_\mu\phi\nabla_\nu\phi-V(\phi)\right]\,,
\label{eq:sf_action}
\end{align}
where $g\equiv\rm{Det}(g_{\mu\nu})$, 
$M_{\mathrm{Pl}}\equiv (8\pi G)^{-1/2}$ is the reduced 
Planck mass in units of $\hbar=c=1$, and $\phi$ is the 
inflaton field. Throughout the paper, we use the 
signature $(-,+,+,+)$ and the flat Friedmann-Lemaître-Robertson-
Walker (FLRW) background metric as an invariant under gauge 
transformations. We choose the quartic potential 
$V(\phi) = \lambda\phi^4/4$ as a test case, but most of the 
dynamical features of the system do not depend on this choice. 

\infmapsf

The background inflaton dynamics is determined by the 
standard second-order equation of motion
\begin{align}
\ddot{\phi}+3H\dot{\phi}+ \frac{ \partial V(\phi) }{\partial \phi} =0\,,
\label{eq:back_inf}
\end{align}
where dots denote derivatives with respect to the cosmic time $t$. 
The evolution of the scale factor, in terms of the 
Hubble parameter $H\equiv\dot{a}/a$, follows from the 
two Friedmann equations
\begin{subequations}
\begin{align}
&3M^2_{\mathrm{Pl}}H^2=\frac{\dot{\phi}^2}{2}+V(\phi)\,,
\label{eq:Fried_1}\\
&-2M^2_{\mathrm{Pl}}\dot{H}=\dot{\phi}^2\,.
\label{eq:Fried_2}
\end{align}
\end{subequations}

In the slow-roll regime, inflation comes to an end 
when the parameter $\epsilon=\dot{\phi}^2/(2H^2)$ equals one. 
\Figref{fig:inf_map_lp4} shows a map of initial 
conditions: each pixel represents a different 
initial choice of field values, and the color (indicated in the colorbar to the right) shows the corresponding number of 
e-folds of inflation, dubbed by $N_{\rm{inf}}$. The map is approximately symmetric about the axis $\phi=0$. 
In addition to this map, we display a few field trajectories 
that rapidly flow toward the inflationary attractor. 
The slow-roll approximation restricts the range of allowed 
field configurations to a wedge (hatched with thin green lines), 
bounded by the dotted green curves. While the restriction has no 
significant consequences for standard slow-roll models, 
it may preclude interesting field dynamics in scenarios where 
potential instabilities arise 
\cite{Martin:2014kja, Bhattacharya:2022fze}.

In the case of gauge-invariant scalar perturbations $(\zeta)$, 
these arise from 
expanding the action in \eqref{eq:sf_action} up to second order 
in the $\delta\phi=0$ gauge
\begin{align}
S_2=\frac{1}{2}\int d\tau~d^3\mathbf{x} \left(\frac{\phi'^2}{2H^2}\right)\left[\zeta'^2-(\nabla^i\zeta)(\nabla_i\zeta)\right]\,,  \label{eq:act_zeta}  
\end{align}
where $\tau$ is the conformal time and primes denote derivatives 
with respect to it. Latin indices are reserved for cartesian 
coordinates, and are raised and lowered by the Kronecker delta. 
To introduce canonical normalization, Mukhanov-Sasaki 
variables are defined as $v=\phi'\zeta/H$, which enable us 
to rewrite the action as
\begin{align}
S_2 = \frac{1}{2}\int~d\tau~d^3\mathbf{x}\left[v'^2 - 
(\nabla_i v)(\nabla^i v)-\frac{z''}{z} v^2\right]\,,
\label{eq:ms_action}
\end{align}
where $z\equiv (\sqrt{2}M_{\mathrm{Pl}}H)^{-1}\phi'$. The 
corresponding Hamiltonian can be found by performing a 
Legendre transformation of the action
\begin{align}
H_{\rm sys} = \frac{1}{2}\int~d^3\mathbf{x}\left[\pi^2 + 
(\nabla_i v)(\nabla^i v)+\frac{z''}{z} v^2\right]\,,
\label{eq:ms_hamiltonian}
\end{align}
where the canonical momentum is given by 
$\pi\equiv\delta S_2/\delta v'$ as usual. In Fourier 
space, the equations of motion for each of the Mukhanov-Sasaki 
modes are
\begin{subequations}
\begin{align}
&v'_{\mathbf{k}}=\pi_{\mathbf{k}}
\label{eq:sf_eom_v}\,,\\
&\pi'_{\mathbf{k}}=-\left(k^2-\frac{z''}{z}\right)v_{\mathbf{k}}\,.
\label{eq:sf_eom_p}
\end{align}    
\end{subequations}
From this equation, it is clear that the dynamics of the 
background field $\phi$ has an active role in the evolution 
of the effective oscillation frequency. 

Primordial curvature fluctuations emerge as quantum-mechanical 
perturbations of the inflaton field. Therefore, we proceed with the 
canonical quantization of each Fourier mode, in which the 
canonical variables are promoted to operators. Thus, the 
Hamiltonian becomes

\begin{align}
\hat{H}_{\rm sys} = \int_{\mathbb{R}^{3+}}d^3\mathbf{k}~\left[\hat{\pi}_{\mathbf{k}}\hat{\pi}^\dagger_{\mathbf{k}}+\hat{v}_{\mathbf{k}}\left(k^2-\frac{z''}{z}\right)\hat{v}^\dagger_{\mathbf{k}}\right]\,,
\label{eq:non_her_ham}
\end{align}
where the integration region $\mathbb{R}^{3+}$ represents the 
upper half of the Fourier sphere. Momentum and field operators obey 
the canonical commutation relation
\begin{align}
\left[\hat{v}_{\mathbf{k}},\hat{\pi}^\dagger_{\mathbf{k}'}\right]=i\delta^{(3)}(\mathbf{k}-\mathbf{k}')\,,
\label{eq:st_q_rel}
\end{align}
as stated in \cite{Martin:2021znx}, the main problem with this 
representation is that operators are not Hermitian. To ensure 
Hermiticity, it is convenient to build a new set of field operators
\begin{align}
    \hat{v}_{\mathrm{R}} \equiv \frac{\hat{v}_{\mathbf{k}} + \hat{v}^{\dagger}_{\mathbf{k}}}{\sqrt{2}} \quad , \quad \hat{v}_{\mathrm{I}} \equiv \frac{\hat{v}_{\mathbf{k}} - \hat{v}^{\dagger}_{\mathbf{k}}}{i\sqrt{2}},
\label{eq:RI_partition}
\end{align}
where the labels $\mathrm{R}$ and $\mathrm{I}$ dub 
the corresponding real and imaginary parts of the field. These 
definitions extend naturally to the conjugate momentum. As for 
the dynamics of the operators $\hat{v}_{\rm{R}}$ and 
$\hat{v}_{\rm{I}}$, their evolution is governed by 
Eqns.~\eqref{eq:sf_eom_v} and \eqref{eq:sf_eom_p}, as in the case 
of the original field operators. The canonical commutation rule 
in \eqref{eq:st_q_rel} imposes similar relations for the real and 
imaginary components
\begin{align}
    \left[\hat{v}_{\mathrm{S}} (\mathbf{k}) , \hat{\pi}_{\mathrm{S'}}(\mathbf{k^{\prime}})\right]=
    i\delta_{\mathrm{S S^{\prime}}}\delta^{(3)}\left(\mathbf{k}-\mathbf{k^{\prime}}\right),
\label{eq:CanoConmRI}
\end{align}
where the indices $\mathrm{S}, \mathrm{S}^{\prime} \in \{\mathrm{R}, 
\mathrm{I}\}$. Notice that the separation into real and imaginary 
parts does not introduce any new degrees of freedom in the system.
Therefore, we can recast the Hamiltonian operator in 
terms of the real and imaginary field components as
\begin{align}
\hat{H}_{\rm sys} = \frac{1}{2}\sum_{\rm{S}\in\{\rm{R},\rm{I}\}}\int_{\mathbb{R}^{3+}}d^3\mathbf{k}\left[\hat{\pi}_{\rm{S}}\hat{\pi}^\dagger_{\rm{S}}+\hat{v}_{\rm{S}}\left(k^2-\frac{z''}{z}\right)\hat{v}^\dagger_{\rm{S}}\right]\,,
\label{eq:her_ham_sf}
\end{align}
where we omitted the $k$-dependence to shorten the notation. With 
respect to these canonical variables, the Hamiltonian -- together 
with the other field operators and their combinations-- is Hermitian, 
which guarantees that its expectation values are real. Furthermore, 
it is important to consider that Minkowski vacuum minimizes the 
energy and variance for all values of $k$, and in consequence, it 
represents the state of quantum fluctuations at subhorizon scales and defines 
a well-suited set of initial conditions. 

At second order in perturbative expansion the Hamiltonian is 
quadratic in its canonical coordinates, so all of the information of the system is 
contained in two-point correlation functions. Consequently, 
Gaussian states provide a complete description of the state of 
vacuum fluctuations. Moreover, it is well-known that Gaussian states 
remain Gaussian under linear evolution, and the deformations 
of the Wigner ellipse\footnote{In the Wigner formalism, this 
ellipse is defined by the bilinear form in the Gaussian 
Wigner quasi-probability distribution, and can be trivially 
extended to an arbitrary number of fields, see 
\cite{Martin:2015qta,Martin:2021qkg,Colas:2021llj} for 
further details.} -- which encloses all possible 
configurations of a Gaussian state -- are limited to 
shears, contractions, expansions and rotations. 
With this in mind, the information of each mode is 
contained in the following state vector of operators
\begin{equation}
\hat{\mathbf{R}}=\left(\hat{v}_{\mathrm{R}},\hat{\pi}_{\mathrm{R}},\hat{v}_{\mathrm{I}}, \hat{\pi}_{\mathrm{I}}\right)^{\rm T}\,,
\label{eq:state_vector_sf}
\end{equation}
and all of the two-point correlation functions are encoded in the 
following matrix of the Minkowski vacuum expectation values
\begin{align}
\langle\hat{\mathbf{\Sigma}}\rangle \equiv \left\langle \left\{\hat{\mathbf{R}},\hat{\mathbf{R}}\right\}\right\rangle = \langle\hat{\mathbf{\Sigma}}^{\mathrm{R}}\rangle\oplus \langle\hat{\mathbf{\Sigma}}^{\mathrm{I}}\rangle\,,
\label{eq:cov_sf}
\end{align}
where the symbol $\{\cdot~,\cdot\}$ denotes the anticommutator acting on each 
component of the operator's state vector $\hat{\mathbf{R}}$. The direct sum 
in this last identity indicates that cross-correlations between real
and imaginary components cancel, which reduces $\langle\hat{\mathbf\Sigma}\rangle$ to a 
block-diagonal matrix containing two $2\times 2$ submatrices for 
the real and imaginary parts. 
The lengths of the semimajor and semiminor axes of the Wigner 
ellipse are the eigenvalues of the covariance matrix, 
while the rotation angle is determined by their corresponding 
eigenvectors. Each of the covariance matrices for the real and 
imaginary parts, $\langle\hat{\mathbf{\Sigma}}^{\rm S}\rangle$ for $\rm S\in\{R,I\}$, has the following 
form
\begin{equation}
    \langle\hat{\mathbf{\Sigma}}^{\mathrm{S}}\rangle = \left(\begin{array}{cc}
\langle\hat{\Sigma}^{\mathrm{S}}_{(v v)}\rangle & \langle\hat{\Sigma}^{\mathrm{S}}_{(v p)}\rangle \\
    \langle\hat{\Sigma}^{\mathrm{S}}_{(v p)}\rangle & \langle\hat{\Sigma}^{\mathrm{S}}_{(p p)}\rangle
    \end{array}\right) \,.
\label{eq:Covariance_Matrix}
\end{equation}
Explicitly, each of these matrix elements is computed from the 
vacuum expectation values of the following operators 
\begin{subequations}
\begin{align}
&\hat{\Sigma}^{\mathrm{S}}_{(vv)}= \left\{\hat{v}_{\mathrm{S}}(\mathbf{k'})\,,\hat{v}_{\mathrm{S}}(\mathbf{k})\right\}\,,\label{eq:cov_comp_vv}\\
&\hat{\Sigma}^{\mathrm{S}}_{(vp)}=\left\{ \hat{v}_{\mathrm{S}}(\mathbf{k'})\,,\hat{\pi}_{\mathrm{S}}(\mathbf{k})\right\}\,,
\label{eq:cov_comp_vp}\\
&\hat{\Sigma}^{\mathrm{S}}_{(pp)}=\left\{\hat{\pi}_{\mathrm{S}}(\mathbf{k'})\,,\hat{\pi}_{\mathrm{S}}(\mathbf{k})\right\}\,,\label{eq:cov_comp_pp}
\end{align}
\end{subequations}
where $\rm S\in \{R,I\}$ and $k=|\mathbf{k}|$. Therefore, 
it is clear that these operators correspond to the 
block-diagonal pieces of the anticommutator 
$\{\hat{\mathbf{R}},\hat{\mathbf{R}}\}$ defined 
in Eq.~\eqref{eq:cov_sf}. Standard commutation rules allow us 
to assign a covariance matrix for each $k$-mode, 
which yields
\begin{subequations}
\begin{align}
&\langle\hat{\Sigma}^{\mathrm{S}}_{(vv)}\rangle= 2\left\langle\hat{v}_{\mathrm{S}}(\mathbf{k}),\hat{v}_{\mathrm{S}}(\mathbf{k})\right\rangle\,,\label{eq:cov_vv_sf}\\
&\langle\hat{\Sigma}^{\mathrm{S}}_{(vp)}\rangle=\left\langle \hat{v}_{\mathrm{S}}(\mathbf{k})\,,\hat{\pi}_{\mathrm{S}}(\mathbf{k})\right\rangle+\left\langle \hat{\pi}_{\mathrm{S}}(\mathbf{k})\,,\hat{v}_{\mathrm{S}}(\mathbf{k})\right\rangle\,,
\label{eq:cov_vp_sf}\\
&\langle\hat{\Sigma}^{\mathrm{S}}_{(pp)}\rangle=2\left\langle\hat{\pi}_{\mathrm{S}}(\mathbf{k})\,,\hat{\pi}_{\mathrm{S}}(\mathbf{k})\right\rangle\,,\label{eq:cov_pp_sf}
\end{align}
\end{subequations}
after excluding overall mode fixing factors of 
$\delta^{(3)}(\mathbf{k}-\mathbf{k}')$. 
From now on, the index $\rm S$ can be fixed to only refer to  
the real or the imaginary part of the solution, without losing 
any information after this choice.
We can derive equations of motion for the expectation values by 
taking derivatives 
\begin{subequations}
\begin{align}
&\frac{\rm d}{\rm d\tau}\langle\hat{\Sigma}^{\mathrm{S}}_{(vv)}\rangle= 2\langle\hat{\Sigma}^{\mathrm{S}}_{(vp)}\rangle\,,\label{eq:eom_cov_vv_sf}\\
&\frac{\rm d}{\rm d\tau}\langle\hat{\Sigma}^{\mathrm{S}}_{(vp)}\rangle=\langle\hat{\Sigma}^{\mathrm{S}}_{(pp)}\rangle-\left(k^2-\frac{z''}{z}\right)\langle\hat{\Sigma}^{\mathrm{S}}_{(vv)}\rangle\,,
\label{eq:eom_cov_vp_sf}\\
&\frac{\rm d}{\rm d\tau}\langle\hat{\Sigma}^{\mathrm{S}}_{(pp)}\rangle=-2\left(k^2-\frac{z''}{z}\right)\langle\hat{\Sigma}^{\mathrm{S}}_{(vp)}\rangle\,,\label{eq:eom_cov_pp_sf}
\end{align}
\end{subequations}
or, equivalently, by replacing the covariance matrix operators 
(for every fixed $\mathbf{k}$) defined in 
Eqns.~(\ref{eq:cov_comp_vv}--\ref{eq:cov_comp_pp}) in the 
Liouville equation
\begin{align}
\frac{\rm d}{\rm d\tau}\langle\hat{\mathbf{\Sigma}}^{\mathrm{S}}\rangle = i\left\langle\left[\hat{H}_{\rm sys},\hat{\mathbf{\Sigma}}^{\mathrm{S}}\right]\right
\rangle\,.
\label{eq:liouville}
\end{align}
Equations (\ref{eq:eom_cov_vv_sf}--\ref{eq:eom_cov_pp_sf}) can 
be combined into a single third-order transport equation for 
the covariance component $\langle\hat{\Sigma}^{\mathrm{S}}_{(vv)}\rangle$:
\begin{align}
&\frac{\rm d^3}{\rm d\tau^3}\langle\hat{\Sigma}^{\mathrm{S}}_{(vv)}\rangle+4\left(k^2-\frac{z''}{z}\right)\frac{\rm d}{\rm d\tau}\langle\hat{\Sigma}^{\mathrm{S}}_{(vv)}\rangle\label{eq:sf_trans_no_acc}\\
&+2\left[\left(k^2-\frac{z''}{z}\right)'\langle\hat{\Sigma}^{\mathrm{S}}_{(vv)}\rangle\right]=0
\nonumber\,,
\end{align}
which only holds if the real and imaginary modes are solutions 
to the mode equations in Eqns.~\eqref{eq:sf_eom_v} and \eqref{eq:sf_eom_p}, and also implies that it can be order-reduced.

\ellevolsf

The interpretation of the averaged covariance matrix elements is 
crucial, since the Wigner ellipse is defined by this 
Hermitian positive-definite matrix. This ellipse encloses the mode 
configurations accessible for each value of $k$. Moreover, 
the determinant of $\langle\hat{\mathbf{\Sigma}^{\rm S}}\rangle$
\begin{align}
\mathcal{S} \equiv \rm{det}\langle\hat{\mathbf{\Sigma}}^{\rm S}\rangle\,,
\label{eq:det_sf}
\end{align}
is the area of the Wigner ellipse and, in the case of an isolated 
system, needs to be conserved throughout the entire mode evolution.

In \Figref{fig:ell_evol_sf}, we present an evolution scheme of modes 
that takes advantage of the invariance of the vacuum state on deep 
subhorizon scales \cite{GalvezGhersi:2016wbu, GalvezGhersi:2018haa}. 
This approximate symmetry allows us 
to avoid evolving high-frequency modes over long timescales, during 
which the state remains effectively unchanged. The relation 
$k=k_{\rm phys} a$ enables us to initialize modes (plotted in dashed 
black lines) with different comoving wavevectors after $N$ 
e-folds of background expansion. The end of mode evolution is 
denoted in the figure by the dashed green vertical, which denotes 
when the slow-roll parameter $\epsilon$ equals to one.   
The side insets of the figure illustrate the deformation of the 
Wigner ellipse for an arbitrarily chosen mode as it crosses the 
horizon. The ellipse squeezing process is characterized by the 
exponential growth of the semimajor axis and a corresponding 
contraction of the semiminor axis. Hence, horizon crossing is 
an effective way to produce squeezed states without breaking 
unitary evolution. However, this behavior is a complication for 
straightforward numerical implementations of 
Eqns.~(\ref{eq:eom_cov_vv_sf}--\ref{eq:eom_cov_pp_sf}), which are 
prone to develop instabilities if the conservation of $\mathcal{S}$ 
(\ie the area of the ellipse) is not carefully preserved during the 
exponential squeezing phase. Such instabilities can be exacerbated 
by state deformations induced by decoherence, or by increasing 
the number of degrees of freedom.

With the initial part of the formalism in place, it is possible 
to compute the power spectrum of primordial curvature fluctuations 
$\mathcal{P}_{\zeta}(k)$ by returning to the gauge-invariant 
curvature fluctuation $\hat{\zeta}_{\mathbf{k}}$
\begin{align}
\mathcal{P}_{\zeta}(k)=\frac{k^3}{4\pi^2}\left(\frac{H^2}{a^2\dot{\phi}^2}\right)\langle\hat{\Sigma}^{\mathrm{S}}_{(vv)}\rangle\,.
\label{eq:ps_sf}
\end{align}
While not practical, the corresponding transport equation for the 
curvature power spectrum can be obtained by replacing this definition 
in Eq.~\eqref{eq:sf_trans_no_acc}.

\subsection{Two-point correlators of a system coupled to an 
environment}
\label{subsec:open_sf}

Thus far, we have reviewed the fundamentals of inflation 
driven by a single scalar field. In this subsection, we follow 
the open system formalism developed in 
\cite{Martin:2016qta, Martin:2018lin, Martin:2021qkg, Martin:2021znx, Burgess:2022nwu} 
to examine the simplest form of coupling between the field 
fluctuation modes and an environment, and to analyze its impact 
on the dynamics of inflationary two-point correlators. Thus, to 
provide a preliminary notion of such an interaction, we write the 
form of the total Hamiltonian
\begin{align}
\hat{H}_{\rm tot}=\hat{H}_{\rm sys}\otimes\hat{\mathbbm{1}}_{\rm env}+
\hat{\mathbbm{1}}_{\rm sys}\otimes\hat{H}_{\rm env}+g\hat{H}_{\rm int}\,,
\label{eq:h_tot_sf}
\end{align}
where $\hat{H}_{\rm sys}$ is the system Hamiltonian introduced in 
Eq.~\eqref{eq:her_ham_sf}, $g$ is the coupling constant and 
$\hat{H}_{\mathrm{int}}$ encodes the interaction between the system 
and the environment. We assume that the interaction term is 
separable (as the rest of the Hamiltonian). Hence, it can be 
cast in the form
\begin{align}
\hat{H}_{\mathrm{int}}(\tau) = 
\int_{\mathbb{R}^{3+}}d^3\mathbf{k}~\hat{I}_{\mathbf{k}}(\tau) 
\otimes \hat{E}_{\mathbf{-k}}(\tau)\,,
\label{eq:ham_int}
\end{align}
where $\hat{I}_{\mathbf{k}}$ is an operator acting on the system 
states, while $\hat{E}_{\mathbf{-k}}$ acts on the environment states. 
The effects of such interactions are confined to specific bands of
length scales, and their spatial realizations are consistent with 
either fine- or coarse-graining. 
The interaction operators can also be decomposed into their real 
and imaginary parts in the same way as the field and momentum 
operators.

As stated in the previous subsection, the complete characterization 
of a Gaussian state is given by the expectation values of the 
covariance matrix operators. Obtaining these requires tracing over 
the environmental degrees of freedom to construct the reduced density operator
\begin{align}
\hat{\rho}_{\rm sys}=\rm{Tr}_{\rm env}\left[\hat{\rho}_{\rm tot}\right]\,,
\label{eq:reduced_density}
\end{align}
which serves as a starting point for evaluating the expectation values of the operators defined in Eqns.~(\ref{eq:cov_comp_vv}--\ref{eq:cov_comp_pp}) 
\begin{align}
\langle\hat{\mathbf\Sigma}^{\rm S}\rangle=\rm{Tr}\left[\hat{\rho}_{\rm sys}\hat{\mathbf\Sigma}^{\rm S}\right]\,,
\label{eq:exp_sf}
\end{align}
thereby yielding the quantities that fully specify the Gaussian state.

Under the assumption of a nearly stationary environment,
we consider that the evolution of covariance matrix is governed 
by the Lindblad equation in Fourier domain (see \cite{Martin:2018zbe} 
for a complete derivation)
\begin{widetext}
\begin{align}
        \frac{\mathrm{d}}{\mathrm{d} \tau}  \langle\hat{\mathbf{\Sigma}}^{\mathrm{S}}\rangle  = - i  \left\langle[\hat{\mathbf{\Sigma}}^{\mathrm{S}}, 
        \hat{H}_{\mathrm{sys}}]\right\rangle 
        - (2\pi^3)^{1/2} \Gamma \int_{\mathbb{R}^{3+}}  d^3\mathbf{k}~\tilde{C}_E(\mathbf{k},\tau)  \left\langle\left[ [\hat{\mathbf{\Sigma}}^{\mathrm{S}}, \hat{I}^{\mathrm{S}}_{\mathbf{k}}], \hat{I}^{\mathrm{S}}_{\mathbf{k}} \right]\right\rangle \,,
        \label{eq:LME_sf}
\end{align}
\end{widetext}
where $\Gamma \equiv 2g^2 \tau_{\mathrm{c}}$ is the decay
rate considering the autocorrelation time $\tau_{\rm c}$ to be small 
compared to the characteristic timescales of the environment. 
The quantity $\tilde{C}_E(\tau, k)$ denotes the power spectrum of the 
environmental interaction operator $\hat{E}$, i.e., 
$\langle|\hat{E}_{\mathbf{k}}(\tau)|^2\rangle$, assumed to be 
statistically homogeneous. This term results from tracing over the 
eigenstates of $\hat{E}_{\mathbf{-k}}$, and acts dissipatively 
due to the hermiticity of all the operators involved. A preliminary 
physical interpretation is that the environment's power spectrum 
effectively acts as a source term for the system's power spectrum.
Obtaining solutions for the general form of the interaction operators 
is a challenging task. However, in the regime where the Wigner 
function is expected to remain Gaussian throughout the evolution, the 
interaction operator reduces to a linear combination of the field and 
its conjugate momentum
\begin{align}
\hat{I}_{\mathbf{k}}^{\rm S} = \alpha(\tau)\hat{v}_{\mathbf{k}}^{\rm S}+\beta(\tau)\hat{\pi}_{\mathbf{k}}^{\rm S}\,,
\label{eq:gen_lin}
\end{align}
which greatly simplifies the derivation of transport equations 
for the covariance matrix. In this case, it suffices to solve the 
particular case $\alpha(\tau)=1$ and $\beta(\tau)=0$ 
to obtain the transport equations for a general linear combination.
Fixing $\hat{I}_{\mathbf{k}}^{\rm S} = \hat{v}^{\rm S}_{\mathbf{k}}$, 
and substituting the operator definitions from Eqns.~(\ref{eq:cov_comp_vv}--\ref{eq:cov_comp_pp}) yields
\begin{subequations}
\begin{align}
&\frac{\rm d}{\rm d\tau}\langle\hat{\Sigma}^{\mathrm{S}}_{(vv)}\rangle= 2\langle\hat{\Sigma}^{\mathrm{S}}_{(vp)}\rangle\,,\label{eq:eom_cov_vv_sf_def}\\
&\frac{\rm d}{\rm d\tau}\langle\hat{\Sigma}^{\mathrm{S}}_{(vp)}\rangle=\langle\hat{\Sigma}^{\mathrm{S}}_{(pp)}\rangle-\left(k^2-\frac{z''}{z}\right)\langle\hat{\Sigma}^{\mathrm{S}}_{(vv)}\rangle\,,
\label{eq:eom_cov_vp_sf_def}\\
&\frac{\rm d}{\rm d\tau}\langle\hat{\Sigma}^{\mathrm{S}}_{(pp)}\rangle=-2\left(k^2-\frac{z''}{z}\right)\langle\hat{\Sigma}^{\mathrm{S}}_{(vp)}\rangle +2\mathcal{F}(k,\tau)\,,\label{eq:eom_cov_pp_sf_def}
\end{align}
\end{subequations}
where $\mathcal{F}(k,\tau)\equiv(2\pi)^{3/2}\Gamma \tilde{C}_E(k,\tau)$.
After assigning a wavevector $\mathbf{k}$, it is safe to omit 
writing $\delta^{(3)}(\bf k-k')$ in the last term at the right hand side of 
\eqref{eq:eom_cov_pp_sf_def}. The case of a general linear 
combination given by Eq.~\eqref{eq:gen_lin} provides a useful 
framework for exploring momentum-dependent couplings to an 
environment. Such couplings are particularly relevant when 
constructing effective inflationary models with kinetic deformations 
\cite{Armendariz-Picon:1999hyi}. In these and other scenarios, 
one finds that the corresponding transport equations retain the same 
form as in Eqns.~(\ref{eq:eom_cov_vv_sf_def}--\ref{eq:eom_cov_pp_sf_def}) provided that the following 
variable redefinitions are applied
\begin{widetext}
\begin{subequations}
\begin{align}
&\langle\hat{\Sigma}^{\mathrm{S}}_{(vv)}\rangle\rightarrow \overline{\langle\hat{\Sigma}^{\mathrm{S}}_{(vv)}\rangle} = \langle\hat{\Sigma}^{\mathrm{S}}_{(vv)}\rangle\,,\label{eq:id_vv_gauge}\\
&\langle\hat{\Sigma}^{\mathrm{S}}_{(vp)}\rangle\rightarrow \overline{\langle\hat{\Sigma}^{\mathrm{S}}_{(vp)}\rangle} = \langle\hat{\Sigma}^{\mathrm{S}}_{(vp)}\rangle+\beta^2(\tau)\mathcal{F}(k,\tau)\,,\label{eq:id_vp_gauge}\\
&\langle\hat{\Sigma}^{\mathrm{S}}_{(pp)}\rangle\rightarrow 
\overline{\langle\hat{\Sigma}^{\mathrm{S}}_{(pp)}\rangle} = 
\langle\hat{\Sigma}^{\mathrm{S}}_{(pp)}\rangle - 2\alpha(\tau)\beta(\tau)\mathcal{F}(k,\tau)+
\frac{\rm d}{\rm 
d\tau}\left[\beta^2(\tau)\mathcal{F}(k,\tau)\right]\,,\label{eq:id_pp_gauge}\\
&\mathcal{F}\rightarrow \overline{\mathcal{F}}(k,\tau) = 
\left[\alpha^2(\tau)+\beta^2(\tau)\left(k^2-
\frac{z''}{z}\right)\right]\mathcal{F}(k,\tau)-
\frac{\rm d}{\rm 
d\tau}\left[\alpha(\tau)\beta(\tau)\mathcal{F}(k,\tau)\right]+
\frac{1}{2}\frac{\rm d^2}{\rm d\tau^2}\left[\beta^2(\tau)\mathcal{F}(k,\tau)\right]\,.\label{eq:id_F_gauge}
\end{align}    
\end{subequations}
Therefore, it is possible to consider the same transport equation 
and perform an additional change of variables to study 
other forms of coupling. 
As in the case of isolated systems (\eg 
Eq.~\eqref{eq:sf_trans_no_acc}), it is possible to condense the transport equations with respect to the redefined variables into a single third-order differential equation for the field correlator
\begin{align}
\bigg\{\frac{\rm d^3}{\rm d\tau^3}+4\left(k^2-\frac{z''}{z}\right)\frac{\rm d}{\rm d\tau}+2\left(k^2-\frac{z''}{z}\right)'\bigg\}\langle\hat{\Sigma}^{\mathrm{S}}_{(vv)}\rangle=4\mathcal{F}(k,\tau)
\,,\label{eq:sf_trans_acc}
\end{align}
\end{widetext}
where we omitted the bars for conciseness 
in the notation. This expression is consistent with the notion of the environment's power 
spectrum acting as a source term for the power spectrum of the 
system. Furthermore, if we compute the time derivative of the covariance 
matrix using the transport equation, we can gain a deeper understanding into the role 
of this source term
\begin{align}
\frac{\rm d}{\rm d\tau}\mathcal{S} = 2\langle\hat{\Sigma}^{\mathrm{S}}_{(vv)}\rangle\mathcal{F}(k,\tau)\,,
\label{eq:det_evol_sf}
\end{align}
which implies that $\mathcal{F}(k,\tau)$ controls the size of the 
Wigner ellipse. Similarly, the results in \cite{Martin:2021znx}
show that the source term affects the evolution of all squeezing 
parameters, albeit in distinct ways. Among these deformations, the 
dominant effect is the variation in the enclosed area, which offers the 
most direct signature of information exchange with the 
environment.\footnote{From a thermodynamic standpoint, such variations 
modify the number of accessible states and, consequently, the production 
of entropy for the system.}. 

\sfaccidents

So far, previous studies have proposed different forms of the 
source term $\mathcal{F}(k,\tau)$ following 
diverse physical motivations. A conservative choice
is to restrict its action to 
the ultraviolet regime \cite{Martin:2018zbe}, thereby avoiding low-frequency observational 
signatures. 
In the extreme ultraviolet limit, this choice 
can also serve as a tool to explore implications 
of the Trans-Planckian Censorship Conjecture (TCC)
\cite{Bedroya:2019snp, Brandenberger:2021pzy}. 
Extensions that incorporate infrared effects -- by 
considering a window function instead of a 
smoothened step -- have also been used to study CMB 
constraints on real-space entanglement 
\cite{Martin:2021qkg}. In contrast, our objective 
is to generate a flexible framework for generating 
events in a broad range of frequencies and 
durations. As a preliminary step, we recall that each physical 
wavelength $\ell$ observed after $N$ e-folds of inflation corresponds 
to a comoving wavenumber $k=e^N/\ell$. Similarly, for 
subhorizon evolution, the conformal time is given by $\tau=\int H^{-1}e^{-N}dN$ as a function of $N$. This allows us to recast 
the source term $\mathcal{F}(k,\tau)$ with respect to $\ln \ell$ and $N$, 
which naturally captures physical scales and their evolution during 
inflation. We then propose the following functional form
\begin{align}
&\mathcal{F}(k,\tau)=\label{eq:tiling_source}\\
&\sum^{n_{\ell}}_{j=1}\sum^{n_{N}}_{i=1} A(\ln\ell_j,N_i)~T\left(\frac{N-N_i}{\Delta N_{ij}}\right)\kappa\left(\frac{\ln\ell-\ln\ell_j}{\Delta \ell_{ij}}\right)\,,\nonumber
\end{align}
which partitions the region between the horizon and 
constant physical wavelength into an array of 
$n_{\ell}\times n_{N}=n_{\rm tiles}$
events distributed like ``tiles'' in that region of the 
$(N,\ln\ell)$ plane. The amplitude of each tile, $A(\ln\ell_j,N_i)$, is 
given by
\begin{align}
A(\ln\ell_j, N_i)=\frac{\exp[(p-1)N_i]}{(2\pi^3)^{1/2}\ell_j^2}\alpha_{\Gamma}(\ln\ell_j,N_i)\,,
\label{eq:amp_tile}
\end{align}
where $\alpha_{\Gamma}(\ln\ell_j,N_i)$ is a 
dimensionless function specifying the strength 
of the environmental effect for the tile $(i,j)$. This 
expression is obtained from the amplitude prescription for a single 
accident given, for example, in \cite{Martin:2018zbe}, after 
transforming to the $(N,\ln\ell)$ variables.
Here, $\ln\ell_j$ denotes the logarithm of the 
physical length scale at which the 
accident is most effective, $N_i$ is the 
number of e-folds where the tile is located,
and $p$ controls the steepness of the event. 
It is important to emphasize that the tiling formalism is not intended to model a specific microscopic environment, but rather to span a space of effective open-system histories consistent with Gaussian evolution. Nevertheless, tile arrays can be used to mimic concrete physical situations. For example, sharp background trajectory turns \cite{Achucarro:2014msa,Bhattacharya:2022fze} may be represented by 
transient tiles, while heavy-field thresholds \cite{Cespedes:2012hu} in natural or chaotic models can be replicated by localized tile arrays. In this sense, the tiling approach should be viewed as a diagnostic tool rather than a first-principles model of environmental interactions during inflation.
A key feature of our approach is that 
$\alpha_{\Gamma}(\ln\ell_j,N_i)$ may take positive 
and negative values, representing, respectively an 
increase or a decrease in the number of available 
states encoded in each Wigner ellipse. For the evolution 
of Gaussian states described in 
Eqns.~(\ref{eq:eom_cov_vv_sf_def}--\ref{eq:eom_cov_pp_sf_def}), 
negative values 
$\alpha_{\Gamma}$ do not violate the positivity 
of the Wigner function, except when the covariance 
matrix becomes singular. The functions $T(N)$ and $\kappa(\ln\ell)$ 
behave as window functions centered at $N_i$ and $\ln\ell_j$, 
respectively, with the widths of the corresponding length and time 
bands for the tile $(i,j)$, given by $\Delta N_{ij}$ and $\Delta \ell_{ij}$.

Using the same mode injection scheme depicted in \Figref{fig:ell_evol_sf}, in \Figref{fig:scheme_acc_sf} 
we illustrate an example configuration of decoherence accidents 
consistent with our proposal for the source term 
$\mathcal{F}(k,\tau)$ in Eq.~\eqref{eq:tiling_source}. 
Such a setup offers several possibilities of interest, for 
example: (1) it reformulates the mode-evolution problem as an 
analogue of a scattering process, in which each mode may undergo 
multiple accidents throughout its evolution, with the primary 
interest placed on the characteristics of the resulting output state; (2) it enables the preparation of arbitrary 
initial Gaussian states to investigate the emergence of features 
in the primordial power spectrum of curvature fluctuations across 
different frequencies; (3) and it accommodates more general 
configurations that capture the effects of an evolving environment 
across different scales. With these possibilities in mind, we must 
design optimized numerical routines that guarantee stable solutions 
to the transport equations without compromising computational 
efficiency. The implementation and validation of our approach are 
detailed in the following subsection.

\Figref{fig:scheme_acc_sf} reveals that, in principle, it is possible to configure decoherence sequences containing accidents in which $\alpha_\Gamma$ becomes negative. This regime admits several interpretations, ranging from apparent violations of the uncertainty principle to effective theories defined on specific projections of the Wigner ellipse rather than in the complete phase-space volume. These and other scenarios may also be incorporated with the purpose of controlling initial state deformations in a more precise way.

\subsection{Implementation of fast and slow component separation 
for single field models}
\label{subsec:FSCS}
Until now, we summarized the main steps required to derive 
the transport equations in (\ref{eq:eom_cov_vv_sf_def}--\ref{eq:eom_cov_pp_sf_def}), which independently evolve each 
Fourier mode of the two-point correlators in the presence 
of decoherence events. These equations can be recast in terms of
the original correlators 
\begin{widetext}
\begin{subequations}
\begin{align}
 &\frac{\rm d}{\rm d\tau} \langle\hat{v}_{\rm S},\hat{v}_{\rm S}\rangle=\langle\hat{v}_{\rm S},\hat{\pi}_{\rm S}\rangle  +\langle\hat{\pi}_{\rm S},\hat{v}_{\rm S}\rangle\label{eq:eom_acc_sf_vv}\,,\\
 &\frac{\rm d}{\rm d\tau} \left[\langle\hat{v}_{\rm S},\hat{\pi}_{\rm S}\rangle+\langle\hat{\pi}_{\rm S},\hat{v}_{\rm S}\rangle\right]=2\left[\langle\hat{\pi}_{\rm S},\hat{\pi}_{\rm S}\rangle  -\left(k^2-\frac{z''}{z}\right)\langle\hat{v}_{\rm S},\hat{v}_{\rm S}\rangle\right]\,,\label{eq:eom_acc_sf_vp}\\
 &\frac{\rm d}{\rm d\tau} \langle\hat{\pi}_{\rm S},\hat{\pi}_{\rm S}\rangle=-\left(k^2-\frac{z''}{z}\right)\left[\langle\hat{v}_{\rm S},\hat{\pi}_{\rm S}\rangle+\langle\hat{\pi}_{\rm S},\hat{v}_{\rm S}\rangle\right]+\mathcal{F}(k,\tau)\,.\label{eq:eom_acc_sf_pp}
\end{align}
\end{subequations}
\end{widetext}
A crucial aspect in our analysis is that all the 
operators are Hermitian, and their corresponding 
expectation values are real. This allows  
substituting the quantum field operators by  
rescaled Gaussian random variables spanning 
the spectrum of real field values. In the case of 
$\hat{v}_{\rm S}$, this substitution yields 
\begin{align}
\hat{v}_{\rm S} \rightarrow L_k\tilde{\chi}_{\rm S}\,,
\label{eq:v_sf_rv}
\end{align}
where the quantity $\tilde{\chi}$ is a unit Gaussian random variable, and in 
the isolated case, each possible expectation value is a solution 
to the mode equations in (\ref{eq:sf_eom_v}--\ref{eq:sf_eom_p}).\footnote{Note that these 
substitutions concern only the real expectation 
values inside the Wigner ellipse, not the usual 
complex mode expansion of quantum field theory.} 
By direct substitution in the Lindblad equation, it is 
straightforward to check that this equations also govern the 
dynamics of one-point functions.  
$L_k$ is a real rescaling factor, which is consistent with the field 
amplitude. In the same way, we compute the corresponding 
form of the momentum operator 
\begin{align}
\hat{\pi}_{\rm S} \rightarrow L_k'\tilde{\chi}_{\rm S}+L_k\tilde{\chi}_{\rm S}'\,,
\label{eq:p_sf_rv}
\end{align}
where, in contrast with previous work -- \eg \cite{Sasaki:1995aw} --, we 
consider dynamical random variables. Normalization of random 
variables introduces a gauge condition 
\begin{align}
\langle\tilde{\chi}_{\rm S},\tilde{\chi}_{\rm S}\rangle =1\,,
\label{eq:gauge_sf}
\end{align}
which, when replaced in Eq.~\eqref{eq:eom_acc_sf_vv} together with 
the field ansatz in \eqref{eq:v_sf_rv} and \eqref{eq:p_sf_rv}, 
yields the following consistency relation
\begin{align}
\frac{\rm d}{\rm d\tau}\langle\tilde{\chi}_{\rm S},\tilde{\chi}_{\rm S}\rangle = \langle\tilde{\chi}'_{\rm S},\tilde{\chi}_{\rm S}\rangle+\langle\tilde{\chi}_{\rm S},\tilde{\chi}'_{\rm S}\rangle=0\,,
\label{eq:gauge_fixing_sf}
\end{align} 
which imposes the gauge condition at all times. 
This normalization condition ensures that the
field correlator $\langle\hat{v}_{\rm S},\hat{v}_{\rm S}\rangle$ 
remains equal to $L_k^2$ at all times. In a way reminiscent of mode 
evolution for isolated systems, the evolution of $L$ alone is sufficient to determine this correlator. At the same time, 
as noted in the previous subsection, numerically evolving the 
field correlators via Eqns.~(\ref{eq:eom_acc_sf_vv}--\ref{eq:eom_acc_sf_pp}) can develop instabilities. 
A natural next step is therefore to reformulate the transport 
equations in terms of $L_k$ and other derivatives of the 
random-variable correlators. To proceed with this step, we substitute 
into Eq.~\eqref{eq:eom_acc_sf_vp} the field redefinitions, the gauge 
condition in \eqref{eq:gauge_sf}, and its consistency relation in 
\eqref{eq:gauge_fixing_sf}, obtaining
\begin{align}
\left[\frac{\rm d^2}{\rm d\tau^2}+\left(k^2-\frac{z''}{z}\right)-\langle\tilde{\chi}'_{\rm S},\tilde{\chi}'_{\rm S}\rangle\right]L_k=0\,.
\label{eq:eom_L_sf}
\end{align}
To close the system, we need to find the equation of motion for the 
correlator $\langle\tilde{\chi}'_{\rm S},\tilde{\chi}'_{\rm S}\rangle$. This follows from substituting the above relations, together with 
Eq.~\eqref{eq:eom_L_sf}, into Eq.~\eqref{eq:eom_acc_sf_pp}:
\begin{align}
\frac{\rm d}{\rm d\tau}\langle\tilde{\chi}'_{\rm S},\tilde{\chi}'_{\rm S}\rangle = -\frac{4L_k'}{L_k}\langle\tilde{\chi}'_{\rm S},\tilde{\chi}'_{\rm S}\rangle+\frac{1}{L_k^2}\mathcal{F}(k,\tau)\,.
\label{eq:eom_xp_xp_sf}
\end{align}
We thus arrive at a second-order system of 
ordinary differential equations -- rather than the third-order 
equation in Eq.~\eqref{eq:sf_trans_acc} -- representing a minor 
modification to the Ermakov-Pinney nonlinear equation used in 
\cite{GalvezGhersi:2018haa} to evolve correlators, 
and is analog to a mode decomposition in polar 
coordinates. Notably, although real Gaussian random variables commute, this property 
is not required for deriving Eqns.~\eqref{eq:eom_L_sf} and 
\eqref{eq:eom_xp_xp_sf}. 

\sfrndm

A complete description of the dynamical system 
requires initial conditions, set at the surface of 
constant physical wavelength
\begin{subequations}
\begin{align}
&L_k|_{\tau=\tau_0} = \left[4\left(k^2-\frac{z''}{z}\right)\right]_{\tau=\tau_0}^{-1/4}\label{eq:L_sf_init_cond}\,,\\
&L_k'|_{\tau=\tau_0} = 0\,,\label{eq:Lp_sf_init_cond}\\
&\langle\hat{\chi}'_{\rm S},\hat{\chi}'_{\rm S}\rangle|_{\tau=\tau_0}=k^2-\frac{z''}{z}\bigg|_{\tau=\tau_0}\,,
\end{align}
\end{subequations}
which are built to match standard Minkowski vacuum deep 
inside the horizon. By construction, the choice for the ``velocity'' correlators $\langle\tilde{\chi}'_{\rm 
S},\tilde{\chi}'_{\rm S}\rangle$ makes the 
oscillation frequency in Eq.~\eqref{eq:eom_L_sf} 
vanish exactly at $\tau=\tau_0$. Moreover, these correlators evolve according to
first-order differential equations, making them 
computationally inexpensive to solve. 
This combination -- the suppression of the net 
oscillation frequency and the reduced 
cost of computing random-variable velocity correlator dynamics -- damps oscillations in the evolution of $L_k$,
and permits larger time steps when resolving field 
correlators on deep subhorizon scales. With 
these initial conditions, we obtain the evolution of the velocity correlators 
\begin{align}
&\frac{\langle\tilde{\chi}'_{\rm 
S},\tilde{\chi}'_{\rm S}\rangle}{\langle\tilde{\chi}'_{\rm 
S},\tilde{\chi}'_{\rm S}\rangle|_{\tau_0}} = \frac{L^4_k|_{\tau_0}}{L^4_k}\left[1+\Delta_p(k,\tau)\right]\,,\label{eq:phase_vel_sf}\\
&\Delta_p(k,\tau)= \int^{\tau}_{\tau_0}d\tau' \frac{ L^2_{k} \mathcal{F}(k,\tau')}{L^4_k|_{\tau_0}  \langle\tilde{\chi}'_{\rm 
S},\tilde{\chi}'_{\rm S}\rangle|_{\tau_0}}\,,\nonumber
\end{align}
where $\Delta_p(k,\tau)$ captures the contribution from the particular solution of Eq.~\eqref{eq:eom_xp_xp_sf}, that the source term generates. From this result, we derive the equation of motion for the 
correlation factors $L_k$:
\begin{align}
&\bigg[\frac{\rm d^2}{\rm d\tau^2}+\left(k^2-\frac{z''}{z}\right)-\frac{L^4_k|_{\tau_0}}{L^4_k}\langle\tilde{\chi}'_{\rm 
S},\tilde{\chi}'_{\rm S}\rangle|_{\tau_0}\label{eq:eom_L_sf_EFT}\\
&-\frac{L^4_k|_{\tau_0}}{L^4_k}\langle\tilde{\chi}'_{\rm 
S},\tilde{\chi}'_{\rm S}\rangle|_{\tau_0}\Delta_p(k,\tau)\bigg]L_k=0
\nonumber
\,.
\end{align}
The first line reproduces the usual operator 
form of the Ermakov-Pinney equation in the case of 
isolated systems. The extra 
term proportional to $\Delta_p$ actively modifies the dynamics of $L_k$: it may not only reproduce 
effects usually attributed to interactions with 
external fields, but it can also mimic changes in 
the background geometry, such as those arising from 
the expansion history, via a wavenumber-dependent 
redefinition of $z''/z$ 
\begin{align}
&\bigg[\frac{\rm d^2}{\rm d\tau^2}+\left(k^2-\frac{z_{\rm eff}''}{z_{\rm eff}}\right)-\frac{L^4_k|_{\tau_0}}{L^4_k}\langle\tilde{\chi}'_{\rm 
S},\tilde{\chi}'_{\rm S}\rangle|_{\tau_0}\bigg]L_k=0
\nonumber\,,\\
&\frac{z_{\rm eff}''}{z_{\rm eff}}\equiv \frac{z''}{z}+\frac{L^4_k|_{\tau_0}}{L^4_k}\langle\tilde{\chi}'_{\rm 
S},\tilde{\chi}'_{\rm S}\rangle|_{\tau_0}\Delta_p(k,\tau)\label{eq:eff_zpp_z}\,.
\end{align}
In this way, the open-systems framework provides a versatile tool to 
build effective descriptions of inflationary dynamics 
\cite{Salcedo:2024smn}. The dependence of the last expression on $L_k$
suggests that the description is scale-dependent and may also 
be potentially affected by backreaction effects.

Based on the description of the system's dynamical
variables, we recast the form of the covariance 
matrix $\langle \Sigma^{\rm S}\rangle$ 
-- as defined in Eq.~\eqref{eq:Covariance_Matrix} --
in terms of the correlation amplitude $L_k$ and 
the velocity correlators for an arbitrary $k$
\begin{align}
\langle \hat{\mathbf{\Sigma}}^{\rm S}\rangle &= \left[\begin{array}{cc}
\langle\hat{\Sigma}^{\rm S}_{(v v)}\rangle & \langle\hat{\Sigma}^{\rm S}_{(v p)}\rangle \\
    \langle\hat{\Sigma}^{\rm S}_{(v p)}\rangle & \langle\hat{\Sigma}^{\rm S}_{(p p)}\rangle
    \end{array}\right]\label{eq:cov_mat_sf_L}\,,\\ 
    &= 2\left[\begin{array}{cc}
L^2_k & L_kL'_k \\
    L_kL'_k & (L'_k)^2+L^2_k\langle\tilde{\chi}'_{\rm 
S},\tilde{\chi}'_{\rm S}\rangle\end{array}\right]\,,\nonumber
\end{align} 
and find its determinant
\begin{align}
\mathcal{S}= 4L^4_k\langle\tilde{\chi}'_{\rm 
S},\tilde{\chi}'_{\rm S}\rangle\,,
\label{eq:det_sf_acc}
\end{align}
which, in the absence of decoherence events, 
needs to be conserved through the entire 
evolution. The derivative of the determinant in
Eq.~\eqref{eq:det_evol_sf} naturally motivates the 
interpretation of the source term
$\mathcal{F}(k,\tau)$ as an analogue of a torque in 
classical mechanics,
responsible for inducing a sudden change in the 
angular momentum. In the present context, the 
counterpart of the angular momentum (squared)
is given by the determinant of the covariance 
matrix. 

After setting up the equations of motion, 
initial conditions and the array 
of decoherence events represented as ``tiles'', the results in \Figref{fig:imp_rndm_acc_sf} present 
the mode evolution scheme for an arbitrary array of decoherence and recoherence events  (panel (a)). The effects 
of this configuration are further represented in panel (b), in the form of features in the power spectrum of primordial fluctuations, where it is compared with the accident-free case. 
After selecting a few representative modes from the injection scheme in panel (a), panel (c) 
displays modifications in the mode evolution at fixed values of $k$, which arise directly from the introduction of accidents at subhorizon scales. To check if mode evolution is stable, panel 
(d) illustrates the evolution of 
the Wigner ellipse area -- \ie $\mathcal{S}$ --
for each Fourier mode, which also makes evident the gain 
or loss of area as the modes traverse the 
tiles. 

As indicated in Eq.~\eqref{eq:amp_tile}, the magnitude of the accident represented by 
each tile of panel (a) is governed by both the parameter $\alpha_\Gamma$ and the dependence of $A(\ln\ell_j,N_i)$ on the scale factor via $p$. In this specific case, we considered $\alpha_\Gamma\in[-0.37;0.29]$ 
in the low-$k$ regime, and values up to $\alpha_\Gamma=3.3$ in the high-$k$ tail, close to the end of inflation. Negative values of $\alpha_\Gamma$ are of particular interest, as they can be interpreted as modeling  
entropy outflows from the system to the environment. However, these 
scenarios require careful treatment, since
such values -- or the superposition of negative-valued tiles active at different times -- may also drive the determinant towards cancellation, thereby rendering it 
ill-defined. 

To specify the functional forms of $T$ and $\kappa$ in Eq.~\eqref{eq:tiling_source}, we employed cosine tampered window functions. Nonetheless, the implementation is sufficiently general to allow alternative shapes, enabling us to model smooth or sharp accidents across 
the $(N,\ln \ell)$ plane. Regarding the values of $p$, we found $p\sim 3$ adequate to control the relevant accident magnitudes, and thus adopted $p\in[2.8;3.1]$. 
In contrast with the schematic tiling approach suggested 
in \Figref{fig:scheme_acc_sf}, we assigned different positions, widths and heights for each tile in order to demonstrate that the approach is flexible enough to accommodate a 
variety of
possible distributions of decoherence events.

\Figref{fig:imp_rndm_acc_sf}, panel (b) illustrates how different arrays of
tiles can be used to generate distinct features in the power spectrum 
across various ranges of $k$. Such features can be exploited in 
phenomenological or statistical studies. For comparison, the dotted 
black line corresponds to the accident-free case, obtained from the 
nonlinear potential $V(\phi)=\lambda\phi^4/4$ with 
$\lambda=2\times10^{-14}$ and initial conditions chosen to produce 
73 e-folds of inflation (approx.).  
Relative to this reference, the presence of accidents can either 
enhance or suppress the power in specific wavenumber bands (expressed 
in units of $k_{\rm phys}=10^{3}H_0$, with $H_0$ fixed 
as an initial condition). Colored dots, following the color code in 
the legend, indicate the contribution of individual modes to the final 
power spectrum. The numerical implementation permits any other single-field inflationary 
setup to produce features displayed in more intricate models of 
inflation.

\sfcmpnstd

Modifications in the subhorizon evolution of modes for various 
wavenumbers $k$ are presented in \Figref{fig:imp_rndm_acc_sf}, panel (c). 
Decoherence effects displace the modes from their unperturbed 
trajectories, leading to either an amplification or suppression of 
power by several orders of magnitude. For single-field inflation, no 
transfer of power occurs after horizon crossing, and hence no further 
deviations arise. As we will explicitly show by the end of this section, 
the open-system formalism can be used to reproduce specific features 
in the primordial power spectrum, which are commonly attributed to multifield 
models. An additional application of this framework is the controlled 
generation of deformed states, where short evolution timescales combined with 
ad-hoc sequences of decoherence events are sufficient to deform the Wigner ellipse as 
required. Furthermore, the results of this panel provide indirect evidence that 
introducing multiple decoherence events does not necessarily lead to pathological 
behavior in their subsequent mode evolution. This can be understood from 
the definition of $P_\zeta$, which is proportional to $L^2_k$
(from Eqns.~\eqref{eq:ps_sf} and \eqref{eq:cov_mat_sf_L}). Since no complex 
conjugation is involved, the smooth behavior observed in panel (c) directly reflects 
the dynamics of $L_k$ within the fast-slow separation scheme. 
In contrast to the standard mode function approach, where the same smoothness arises 
only from algebraic phase cancellation between complex-conjugate contributions, here 
the suppression of fast oscillations in the spectrum evolution is a dynamical effect.

The evolution of the Wigner ellipse area is plotted in \Figref{fig:imp_rndm_acc_sf}, panel (d) for the 
same $k$-modes selected in the 
mode injection scheme. The reference mode at $\mathcal{S}=1$ 
(labeled as ``no accidents'') 
confirms the conservation of the determinant in the absence of 
accidents. Among the other modes, some exhibit growth while others show
a reduction in the determinant, yet none display signs of numerical 
instabilities through their evolution. As noted earlier, cases with negative $\alpha_{\Gamma}$ (and their superpositions) require careful consideration to account for possible cancellations of 
the velocity correlator $\langle\hat{\chi}'_{\rm S},\hat{\chi}'_{\rm S}\rangle$ in the determinant, since the numerical evolution breaks as soon as $\mathcal{S}$ vanishes.

Intermittent decoherence events may also occur without leaving 
observable traces in the final power spectrum. Nevertheless, they can 
produce transient modifications of the field variance and thereby lead 
to nonlinear effects during inflation, including a possible enhancement 
of non-Gaussianities in models with more complex potentials. 
To remain compatible with the lack of spectral imprints, their influence 
on mode evolution must be reversible, which is now possible since we 
consider accidents with positive or negative values of $\alpha_\Gamma$. 
To configure this type of accidents, it is convenient to reparametrize 
the source term $\mathcal{F}(k,\tau)$ so as to control the specific 
modes in which the accidents are effective,
\begin{align}
&\mathcal{F}(k,\tau)=\label{eq:tiling_source_mod}\\
&\sum^{n_{\ell}}_{j=1}\sum^{n_{N}}_{i=1} A(\ln\ell_j,N_i)~T\left(\frac{k-k_i}{\Delta k_{ij}}\right)\kappa\left(\frac{\ln\ell-\ln\ell_j}{\Delta \ell_{ij}}\right)\,,\nonumber
\end{align}
where $\Delta k_{ij}$ is the width of the wavenumber band for the tile $(i,j)$. This reparameterization has the effect of transforming the square tiles in \Figref{fig:scheme_acc_sf} into parallelograms by changing the argument 
of the window function $T$, and is only a minimal modification of the original prescription in Eq.~\eqref{eq:tiling_source} 
due to the definition of comoving wavenumbers $k_i=k_{\rm phys}e^{N_i}$.
As a next step, we construct an array of decoherence and recoherence 
events with varying durations, each spanning distinct wavelength bands. 
To render the overall transformation reversible, each accident needs to 
be paired with an anti-accident, in which $\alpha_\Gamma$ takes the 
opposite sign, so that the determinant is reset to its initial value. 
In \Figref{fig:cmpnstd_acc_sf}, we illustrate a concrete 
implementation of this idea through an arbitrary configuration 
of reversible decoherence events, where the accidents ($\rm A$) and 
anti-accidents ($\rm \bar A$) are distributed across 
the mode injection scheme. Panel (a) shows the distribution of 
accident/anti-accident pairs and the $k$-modes that traverse them. 
Panel (b) depicts how the original configuration of accidents deforms 
the curvature power spectrum, and how this distortion is suppressed by 
the anti-accidents, restoring the spectrum to its original form. Panels 
(c1) and (c2) show the evolution of the squeezing 
parameter variations $\Delta r_k$ and $\Delta \varphi_k$, defined 
with respect to their corresponding accident-free values $r^{(0)}_k$ and 
$\varphi^{(0)}_k$, with the purpose of evaluating perturbations in the 
shape of the Wigner ellipse. Panel (d) shows the evolution of the determinant 
$\mathcal{S}$, which is consistent with the cancellation of the 
effects produced by accident tiles $(\rm A)$. 

The premise in designing the anti-accident is that its setup requires knowing in advance how the 
original decoherence event modifies the covariance matrix determinant, so that this modification can be 
cancelled. There is not a unique way to cancel the effects of a decoherence event, for simplicity we assumed that the 
anti-accidents in \Figref{fig:cmpnstd_acc_sf}, panel (a), share the same duration and act over the same 
span of frequencies as the corresponding accidents. After this simplification, $\alpha_\Gamma$ is the only parameter that needs to be determined to complete the configuration of a 
counterterm. Knowing the value of $\mathcal{S}$ after crossing the original accident at the value of $k$ where 
its effect is maximal, we determined the amplitude of the anti-accident by bisection, taking advantage of the 
fact that the numerical routine is sufficiently fast to evaluate each iteration without a significant 
computational cost, needing only a few iterations to converge. 
Some of the additional parameters in this configuration that can be tuned without producing major alterations 
in the final outcome are the time interval separating an accident from its anti-accident (set to zero in 
our case), the duration of the anti-accident (which can shift its required amplitude) or the shape of the time 
and frequency windows (which has to be equally changed for both the accident and the anti-accident). 

\flowersf

The red curve in \Figref{fig:cmpnstd_acc_sf}, panel (b), shows that the shape of the power 
spectrum is indistinguishable from the case free of accidents. The contrast with the accident-only case 
(plotted with a black dotted curve) confirms that the anti-accidents $(\bar{\rm A})$ restore the spectrum to 
its unperturbed form (which is also visible in the black dotted curve in \Figref{fig:imp_rndm_acc_sf}, panel 
(b)). Although not shown here, we verified that the mode evolution is modified only during the time interval in 
which the accident/anti-accident pair is active. Although the 
impact of these reversible decoherence events is not visible in the 
primordial power spectrum, its effects are visible for higher 
moments of the distribution at the level in which its dynamics is 
treated at nonlinear level. We leave the study of potential effects in the bispectrum for a future project.

As mentioned before, the evolution of correlators fully characterizes deformations of the Wigner ellipse 
through the squeezing parameters
\begin{align}
&\cosh{(2r_k)}=\frac{1}{2}\left[\frac{k^2+\langle\tilde{\chi}'_{\rm 
S},\tilde{\chi}'_{\rm S}\rangle+\left(\frac{L'_k}{L_k}\right)^2}{k\langle\tilde{\chi}'_{\rm 
S},\tilde{\chi}'_{\rm S}\rangle^{1/2}}\right]\,,\label{eq:squeezing_sf}\\
&\tan{(2\varphi_k)}=\frac{2kL'_kL_k}{k^2L_k^2-(L'_k)^2-L^2_k\langle\tilde{\chi}'_{\rm 
S},\tilde{\chi}'_{\rm S}\rangle}\,,\label{eq:angle_sf}
\end{align}
which are only different from the standard definitions by a rescaling factor of $k$. Using the evolution of 
field correlators at fixed wavenumbers, \Figref{fig:cmpnstd_acc_sf}, panels (c1) and (c2), illustrate 
how the squeezing parameters evolve away from their accident-free values as the $k$-modes pass through the 
accident/anti-accident pairs, and continue evolving up to the end of inflation.\footnote{More explicitly, deviations are measured with respect to the squeezing parameters $\varphi^{(0)}_k$ and $r^{(0)}_k$, which arise from computing the evolution of correlators in the accident-free scenario, defined by $\mathcal{F}(k,\tau)=0$.} From both subpanels, it is clear that the most significant 
effects occur when the modes traverse each individual (anti-) accident within the pair. 
As an ancillary file attached to this paper \cite{supplemental2}, 
we include an animation showing the evolution of the Wigner ellipse and its projections 
over the normalized axes $k^{-1/2}\pi_k$ and $k^{1/2}v_k$, as an arbitrary $k$-mode traverses a decoherence event to produce a squeezed state. While arbitrary arrays of decoherence events 
can be used to generate squeezed states, they are more effective at inducing variations in the area of 
the Wigner ellipse than at producing strong squeezing. 

Inset plots reveal that the variations in the phase $\Delta\varphi_k$ (visible in panel (c1), in solid light blue) track the time 
dependence of the source term amplitude, while the squeezing parameter $\Delta r_k$ (in panel (c2), in solid blue) develops 
peaks that fade once the (anti-) accident is turned off. Another characteristic feature visible in the 
insets is the appearance of oscillations in both squeezing parameters immediately after the modes cross the 
accident/anti-accident pairs. These oscillations are sourced by small variations in the eccentricity of the 
Wigner ellipse, which induce a rotation phase that manifests as a slight ``wobble''. 
This wobble gradually fades as the modes approach horizon crossing.
Such a behavior is not exclusive of the accident/anti-accident pair, since we 
observe that it also manifests after crossing individual accidents, regardless if $\alpha_\Gamma$ is positive 
or negative.

The evolution of the Wigner ellipse area is plotted in \Figref{fig:cmpnstd_acc_sf}, panel 
(d)). It demonstrates that the determinant $\mathcal{S}$ can vary by several orders of magnitude before  
returning to a unit circle. For comparison, dotted lines correspond to the scenarios without anti-accidents, 
in which the determinant settles at a final value different from the original. This contrasts with the behavior 
of the squeezing parameters, whose 
deformations essentially vanish once the 
accidents are no longer active. The widths 
of the peaks visible in the determinant 
evolution depend not only on the smoothness 
of the accidents, but also on the time 
interval separating an accident from its 
corresponding anti-accident. 

\mfreprod

Thus far, we have set initial 
conditions on a surface of constant 
$k_{\rm phys}$ to match the Minkowski 
vacuum, with the objective of studying 
deformations of a unit circle. In future 
work, we may also consider squeezed states 
to explore more general scenarios. This case 
requires careful attention, since the computational 
cost increases due to the rotation velocity 
of the ellipse, which doubles the 
mode-oscillation frequency and is not 
suppressed by the fast-frequency separation 
scheme introduced here. However, we have 
checked that the system can still be solved 
accurately without introducing instabilities 
in the determinant evolution, using the 
equations of motion for 
$\langle\tilde{\chi}'_{\rm S},\tilde{\chi}'_{\rm S}\rangle$ and $L_k$, given in 
Eqns.~\eqref{eq:eom_L_sf} and 
\eqref{eq:eom_xp_xp_sf}.

The numerical evolution remains stable 
enough to accommodate arbitrarily large 
arrays of tiles at minimal computational 
cost, allowing us to implement accidents 
acting at different times during mode 
evolution and across various $k$ ranges. 
As illustrated in the left panel of 
\Figref{fig:flwr_pwr}, we return to the 
original source-term parameterization  of 
Eq.~\eqref{eq:tiling_source} (with squared tiles) to 
collocate a $48\times48$ array of 
decoherence and recoherence events arranged in a sunflower pattern. The accident distribution is loaded 
into the code as a bitmap, and amplitudes 
must be chosen carefully to avoid spurious 
determinant cancellations, an issue avoided 
by restricting $\alpha_\Gamma>0$. The computational implementation  
permits to use either the square or the parallelogram tile configurations. 
Computational efficiency is achieved by 
updating the source term only for the decoherence (and recoherence)
tiles that contribute effectively in the evolution of each mode, 
making the scheme scalable. The smaller 
array located in the upper-right corner of 
the sunflower pattern illustrates the 
duration and wavelength 
coverage of individual accidents. Despite 
the apparent smallness of each tile, their 
interaction times exceed the mode-oscillation 
scale. These are still consistent with the 
timescales of the field and environment 
operators in which the form of the Lindblad 
equation in Eq.~\eqref{eq:LME_sf} is  
valid. 

With these considerations, this setup enables the study of 
decoherence in close analogy with a scattering problem 
involving measurements of the ``out'' states from multiple targets.
Such arrays of decoherence events can also be viewed as analogues of arbitrary 
pumping protocols in quantum optics, deforming the Wigner ellipse through changes in its area and degree 
of squeezing, and thereby inducing deformations in the curvature spectrum.

In the right panel of \Figref{fig:flwr_pwr},
we depict the deformations of the primordial 
power spectrum caused by the sunflower  
configuration of tiles displayed in the 
left panel. As in 
\Figref{fig:imp_rndm_acc_sf}, panel (b), our 
results show that it is possible to produce 
features that may either enhance or suppress 
power in any band of wavenumbers. This behavior is 
consistent with the effective field theory 
approach to inflation and is particularly 
relevant for mimicking effects of radiatively 
stable models of inflation that exhibit 
different forms 
of features (\eg \cite{Dimopoulos:2005ac, Achucarro:2014msa, Konieczka:2014zja, Renaux-Petel:2015mga, Bhattacharya:2022fze}). It should be emphasized that, 
although some of the power spectra depicted in this work are not consistent 
with current observational constraints, their role is to exemplify the flexibility 
of the framework in accommodating a wide variety of feature profiles at different 
wavelengths.

The results shown in Figures \ref{fig:cmpnstd_acc_sf} and \ref{fig:flwr_pwr} suggest 
that decoherence accidents can be configured to reconstruct the primordial power 
spectra of more intricate inflationary models. As an example of this, in 
\Figref{fig:mf_inst_reprod}, to illustrate the diagnostic power of this formalism, we construct a sequence of decoherence events designed to 
reproduce the characteristic features of a multifield inflationary model mode by mode. 
The power spectrum of such a model exhibits multiple deformations induced by 
instabilities sourced by the field-space curvature. As an example, we produced 
the spectrum of a two-field model where the background dynamics is governed by the
potential
\begin{align}
V(\phi,\sigma)=\frac{\lambda}{4}\phi^4+\frac{g}{2}\phi^2\sigma^2\,,
\label{eq:double_lp4_sfs}
\end{align}
where $\lambda=10^{-14}$ and $g=2\lambda$ are the model parameters. In addition to this, the background trajectories deviate from their convergence path toward 
the attractor due to the following field-space diagonal metric
\begin{align}
h_{AB}=\delta^\phi_A\delta^\phi_B+A_{\sigma}\sin\left(\frac{\phi}{f_{\phi}}\right)\delta^\sigma_A\delta^\sigma_B\,,
\label{eq:met_sf}
\end{align}
where chose $A_\sigma = 0.75$ and $f_\phi = 10^{-1} M_{\rm Pl}$ as the metric parameters. 
This field-space geometry causes the background trajectories to oscillate rather than converge 
monotonically. Consequently, these oscillations induce time-dependent modifications to the 
effective oscillation frequency in the mode equations through the associated field-space curvature 
\cite{Renaux-Petel:2015mga}.
Using the parallelogram parameterization of tiles introduced in 
Eq.~\eqref{eq:tiling_source_mod}, we adopt the same quartic single-field potential used 
throughout this section as a baseline and modify the mode evolution through the 
inclusion of decoherence events. Following the same procedure used for constructing 
accident/anti-accident pairs, and taking the power of each $k$-mode of the multifield 
spectrum as the target, we apply a bisection method to determine the values of 
$\alpha_\Gamma$ required to reproduce the desired power. The left panel of this figure 
shows the pumping protocol constructed for this task, which is not unique. 
The inset shows a two-layer structure of overlapping tiles: the first layer of tiles 
provides a crude approximation to the target power 
value, while the second layer refines the result. The use of overlapping tiles improves 
the smoothness of the reconstructed spectrum. 

The right panel compares the 
curvature spectrum obtained from this deformation procedure with both the original and 
target spectra. The results shown in this figure illustrate
degeneracies: initial-state deformations in a simple inflationary scenario and 
dynamical features of a more intricate model can, in principle, produce (nearly) 
indistinguishable primordial power spectra. 
Although the resulting spectrum exhibits small but non-negligible 
deviations from the multifield spectrum, this indicates that more precise convergence 
techniques 
-- such as those based on machine learning algorithms -- may be required to reduce 
these discrepancies to machine precision.  Rather than providing a new explanation for observed features, these results underscore the difficulty of uniquely attributing spectral signatures to background dynamics in the absence of independent constraints on environmental couplings, and highlight the limitations of inference based solely on the power spectrum.

\section{Numerical implementation of decoherence in 
multifield inflation}
\label{sec:Multi_deco}

The objective of this section is to apply the open-system 
formalism introduced in \secref{sec:OQSF} to a linearized 
$\mathcal{N}$-field system, where fields interact among themselves while the system is also coupled to an environment. 
We also extend 
the fast-slow separation scheme to solve 
the transport equations for the two-point correlation 
functions of multiple fields and their 
conjugate momenta. One of the main goals is to capture the 
cross-correlations between fields to observe how decoherence 
effects propagate across them. Consequently, these effects 
are also reflected in the evolving projections of the 
2$\mathcal{N}$-dimensional Wigner ellipse.

\backmultinlin

\subsection{Review of perturbation dynamics for multifield inflation}
\label{subsec:Lindblad_multi}

As a first step, it is essential to describe the evolution of a multifield inflationary 
model within the framework of first-order perturbation theory. Field dynamics 
is governed by the action
\begin{align}
    S=\int \mathrm{~d}^4x\sqrt{-g}\left(\frac{M_{\text{Pl}}^2}{2}R-\frac{1}{2}\nabla_\mu\phi_{A}\nabla^\mu\phi^{A}-V(\phi_{A})\right)\,,
    \label{eq:action_mf}
\end{align}
which is a straightforward generalization of the single-field action in Eq.~\eqref{eq:sf_action}. Here, capital Latin indices denote each of the 
$\mathcal{N}$ scalar fields, which are raised and lowered  
by the field metric $h_{AB}$. In general, this field-space metric 
is a function of the background fields. 
As for the equations 
of motion for the background fields, these are also 
an extension of the single-field scenario 
\begin{align}
\ddot{\phi}_A+3H\dot{\phi}_A+\frac{\partial V(\phi_B)}{\partial\phi_A}=0\,,
\label{eq:bck_mf}
\end{align}
Throughout this section, we focus on the 
two-fields nonlinear potential used by the end of the last 
section in Eq.~\eqref{eq:double_lp4_sfs}, which serves as a test case for the
transport evolution scheme of two-point 
correlators. In \Figref{fig:back_multi_lp4}, 
we show the evolution of several
background trajectories, together with a map 
of the number of e-folds of inflation 
generated for different initial conditions 
with $\lambda=10^{-14}$ and $g=2\lambda$ as 
model parameters. 
All trajectories start with sufficiently high 
values of the potential, with vanishing 
initial velocities 
$(\dot{\phi}=\dot{\sigma}=0)$, and evolve 
until the end of inflation.
As in the single-field case, we identified 
the end of inflation by the slow-roll 
condition $\epsilon=1$.

As for the dynamics of field perturbations,
we adopt spatially flat spacetime coordinates,
defined as a deformation of the flat FLRW metric,
\begin{align}
g_{\mu\nu}=-(1+2A)\delta^t_\mu\delta^t_\nu-2a^2B_{,i} \delta^t_\mu\delta^i_\nu+a^2\delta_{ij}\delta^i_\mu\delta^j_\nu\,,
\label{eq:sf_coord}
\end{align}
where $A$ and $B$ are scalar metric 
perturbations. In these coordinates, the field
perturbations $\delta\phi_A(\mathbf{x},t)$ evolve 
according to the action in Eq.~\eqref{eq:action_mf} expanded up to second 
order:
\begin{align} 
    S_{2} =& \frac{1}{2} \int \mathrm{d}^{4}x \, a^{3} \bigg( \dot{\delta\phi_{A}} \dot{\delta\phi^{A}} -\frac{1}{a^{2}}  \nabla^i\delta\phi_{A} \nabla_i\delta\phi^{A} \label{eq:action-perturbations}\\
    &- \mathcal{M}^{AB} \delta\phi_{A} \delta\phi_{B}\bigg) \nonumber\,,
\end{align}
where we restrict to the case 
of a canonical kinetic term $h_{AB}=\delta_{AB}$, 
so the placement of indices becomes irrelevant. 
With a flat field space metric, the 
effective mass matrix $\mathcal{M}^{AB}$ reduces 
to
\begin{align}
    \mathcal{M}^{AB} \equiv \frac{\partial^2 V}{\partial\phi^{A}\partial\phi^{B}}  - \frac{1}{a^{3}M_{\mathrm{Pl}}^{2}} \frac{\rm d}{\rm dt} \left( \frac{a^{3}}{H} \dot{\phi}^{A} \dot{\phi}^{B} \right) \,.
    \label{eq:m_orig_mf}
\end{align}
In the same way as in the treatment 
of perturbations in the single-field case, 
it is convenient to perform a Fourier mode 
decomposition on constant time hypersurfaces 
$\delta\phi^A(\mathbf{x},t)=(2\pi)^{-3/2}\int d^3\mathbf{k}~e^{i\mathbf{k}\cdot\mathbf{x}}\Phi_\mathbf{k}^A(t)$. In Fourier domain and in 
conformal time, the action for the perturbation 
modes reads 
\begin{align}
S_2=&\frac{1}{2}\int \mathrm{d}\tau~\mathrm{d}^3\mathbf{k}~a^2\bigg[\Phi_{\mathbf{k}}'^A\Phi_{\mathbf{-k}}'^A-\label{eq:ac_fourier_mf}\\
&a^2\left(\frac{k^2}{a^2}\delta_{AB}-\mathcal{M}_{AB}\right)\Phi_{\mathbf{k}}^A\Phi_{\mathbf{-k}}^B\bigg]\,.
\nonumber
\end{align}
Furthermore, we can perform an additional 
variable redefinition $v_{\mathbf k}^A\equiv a\Phi_{\mathbf k}^A$ to recast the action in 
a form amenable for canonical scalar field 
quantization
\begin{align}
S_2=\frac{1}{2}\int \mathrm{d}\tau~\mathrm{d}^3\mathbf{k}~\left[v_{\mathbf{k}}'^A v_{\mathbf{-k}}'^A-\Omega^2_{AB}v_{\mathbf{k}}^A v_{\mathbf{-k}}^B\right]\label{eq:ac_fourier_can_mf}\,,
\end{align}
where the effective oscillation frequency matrix $\Omega^2$ is given by 
\begin{align}
    \Omega^{2}_{AB}\equiv \left( k^{2} - \frac{a^{\prime \prime}}{a} \right) \delta_{AB} + a^{2} \mathcal{M}_{AB}.
\label{eq:eff_freq_mf}
\end{align}
which makes the resulting action in 
Eq.~\eqref{eq:ac_fourier_can_mf} different from 
Eq.~\eqref{eq:ac_fourier_mf} only by a total 
conformal time derivative. With the action 
written in canonical form, the first step toward 
field quantization is to construct the 
system's Hamiltonian via a Legendre 
transformation 
\begin{align}
H_{\rm sys}=\int_{\mathbbm{R}^{3+}}\mathrm{d}^3\mathbf{k}~\left[\pi_{\mathbf{k}}^A (\pi_{\mathbf{k}}^A)^\dagger+\Omega^2_{AB}v_{\mathbf{k}}^A (v_{\mathbf{k}}^B)^\dagger\right]\,.
\label{eq:ham_fourier_can_mf}
\end{align}
where, as in the single-field case, the 
integration is performed over the upper half 
of the Fourier sphere, which eliminates 
the distinction between $+\mathbf{k}$ and 
$-\mathbf{k}$ modes. From this point, 
quantization proceeds by promoting the 
canonical field variables to quantum 
operators. As in \secref{sec:OQSF}, the 
Hermiticity of the operators guarantees 
that the expectation values of the fields 
and their conjugate momenta remain real. 
To make this explicit, we introduce the 
following field redefinitions.
\begin{align}
    \hat{v}^{A}_{\rm R} \equiv \frac{\hat{v}^{A} + (\hat{v}^{A})^{\dagger}}{\sqrt{2}} \quad , \quad \hat{v}^{A}_{\rm I} \equiv \frac{\hat{v}^{A} - (\hat{v}^{A})^{\dagger}}{\sqrt{2}i},
\label{eq:re_im_ops_mf}
\end{align}
where the $\rm R$ and $\rm I$ labels refer 
to the ``real'' and ``imaginary'' parts of the 
original mode operators, and the same expressions 
can be extended to their corresponding conjugate 
momenta. With respect to these operators, the 
canonical commutation relations are an extension 
of the single-field case given by 
\begin{align}
\left[\hat{v}_{\rm S}^{A}(\mathbf{k}) , \hat{\pi}_{\rm S'}^{B}(\mathbf{k^{\prime}})\right]=i \delta^{AB} 
\delta_{\mathrm{S}\mathrm{S^{\prime}}}  \delta^{(3)}\left(\mathbf{k}-\mathbf{k^{\prime}}\right),
\label{eq:mf_can_rel}
\end{align}
where the indices $\rm S,S'\in\{R,I\}$. Thus, the 
Hamiltonian can be recast as an Hermitian operator
\begin{widetext}
\begin{align}
\hat{H}_{\rm sys}=\frac{1}{2}\sum_{\rm S\in \{R,I\}}\int_{\mathbbm{R}^{3+}}\mathrm{d}^3\mathbf{k}~\left[\hat{\pi}_{\mathrm{S}}^A(\hat{\pi}_{\mathrm{S}}^A)^\dagger +\Omega^2_{AB}\hat{v}_{\mathrm{S}}^A(\hat{v}_{\mathrm{S}}^B)^\dagger \right]\,.
\label{eq:ham_fin_mf}
\end{align}
\end{widetext}
in the same way as the field operators and their 
conjugate momenta. From now on, we omit the wavevector 
dependence for conciseness in the notation. With the Hermitian 
field operators in place, we can define 
the multifield version of a state vector
\begin{equation}
    \hat{\mathbf{R}}=\left( \hat{v}_{\rm R}^{A},\hat{\pi}_{\rm R}^{B},\hat{v}_{\rm I}^{C},  \hat{\pi}_{\rm I}^{D}\right)^{\mathrm{T}}.
\label{eq:pv_multi}
\end{equation}
where the only purpose of the field indices is to denote that this is an $4\times \mathcal{N}$ array. The covariance matrix elements are defined as the vacuum expectation values of the anticommutator of state vector operators
\begin{align}
\langle\hat{\mathbf{\Sigma}}\rangle \equiv \left\langle \left\{\hat{\mathbf{R}},\hat{\mathbf{R}}\right\}\right\rangle = \langle\hat{\mathbf{\Sigma}}_{\rm R}\rangle\oplus \langle\hat{\mathbf{\Sigma}}_{\rm I}\rangle\,,
\label{eq:cov_mf}
\end{align}
where, as in the single-field case, the last equality implies that cross-correlations 
between the real and imaginary parts vanish. Specifically, each of these 
$2\mathcal{N}\times2\mathcal{N}$ submatrices is computed from the expectation values of 
the following anticommutators of the field operators
\begin{subequations}
\begin{align}
&\left[\hat{\Sigma}_{\rm S}^{(vv)}\right]^{AB} = \{\hat{v}^{A}_{\rm S}, \hat{v}^{B}_{\rm S}\} \label{eq:cov_vv_mf}\,,\\
&\left[\hat{\Sigma}_{\rm S}^{(vp)}\right]^{AB} = \{\hat{v}^{A}_{\rm S}, \hat{\pi}^{B}_{\rm S}\} \label{eq:cov_vp_mf}\,,\\
&\left[\hat{\Sigma}_{\rm S}^{(pp)}\right]^{AB} = \{\hat{\pi}^{A}_{\rm S}, \hat{\pi}^{B}_{\rm S}\} \label{eq:cov_pp_mf}\,,
\end{align}
\end{subequations}
which implies that the covariance submatrices in Eq.~\eqref{eq:cov_mf}  -- \ie~the real and imaginary parts of the full covariance matrix -- can be structured as follows:
\begin{align}
    \langle\hat{\mathbf\Sigma}_{\rm S}\rangle = \left[\begin{array}{cc}
\left\langle\hat{\Sigma}_{\rm S}^{(vv)}\right\rangle_{AB} & \left\langle\hat{\Sigma}_{\rm S}^{(vp)}\right\rangle_{AB}\vspace{.5em}\\
    \left\langle\hat{\Sigma}_{\rm S}^{(vp)}\right\rangle_{AB}^{\rm T} & \left\langle\hat{\Sigma}_{\rm S}^{(pp)}\right\rangle_{AB}
    \end{array}\right]\,.
\label{eq:cov_mat_mp}
\end{align}
After comparing with Eq.~\eqref{eq:Covariance_Matrix}, it is clear that this
is a straightforward generalization of the single-field scenario.
In the Wigner formalism, these matrices encapsulate the complete 
information of a multifield state for a given $k$-mode. The time evolution for the 
correlators is governed by the Liouville equation
\begin{align}
\frac{\rm d}{\rm d\tau}\langle\hat{\mathbf{\Sigma}}_{\rm S}\rangle=i\left\langle\left[\hat{H}_{\rm sys},\hat{\mathbf{\Sigma}}_{\rm S}\right]\right\rangle\,,
\label{eq:Liouville_mf}
\end{align}
and, upon substitution of the operator definitions in Eqns.~(\ref{eq:cov_vv_mf}--\ref{eq:cov_pp_mf}), we find the corresponding transport equations for each of the 
covariance matrix components
\begin{subequations}
\begin{align}
&\frac{\rm d}{\rm d\tau}\left\langle\hat{\Sigma}_{\rm S}^{(vv)}\right\rangle_{AB} = \mathrm{Sym}^{CD}_{AB}\left\langle\hat{\Sigma}_{\rm S}^{(vp)}\right\rangle_{CD}\label{eq:eq_mov_vv_mf}\,,\\
&\frac{\rm d}{\rm d\tau}\left\langle\hat{\Sigma}_{\rm S}^{(vp)}\right\rangle_{AB} = \left\langle\hat{\Sigma}_{\rm S}^{(pp)}\right\rangle_{AB}-\Omega^2_{AC}\left\langle\hat{\Sigma}_{\rm S}^{(vv)}\right\rangle_{CB} \label{eq:eq_mov_vp_mf}\,,\\
&\frac{\rm d}{\rm d\tau}\left\langle\hat{\Sigma}_{\rm S}^{(pp)}\right\rangle_{AB} = -\mathrm{Sym}^{CD}_{AB}\left\{\Omega^2_{CE}\left\langle\hat{\Sigma}_{\rm S}^{(vp)}\right\rangle_{ED}\right\} \label{eq:eq_mov_pp_mf}\,,
\end{align}
\end{subequations}
where the 4-index operator $(\rm Anti)Sym$ (anti-)symmetrizes the target object. 
Extending our procedures from the single-field section even further, it is possible to 
combine these equations of motion into a set of third-order equations of motion 
\begin{subequations}
\begin{align}
         &\left[ \delta_{AC}\frac{\rm d^3}{\rm d\tau^3 }  + 4\Omega^{2}_{AC} \frac{\rm d}{\rm d\tau }  + 2\left(\Omega^{2}_{AC}\right)^{\prime} \right]\left\langle\hat{\Sigma}_{\rm S}^{(vv)}\right\rangle_{CB}-\nonumber\\
         &2~\mathrm{Antisym}^{CD}_{AE} \left\{ \left\langle\hat{\Sigma}_{\rm S}^{(vp)}\right\rangle_{CD}\right\} \Omega^{2}_{EB}\equiv\Upsilon_{AB}\,,\label{eq:upsilon_mf}\\
         &\mathrm{Sym}^{CD}_{AB} \left[\Upsilon_{CD}\right]=0\,,
\label{eq:third_order_no_env_mf}
\end{align}
\end{subequations}
where, in general, the antisymmetrized cross-correlations in the definition of 
the auxiliary variable $\Upsilon$ do not vanish in the presence of multiple fields. 
It is also worth noting that the definition of $\Upsilon$ preserves the same structure 
as in the single-field case. However, since it is now a matrix relation, matrix 
products couple the different degrees of freedom. As a result, the evolution of the 
correlators is no longer independent component by component, but instead encodes the 
transfer of information and correlations across different field modes. 
Not surprisingly, the appearance of this extra term suggests an 
alternative to mimic multifield effects within a single-field setup: by allowing the 
source term $\mathcal{F}(k,\tau)$ in Eq.~\eqref{eq:sf_trans_acc} to encode the 
cross-correlations with external fields.

\subsection{Transport equations and open-system dynamics for multiple fields}
\label{subsec:Lindblad_multi}

The previous subsection established our conventions and derived 
the transport equations for two-point correlators in the case of 
multiple interacting fields at the perturbative level. 
With this in mind, the objective of this subsection is to 
extend the open-system formalism developed in 
\cite{Martin:2016qta, Martin:2018lin, Martin:2021qkg,Martin:2021znx}, and derive the corresponding 
transport equations for the case in which the multifield system 
develops its first nontrivial corrections through interactions 
with an environment. To achieve this, we consider the same 
Hamiltonian from Eq.~\eqref{eq:h_tot_sf}
\begin{align}
\hat{H}_{\rm tot}=\hat{H}_{\rm sys}\otimes\hat{\mathbbm{1}}_{\rm env}+
\hat{\mathbbm{1}}_{\rm sys}\otimes\hat{H}_{\rm env}+g\hat{H}_{\rm int}\,,
\label{eq:h_tot_mf}
\end{align}
where the system Hamiltonian $\hat{H}_{\rm sys}$, given in Eq.~\eqref{eq:ham_fin_mf}, 
is a sum of quadratic interacting actions. This structure ensures 
that the dynamics of different $k$-modes are independent, while, 
for a given $k$-mode, perturbations from the various field 
components remain coupled. The environment Hamiltonian $\hat{H}_{\rm env}$ is 
chosen such that (a) all of the environment operators 
(including the density matrix in the interaction picture) remain nearly stationary, 
and (b) it is approximately time-independent. Finally, the interaction Hamiltonian 
$\hat{H}_{\rm int}$ is separable into system and environment operators, and thus 
retains the same form as in Eq.~\eqref{eq:ham_int}. Therefore, considering that the derivation developed in \cite{Martin:2018zbe} does not constrain the number
of interacting fields in the systems' Hamiltonian, the evolution of 
the submatrices of covariance correlators described in Eq.~\eqref{eq:cov_mat_mp} is 
determined by the following Lindblad equation
\begin{widetext}
\begin{align}
       \frac{\mathrm{d}}{\mathrm{d} \tau}  \langle\hat{\mathbf{\Sigma}}_{\mathrm{S}}\rangle  = - i  \left\langle[\hat{\mathbf{\Sigma}}_{\mathrm{S}}, 
        \hat{H}_{\mathrm{sys}}]\right\rangle 
        - (2\pi^3)^{1/2} \Gamma \int_{\mathbb{R}^{3+}}  d^3\mathbf{k}~\tilde{C}_E(\mathbf{k},\tau)  \left\langle\left[ [\hat{\mathbf{\Sigma}}_{\mathrm{S}}, \hat{I}^{\mathrm{S}}_{\mathbf{k}}], \hat{I}^{\mathrm{S}}_{\mathbf{k}} \right]\right\rangle \,,
        \label{eq:LME_mf}
\end{align}
\end{widetext}
where the coupling constant $\Gamma \equiv 2g^2 \tau_{\mathrm{c}}$ is fixed  
under the assumption that the field interaction timescales 
(with $\tau_{\rm c}$ scaling approximately as the inverse square root of the 
eigenvalues of $\Omega^2$) are much shorter than the characteristic timescales 
of the environment. 

The operator $\hat{I}^{\rm S}_{\mathbf k}$ is limited to act over 
system states, and in the same way as in the single-field scenario, the first 
nontrivial corrections from the extra terms in the Lindblad equation arise by 
considering that such an operator is a linear combination of the field and momentum 
operators
\begin{align}
\hat{I}_{\mathbf{k}}^{\rm S} = \alpha_A(\tau)\hat{v}^{A}_{\rm S}+\beta_B(\tau)\hat{\pi}^{B}_{\rm S}\,,
\label{eq:gen_lin_mf}
\end{align}
where there is no difference if we pick either the real or the imaginary 
parts of the field, since the dynamical results are the same given our choice of 
covariance operators. Fixing $\hat{I}_{\mathbf k}^{\rm S} = \alpha_A (\tau)\hat{v}^A_{\rm S}$, the evolution equations for the covariance matrix components are found by substituting the  definitions from 
Eqns.~(\ref{eq:cov_vv_mf}--\ref{eq:cov_pp_mf}) and the system Hamiltonian from 
Eq.~\eqref{eq:ham_fin_mf} in the Lindblad equation
\begin{widetext}
\begin{subequations}
\begin{align}
    &\frac{\rm d}{\rm d\tau}\left\langle\hat{\Sigma}_{\rm S}^{(vv)}\right\rangle_{AB} = \mathrm{Sym}^{CD}_{AB}\left\langle\hat{\Sigma}_{\rm S}^{(vp)}\right\rangle_{CD}\,,
\label{eq:mf_trans_vv}\\
&\frac{\rm d}{\rm d\tau}\left\langle\hat{\Sigma}_{\rm S}^{(vp)}\right\rangle_{AB} = \left\langle\hat{\Sigma}_{\rm S}^{(pp)}\right\rangle_{AB}-\Omega^2_{AC}\left\langle\hat{\Sigma}_{\rm S}^{(vv)}\right\rangle_{CB} \label{eq:mf_trans_vp}\,,\\
&\frac{\rm d}{\rm d\tau}\left\langle\hat{\Sigma}_{\rm S}^{(pp)}\right\rangle_{AB} = -\mathrm{Sym}^{CD}_{AB}\left\{\Omega^2_{CE}\left\langle\hat{\Sigma}_{\rm S}^{(vp)}\right\rangle_{ED}\right\} +2\mathcal{F}_{AB}\label{eq:mf_trans_pp}\,,
\end{align}
\end{subequations}
where $\mathcal{F}_{AB} \equiv (2 \pi)^{3/2} \Gamma \tilde{C}_E \alpha_A \alpha_B=\mathcal{F}(k,\tau)\alpha_A\alpha_B$. Here the last equality implies that it 
is possible to reuse the tile parameterizations proposed in the single-field case in Eqns.~\eqref{eq:tiling_source} 
or \eqref{eq:tiling_source_mod}. To solve for the case of a general linear combination in Eq.~\eqref{eq:gen_lin_mf}, we find that the corresponding transport equations have the same form as in Eqns.~(\ref{eq:mf_trans_vv}--\ref{eq:mf_trans_pp}) provided that the following variable redefinitions are applied
\begin{subequations}
\begin{align}
&\left\langle\hat{\Sigma}_{\rm S}^{(vv)}\right\rangle_{AB}\longrightarrow \overline{\left\langle\hat{\Sigma}_{\rm S}^{(vv)}\right\rangle}_{AB}=\left\langle\hat{\Sigma}_{\rm S}^{(vv)}\right\rangle_{AB}\label{eq:gauge_vv_mf}\,,\\
&\left\langle\hat{\Sigma}_{\rm S}^{(vp)}\right\rangle_{AB}\longrightarrow \overline{\left\langle\hat{\Sigma}_{\rm S}^{(vp)}\right\rangle}_{AB}=\left\langle\hat{\Sigma}_{\rm S}^{(vp)}\right\rangle_{AB}+(2 \pi)^{3/2} \Gamma \tilde{C}_E \beta_A \beta_B\label{eq:gauge_vp_mf}\,,\\
&\left\langle\hat{\Sigma}_{\rm S}^{(pp)}\right\rangle_{AB}\longrightarrow \overline{\left\langle\hat{\Sigma}_{\rm S}^{(pp)}\right\rangle}_{AB}=\left\langle\hat{\Sigma}_{\rm S}^{(pp)}\right\rangle_{AB}-(2 \pi)^{3/2} \left\{\Gamma \tilde{C}_E~\mathrm{Sym}^{CD}_{AB} \left[ \alpha_C \beta_D\right] + [ \Gamma \tilde{C}_E \beta_A \beta_B ]'\right\}\label{eq:gauge_pp_mf}\,,\\
&\mathcal{F}_{AB}\longrightarrow \overline{\mathcal{F}}_{AB}=\mathcal{F}_{AB}+(2\pi^3)^{1/2}\left\{\Gamma \tilde{C}_E~\text{Sym}^{CD}_{AB} \left[ \Omega^{2}_{ED}\beta_C \beta_E  \right] - \left[\Gamma \tilde{C}_E~\text{Sym}^{CD}_{AB} \left(\alpha_C \beta_D \right)\right]' + \left[\Gamma \tilde{C}_E \beta_A \beta_B\right]''\right\}\,,\label{eq:gauge_F_mf}
\end{align}
\end{subequations}
where primes denote derivatives with respect to conformal time. Since these 
expressions generalize our results in the single-field case presented in 
Eqns.~(\ref{eq:id_vv_gauge}--\ref{eq:id_F_gauge}), it is possible to solve the 
same transport equations in Eqns.~(\ref{eq:mf_trans_vv}--\ref{eq:mf_trans_pp}) and 
invert these variable redefinitions on the solutions to find the results for other 
forms of coupling with the environment.
\end{widetext}

The set of equations in Eqns.~(\ref{eq:mf_trans_vv}--\ref{eq:mf_trans_pp}) forms a 
coupled system of equations of motion for the evolution of the covariance matrix 
under decoherence effects, which are similar to what was derived in 
\cite{Martin:2018zbe, Martin:2021znx}. As in the case free of environmental effects, 
we may attempt to condense these equations into a single set of third-order 
differential equations for the mode correlators 
\begin{align}
\mathrm{Sym}^{CD}_{AB} \left[\Upsilon_{CD}\right]=8\mathcal{F}_{AB}\,,
\label{eq:third_order_mf_dec}
\end{align}
where, for notational simplicity, overlines have been omitted and the auxiliary 
variable $\Upsilon$ is defined as in Eq.~\eqref{eq:upsilon_mf}. As in the case of a 
single field, the term $\mathcal{F}_{AB}$ originates from the double 
commutator term in the Lindblad equation, and acts as a source term evolving
independently from the mode correlators. Nevertheless, since the 
result still depends on the cross-correlations between the field modes and their 
conjugate momenta, these cross-correlations must be determined separately.

The structure of the interaction operator in Eq.~\eqref{eq:gen_lin_mf} suggests that 
decoherence events can modify the dynamics of mode correlators through couplings to any 
of the fields. In particular, we can examine the effects of decoherence 
interactions along the adiabatic $(\parallel)$ or in the isocurvature $(\bot)$ directions, defined as
\begin{subequations}
\begin{align}
&(\hat{v}^A_{\rm S})_{\parallel}\equiv \frac{\dot{\phi}^A\dot{\phi}_C}{|\dot{\phi}|^2}\hat{v}_{\rm S}^C\label{eq:parallel}\,,\\
&(\hat{v}^A_{\rm S})_{\bot}\equiv \left[\delta^A_C-\frac{\dot{\phi}^A\dot{\phi}_C}{|\dot{\phi}|^2}\right]\hat{v}^C_{\rm S}\label{eq:perp}\,,
\end{align}
\end{subequations}
where $|\dot{\phi}|^2$ denotes the norm squared of the background velocities in a 
flat field-space geometry. 
Since the same projection-based definitions extend naturally to the corresponding conjugate 
momenta, the interaction operator $\hat{I}_{\mathbf k}^{\rm S}$ can also be projected along either direction. For example, its projection onto the parallel direction is
\begin{align}
\hat{I}_{\parallel}^{\rm S} &= \alpha_A(\tau)(\hat{v}^{A}_{\rm S})_{\parallel}+\beta_B(\tau)(\hat{\pi}^{B}_{\rm S})_{\parallel}\,,
\label{eq:gen_lin_par}\\
&=\left(\alpha_A(\tau)\frac{\dot{\phi}^A\dot{\phi}_C}{|\dot{\phi}|^2}\right)\hat{v}_{\rm S}^C+\left(\beta_B(\tau)\frac{\dot{\phi}^B\dot{\phi}_D}{|\dot{\phi}|^2}\right)\hat{\pi}_{\rm S}^D\,,\nonumber\\
&=\tilde{\alpha}_C(\tau)\hat{v}^{C}_{\rm S}+\tilde{\beta}_D(\tau)\hat{\pi}^{D}_{\rm S}\,,\nonumber
\end{align}
where  $\tilde{\alpha}_C(\tau)\equiv\alpha_A(\tau)\dot{\phi}^A\dot{\phi}_C/|\dot{\phi}|^2$ and 
$\tilde{\beta}_D(\tau)\equiv\beta_B(\tau)\dot{\phi}^B\dot{\phi}_D/|\dot{\phi}|^2$ specify the transformed time-dependent coefficients in the linear combination.
This shows that the variable redefinitions described in 
Eqns.~(\ref{eq:gauge_vv_mf}--\ref{eq:gauge_F_mf}) are sufficient to find the mode 
correlators for the case of an adiabatic interaction operator. The same reasoning 
applies to the operator projected along the isocurvature direction, or to any other 
combination of both directions.

\subsection{Implementation of a dynamical coloring scheme via Cholesky 
decomposition}
\label{subsec:Cholesky_lind}

Up to this point, we have established that 
Eqns.~(\ref{eq:mf_trans_vv}--\ref{eq:mf_trans_pp}) govern the 
evolution
of the following 
mode correlators

\begin{subequations}
\begin{align}
&\left\langle\hat{\Sigma}_{\rm S}^{(vv)}\right\rangle^{AB} = 2\left\langle\hat{v}^{A}_{\rm S}, \hat{v}^{B}_{\rm S}\right\rangle \label{eq:cov_val_vv_mf}\,,\\
&\left\langle\hat{\Sigma}_{\rm S}^{(vp)}\right\rangle^{AB} = \left\langle\hat{v}^{A}_{\rm S}, \hat{\pi}^{B}_{\rm S}\right\rangle +\left\langle\hat{\pi}^{B}_{\rm S}, \hat{v}^{A}_{\rm S}\right\rangle\label{eq:cov_val_vp_mf}\,,\\
&\left\langle\hat{\Sigma}_{\rm S}^{(pp)}\right\rangle^{AB} = 2\left\langle\hat{\pi}^{A}_{\rm S}, \hat{\pi}^{B}_{\rm S}\right\rangle \label{eq:cov_val_pp_mf}\,,
\end{align}
\end{subequations}
which describe the evolution of a 
Gaussian state subject to an arbitrary 
sequence of decoherence events. The 
objective of this subsection is to 
introduce a fast-slow decomposition 
scheme analogous to the amplitude-phase 
separation procedure discussed in 
subsection \ref{subsec:FSCS}. To this end, based on the approach 
developed in \cite{GalvezGhersi:2016wbu}, we decompose the field and 
momentum variables in the same way as in the single field-case
\begin{subequations}
\begin{align}
&\hat{v}^A_{\rm S}\rightarrow L_{AB}\tilde{\chi}_B\,,\label{eq:v_mf_dec}\\
&\hat{\pi}^A_{\rm S}\rightarrow L'_{AB}\tilde{\chi}_B+L_{AB}\tilde{\chi}'_{B}\,.
\label{eq:pi_mf_dec}
\end{align}
\end{subequations}
Here $L_{AB}$ denotes the elements of 
a real matrix 
encoding the field amplitude, and 
$\tilde{\chi}_{A}$ denotes an element 
from a set of $\mathcal{N}$ 
time-dependent Gaussian random 
variables. In line with the 
Wigner formalism, Gaussian random 
variables 
are sufficient to reproduce a large 
number of field realizations 
consistent with a Gaussian state \cite{Polarski:1995jg}. Statistical 
averages are therefore computed with 
respect to $\tilde{\chi}$ only. 
In terms of these variables, the mode correlator $\left\langle\hat{v}^{A}_{\rm S}, \hat{v}^{B}_{\rm S}\right\rangle$ can 
be expressed as
\begin{align}
\left\langle\hat{v}^{A}_{\rm S}, \hat{v}^{B}_{\rm S}\right\rangle &= L_{AC}L_{BD}\left\langle\tilde{\chi}_{C}, \tilde{\chi}_{D}\right\rangle\label{eq:ch_dec}\\
&=L_{AC}L^{\rm T}_{CB}\,,\nonumber
\end{align}
where we impose the normalization condition 
\begin{align}
\left\langle\tilde{\chi}_{C}, \tilde{\chi}_{D}\right\rangle=\delta_{CD}\,,\label{eq:gauge_fix}
\end{align}
which fixes the variance to unity 
at all times. This matrix factorization defines a coloring  transformation 
\cite{Kessy:2018}, in which the 
stochastic variables $\tilde{\boldsymbol{\chi}}$ 
capture the statistical properties of 
the Gaussian state, while the matrices 
$\boldsymbol{L}$ and $\boldsymbol{L}'$ encode its deterministic 
evolution. As a direct by-product of 
introducing the fast–slow 
decomposition, the original transport 
problem can be reintepreted as a 
dynamical coloring problem. 
In this formulation, deriving the 
equations of 
motion for the coloring factors (\ie the matrix $\boldsymbol{L}$) from 
Eqns.~(\ref{eq:mf_trans_vv}--\ref{eq:mf_trans_pp}) provides a 
direct description of how the 
coloring transformation evolves in 
time. 

As a first step toward obtaining 
dynamical coloring transformations, we 
rewrite the mode correlators from Eqns.~(\ref{eq:cov_val_vv_mf}--\ref{eq:cov_val_pp_mf}) in terms of the coloring matrix $L_{AB}$ and 
the velocity correlators between the Gaussian 
variables and their derivatives 
\begin{subequations}
\begin{align}
 &\left\langle\hat{\Sigma}_{\rm S}^{(vv)}\right\rangle_{AB} = 2 L_{AC}L^{\mathrm{T}}_{CB}\,,
 \label{eq:cov_vv_L}\\
 &\left\langle\hat{\Sigma}_{\rm S}^{(vp)}\right\rangle_{AB} = 2 L_{AC}\left(L^{\mathrm{T}}_{CB}\right)' + 2 L_{BD} \langle{{\tilde{\chi}}^{\prime}_D, \tilde{\chi}_C}\rangle L^{\mathrm{T}}_{CA}\,,
 \label{eq:cov_vp_L}\\
 &\left\langle\hat{\Sigma}_{\rm S}^{(pp)}\right\rangle_{AB}=2 \left( L_{AC} \right)^{\prime} \left(L^{\mathrm{T}}_{CB}\right)' + 
 2 L_{BD} \langle{{\tilde{\chi}}^{\prime}_D, \tilde{\chi}_C}\rangle \left( L^{\mathrm{T}}_{CA}\right)'\nonumber\\
&+ 2 L_{AC} \langle{{\tilde{\chi}}^{\prime}_C, \tilde{\chi}_D}\rangle \left( L^{\mathrm{T}}_{DB}\right)^{\prime} + 
2 L_{AC} \langle{{\tilde{\chi}}^{\prime}_C, {\tilde{\chi}}^{\prime}_D}\rangle L^{\mathrm{T}}_{DB}\,,
\label{eq:cov_pp_L}
\end{align}
\end{subequations}
where, since the analysis is 
restricted to Gaussian states, all 
statistical averages are computed 
through Gaussian integrals, for which 
the ordering of the variables is 
immaterial. The Hermiticity of the 
field operators then guarantees that 
the associated stochastic variables 
can be taken as real-valued, allowing 
us to represent the operators by real 
Gaussian variables and safely omit the 
commutation relations between 
$\tilde{\boldsymbol{\chi}}$ and $\tilde{\boldsymbol{\chi}}'$. As 
anticipated, the substitution of 
Eqns.~(\ref{eq:cov_vv_L}--\ref{eq:cov_pp_L}) in the system of 
transport equations 
in Eqns.~(\ref{eq:mf_trans_vv}--\ref{eq:mf_trans_pp}) yields the 
equations of motion for the dynamical 
coloring transformations. The substitution of 
Eqns.~(\ref{eq:cov_vv_L}--\ref{eq:cov_vp_L}) 
in Eqns.~(\ref{eq:mf_trans_vv}--\ref{eq:mf_trans_vp}) results 
in the following constraint on the  
$\langle\tilde{\chi}'_A,\tilde{\chi}_B\rangle$ correlators:
\begin{align}
\text{Sym}^{CD}_{AB} \left[\langle{\tilde{\chi}}^{\prime}_{C}, \tilde{\chi}_{D}\rangle\right]=0\,,\label{eq:first_der}
\end{align}
which corresponds to the first derivative of the variance-fixing 
conditions in Eq.~\eqref{eq:gauge_fix}, and implies that the correlator 
$\langle\tilde{\chi}'_A,\tilde{\chi}_B\rangle$ is antisymmetric 
(not zero, as in the single-field case) with respect to 
the indices $A$ and $B$. This agrees with the intuition of the 
$\langle\tilde{\chi}'_A,\tilde{\chi}_B\rangle$ correlators as rotation generators 
of the normalized stochastic vector $\tilde{\boldsymbol{\chi}}$. 
Combined with the derivative of this condition 
\begin{align}
\text{Sym}^{CD}_{AB} \left[\langle{\tilde{\chi}}^{\prime \prime}_{C},\tilde{\chi}_{D}\rangle+\langle{\tilde{\chi}}^{\prime}_{C},{\tilde{\chi}}^{\prime}_{D}\rangle\right]=0\,,\label{eq:second_der}
\end{align}
it ensures that the  Gaussian random variables 
are independent, and fixes their variance to one at 
all times. In fact, the condition in 
Eq.~\eqref{eq:gauge_fix} can also be recovered by inserting the mode correlator 
definitions in Eqns.~(\ref{eq:cov_vv_L}--\ref{eq:cov_pp_L}) 
directly in the transport equations (Eqns.~(\ref{eq:mf_trans_vv}--\ref{eq:mf_trans_pp})). The constraints in Eqns.~\eqref{eq:first_der} and 
\eqref{eq:second_der} reveal that the derivatives of the random variable correlators 
are related in such a way that the system can be order-reduced. This feature is 
particularly advantageous when deriving the evolution equations for the coloring 
matrices and auxiliary correlators. Adding Eq.~\eqref{eq:mf_trans_vp} and its 
transpose, and considering again the mode correlators definitions, we find the 
following expression for the $\boldsymbol{L}$ matrices
\begin{align}
        \text{Sym}^{CD}_{AB}\bigg[&\left(L^{-1}\right)_{CE}{L}_{ED}^{\prime \prime} -\langle{\tilde{\chi}}^{\prime}_C,  {\tilde{\chi}}^{\prime}_D\rangle+\left(L^{-1}\right)_{CE}\Omega^2_{EF}L_{FD}\nonumber\\
        &+2(L^{-1})_{CE}{L}_{EF}^{\prime}\langle{\tilde{\chi}}^{\prime}_F, \tilde{\chi}_{D}\rangle\bigg]=0\,,\label{eq:L_mat_cond}
\end{align}
where the term inside the symmetrizer permits the addition of an antisymmetric matrix 
without altering the expression. The inclusion of an antisymmetric matrix fixes the 
shape of the coloring matrix, and in this specific case we add the matrix 
$\mathcal{A}_{CD}$ to suppress the upper-diagonal elements of $\boldsymbol{L}$ ($D > C$):
\begin{align}
\mathcal{A}_{CD}=&-  \left(L^{-1}\right)_{CE}\Omega^2_{EF}L_{FD} +\langle{\tilde{\chi}}^{\prime}_C ,{\tilde{\chi}}^{\prime}_D\rangle\nonumber\\
        &-2(L^{-1})_{CE}\left({L}_{EF}\right)^{\prime}\langle{\tilde{\chi}}^{\prime}_F, \tilde{\chi}_{D}\rangle\,,\label{eq:antisym_A}       
\end{align}
which fixed the coloring factor $L$ to be a lower-triangular Cholesky matrix with $\mathcal{N}(\mathcal{N}+1)/2$ entries at all 
times. Therefore, with this rearrangement of degrees of freedom, the coloring 
transformation defined in Eq.~\eqref{eq:gauge_fix} is based on a Cholesky 
decomposition, which is unique in the case in which $\left\langle\hat{\Sigma}_{\rm S}^{(vv)}\right\rangle_{AB}$ is a positive-definite matrix. Thus, the equation of 
motion for the coloring Cholesky factor has the form
\begin{align}
&{L}_{AB}^{\prime \prime} +2{L}_{AC}^{\prime}\langle{\tilde{\chi}}^{\prime}_C, \tilde{\chi}_B\rangle +L_{AC}\big(\mathcal{A}_{CB} - \nonumber\\
&\langle{\tilde{\chi}}^{\prime}_C, {\tilde{\chi}}^{\prime}_B\rangle\big)+\Omega^2_{AC}L_{CB}=0\,,
\label{eq:eq_mov_L_mf}
\end{align}
where the choice of $\boldsymbol{\mathcal{A}}$ guarantees that the lower-triangular shape of $\boldsymbol{L}$ 
is preserved at all times. In what remains of this section, we will develop the case of a Cholesky decomposition 
but this method can be extended to other coloring transformations, as we will 
show in \Appref{app:sym_sep}. From the last equation, it becomes evident that the equations 
of motion for the random variable correlators 
$\langle{\tilde{\chi}}^{\prime}_B, \tilde{\chi}_C\rangle$ 
and $\langle{\tilde{\chi}}^{\prime}_B, \tilde{\chi}^{\prime}_C\rangle$ are required to 
close the dynamical system for the coloring transformation. Consequently, just as in 
the case for the coloring matrices, we can obtain their corresponding evolution 
equations from the transport system equations. Rewriting Eq.~\eqref{eq:mf_trans_vp} and 
reducing the order of the Cholesky factors using Eq.~\eqref{eq:eq_mov_L_mf} we derive 
the equations governing the dynamics of 
$\langle{\tilde{\chi}}^{\prime}_B, \tilde{\chi}_C\rangle$
\begin{align}
    \frac{\mathrm{d}}{\mathrm{d}\tau}\langle{\tilde{\chi}}^{\prime}_B, {\tilde{\chi}}_C\rangle = \mathcal{A}_{BC}\,,
    \label{eq:chip_chi_mf}
\end{align}
which conserves the antisymmetric form of such correlators anticipated in 
Eq.~\eqref{eq:first_der}. As for the evolution of the velocity correlators, their 
equations of motion can be obtained from Eq.~\eqref{eq:mf_trans_pp}:
\begin{align}
\frac{\mathrm{d}}{\mathrm{d} \tau}\langle{\tilde{\chi}}^{\prime}_A, {\tilde{\chi}}^{\prime}_B\rangle= ~& \text{Sym}^{CD}_{AB}\bigg[-2L^{-1}_{CE}L_{EF}^{\prime}\langle{\tilde{\chi}}^{\prime}_F, {\tilde{\chi}}^{\prime}_D\rangle\nonumber \\
            &-2L^{-1}_{CE}L_{EF}^{\prime}\langle{\tilde{\chi}}^{\prime}_{F}, \tilde{\chi}_{G}\rangle\langle\tilde{\chi}^{\prime}_{G}, {\tilde{\chi}}_{D}\rangle\nonumber \\
            &+\langle{\tilde{\chi}}^{\prime}_{C},{\tilde{\chi}}^{\prime}_{F}\rangle\langle{\tilde{\chi}}^{\prime}_{F},\tilde{\chi}_{D}\rangle -\mathcal{A}_{CE}\langle{\tilde{\chi}}^{\prime}_E,\tilde{\chi}_{D}\rangle\nonumber \\
            &+ \frac{1}{2}    L^{-1}_{CE}\mathcal{F}_{EF} (L^{-1})^{\rm T}_{FD}   \bigg]\label{eq:chip_chip_mf}\,,
\end{align}
which provides equations of motion 
for all of the variables in the 
dynamical system. 

Conservation of the covariance matrix determinant plays a crucial role in verifying 
whether the dynamical system undergoes stable numerical evolution. In terms of the 
covariance block matrices, this determinant can be written as 
\begin{align}
\mathcal{S}  &= \det \big \{\langle\hat{\Sigma}_{\rm S}^{(v v)}\rangle \big \} \det \big \{\langle\hat{\Sigma}_{\rm S}^{(p p)}\rangle \nonumber\\
&- \langle\hat{\Sigma}_{\rm S}^{(v p)} \rangle^{\rm T}{\langle\hat{\Sigma}_{\rm S}^{(v v)}}\rangle^{-1} \langle\hat{\Sigma}_{\rm S}^{(v p)}\rangle \big \}\,,
\end{align}
where, as in the single-field scenario, it is indistinct 
to choose either the real $(\rm S=R)$ or the imaginary 
mode operators $(\rm S=I)$ since these are independent 
parts of the covariance matrix $\langle\hat{\mathbf{\Sigma}}\rangle$. With respect to the coloring 
matrices, the determinant of each submatrix reduces to
\begin{align}
\mathcal{S}=4^{\mathcal{N}} \det \left \{ \boldsymbol{L} \boldsymbol{L}^{T} \right \} \det \left \{ \boldsymbol{L} \boldsymbol{\mathcal{Z}} \boldsymbol{L}^{T} \right \}\,, 
\label{eq:det_mf}
\end{align}
where the matrix $\mathcal{Z}_{AB}$ is defined as
\begin{align}
\mathcal{Z}_{AB}\equiv \langle{\tilde{\chi}}_A^{\prime}, {\tilde{\chi}}^{\prime}_B\rangle 
+ \langle{\tilde{\chi}}_A^{\prime},\tilde{\chi}_C\rangle \langle{{\tilde{\chi}}_C^{\prime},\tilde{\chi}}_B\rangle\,, 
\label{eq:z_def}
\end{align} 
which serves as an auxiliary variable introduced to isolate 
the contributions associated to the amplitudes from those arising from the dynamics of the
random variable correlators. The role of this new variable is reminiscent to the purpose of the velocity correlator
$\langle{\tilde{\chi}}_{\rm S}^{\prime}, {\tilde{\chi}}^{\prime}_{\rm S}\rangle$ in Eq.~\eqref{eq:det_sf_acc} for the single-field case. Importantly, $\boldsymbol{\mathcal{Z}}$ is the only variable 
other than the Cholesky factors that enters explicitly in the calculation of the 
determinant, making it particularly convenient for testing the conservation of 
$\mathcal{S}$. Thus, the equations of motion in Eq.~\eqref{eq:chip_chip_mf} need to be replaced by evolution equations for the new variable
\begin{align}
\frac{\mathrm{d}}{\mathrm{d} \tau} \mathcal{Z}_{AB}&=  \text{Sym}^{CD}_{AB}\bigg[-2L^{-1}_{CE}L'_{EF} \mathcal{Z}_{FD} \label{eq:z_evol}\\
&- \langle{\tilde{\chi}}^{\prime}_C, \tilde{\chi}_{E}\rangle \mathcal{Z}_{ED}+ \frac{1}{2}  L^{-1}_{CE} \mathcal{F}_{EF}\left(L^{-1}\right)^{\rm T}_{FD} \bigg]\,.\nonumber
\end{align}
which are the only equations affected explicitly by the presence of the environment.
Hence, together with Eqns.~\eqref{eq:chip_chi_mf} and \eqref{eq:z_evol}, the dynamics 
of the coloring transformations is governed by the following equations of motion
\begin{widetext}
\begin{align}
 L_{AB}^{\prime \prime}+2L_{AC}^{\prime}\langle{\tilde{\chi}}^{\prime}_C, \tilde{\chi}_B\rangle+L_{AC}\left[\mathcal{A}_{CB} - \mathcal{Z}_{CB} + \langle{\tilde{\chi}}^{\prime}_C, {\tilde{\chi}}_D\rangle \langle{\tilde{\chi}}^{\prime}_D, {\tilde{\chi}}_B\rangle \right]+\Omega^2_{AC}L_{CB}=0\,,\label{eq:fin_eq_L}
\end{align}
\end{widetext}
where, for consistency, the definition of the shape-fixing antisymmetric matrix  
$\mathcal{A}_{BC}$ in \eqref{eq:antisym_A_mod} also needs to be recast in terms the 
auxiliary variable $\mathcal{Z}_{BC}$:
\begin{align}
\mathcal{A}_{BC}=&-  \left(L^{-1}\right)_{BE}\Omega^2_{EF}L_{FC} +\mathcal{Z}_{BC}-\langle{\tilde{\chi}}^{\prime}_B ,{\tilde{\chi}}_D\rangle\langle{\tilde{\chi}}^{\prime}_D ,{\tilde{\chi}}_C\rangle\nonumber\\
        &-2(L^{-1})_{BE}\left({L}_{EF}\right)^{\prime}\langle{\tilde{\chi}}^{\prime}_F, \tilde{\chi}_{C}\rangle\,,\label{eq:antisym_A_mod}       
\end{align}
where $(C>B)$. Eq.~\eqref{eq:fin_eq_L} constitutes a system of $\mathcal{N}
(\mathcal{N}+1)/2$ equations for the entries of the 
lower-triangular Cholesky factors\footnote{It is indistinct to fix the shape of the 
coloring Cholesky factors to be lower or upper-triangular matrices.}. 
As a sanity check, it is necessary 
to count the number of degrees of freedom: there are 
$\mathcal{N}(\mathcal{N}-1)/2$ evolving constraints in Eq.~\eqref{eq:chip_chi_mf} 
associated with the antisymmetric correlators 
$\langle{\tilde{\chi}}^{\prime}_A, {\tilde{\chi}}_B\rangle$, which guarantee the 
preservation of the unit-variance gauge condition in Eq.~\eqref{eq:gauge_fix} and the 
shape of the Cholesky factors. 
As a result of this, 
subtracting the number of constraints from the amount of evolving matrix elements in the coloring scheme, we find $\mathcal{N}$ evolving degrees 
of freedom per 
$k$-mode, matching the number of fields. 

In addition to the dynamical 
coloring scheme interpretation discussed above, this evolution scheme can also be understood as a fast and slow separation of degrees of freedom, since the net oscillation frequency in Eq.~\eqref{eq:fin_eq_L} is suppressed by the counterterm $\langle\tilde{\chi}'_A,\tilde{\chi}'_B\rangle$ in a manner analogous to our results in 
\cite{GalvezGhersi:2016wbu}, figure 8, and to the scale separation scheme discussed 
in subsection \ref{subsec:FSCS} for the single-field scenario. This suppression enables an increase of one or two orders of magnitude in the 
evolution time step relative to a standard mode evolution approach, allowing us to efficiently resolve the dynamics of the mode correlators undergoing decoherence events deep inside the horizon. 

Before presenting the numerical results, it is important to clarify the regime of validity of the transport system and the numerical limitations that arise in certain dynamical situations.
Some care must be taken when applying this implementation, since the original set 
of transport equations in 
Eqns.~(\ref{eq:mf_trans_vv}--\ref{eq:mf_trans_pp}) still forms a third-order system with respect to $\left\langle\hat{\mathbf{\Sigma}}_{\rm S}^{(vv)}\right\rangle$, 
which -- when written as in Eqns.~\eqref{eq:third_order_no_env_mf} and 
\eqref{eq:third_order_mf_dec} -- cannot be 
order-reduced. At super-horizon scales, 
this system exhibits a sign 
flip in the diagonal elements of 
$\boldsymbol{\mathcal{Z}}$ 
-- which are positive, by definition -- 
leading to numerical instabilities that 
manifest particularly in 
scenarios where the background trajectories 
undergo sharp turns. 
Despite this, the framework retains enough flexibility to generate arbitrarily large 
features in the primordial power spectrum, provided that (a) the model potentials 
remain sufficiently smooth and (b) decoherence and recoherence events occur at 
subhorizon scales (as developed in 
\secref{sec:OQSF} for one field). Moreover, 
since part of the purpose of this approach 
is to mimic Gaussian state deformations of 
more complicated dynamical setups using 
smooth, well-behaved models as a starting 
point, these implementation-related challenges do not pose severe 
limitations to achieving this objective. 
Extending this approach to accommodate 
potentials that induce more intricate 
deformations of the background evolution will be the subject of future work.

The dynamical setup requires initial conditions to be complete. As developed in 
\secref{sec:OQSF} for the single-field case, it is natural to impose 
instantaneous minimal variance/energy initial conditions on a hypersurface of 
constant $k_{\rm phys}$ prior to the onset of decoherence events. This prescription  
requires diagonalizing the effective oscillation frequency matrix $\mathbf{\Omega}^2$ at initial 
time $\tau=\tau_0$ in order to determine the initial mode correlators 
\begin{align}
\omega_{AB}|_{\tau=\tau_0}\equiv U_{AC}\Omega^2_{CD}|_{\tau=\tau_0}U^{\rm T}_{CB}\,,
\label{eq:diag_omega}
\end{align}
where $\omega_{AB}$ in the diagonal form of $\mathbf\Omega^2$, and $U$ is the orthonormal 
matrix that diagonalizes $\mathbf\Omega^2$ at $\tau=\tau_0$. It is known that this choice 
of instantaneous minimal energy/variance initial states is equivalent to the 
Minkowski vacuum. From a geometric perspective, this choice restricts the problem to 
the study of a multidimensional Wigner ellipse, which starts as a circle and is later 
deformed through an arbitrary sequence of decoherence events. Thus, the initial mode 
correlators are given by
\begin{subequations}
\begin{align}
    &\left\langle\hat{\Sigma}_{\rm S}^{(v v)}\right\rangle_{AB}\bigg{|}_{\tau=\tau_0} =U^{\rm T}_{AC} \left(\frac{1}{\omega^{1/2}_{CD}}\right) U_{DB}\,,\label{eq:init_corr_vv}\\
    &\left\langle\hat{\Sigma}_{\rm S}^{(v p)}\right\rangle_{AB}\bigg{|}_{\tau=\tau_0} =0\,,\label{eq:init_corr_vp}\\
    &\left\langle\hat{\Sigma}_{\rm S}^{(p p)}\right\rangle_{AB}\bigg{|}_{\tau=\tau_0} =U^{\rm T}_{AC} \omega^{1/2}_{CD} U_{DB}\,.\label{eq:init_corr_pp}
\end{align}
\end{subequations}
Here, $\omega^{1/2}$ denotes the square root of each the 
diagonal matrix elements of $\omega$. 
Although the initial conditions for the $(vp)$ 
mode correlators are set to zero, they are sourced to not vanish 
during the evolution. 

Once the initial mode correlators are set, the corresponding initial values for the 
dynamical coloring transformations follow directly. The initial 
coloring matrix is computed by performing a Cholesky decomposition of the mode correlator in Eq.~\eqref{eq:init_corr_vv}, while the remaining dynamical variables are 
determined from Eqns.~(\ref{eq:init_corr_vv}--\ref{eq:init_corr_pp})
\begin{subequations}
\begin{align}
&L_{AC}L^{\rm T}_{CB}\bigg{|}_{\tau=\tau_0}=\left\langle\hat{\Sigma}_{\rm S}^{(vv)}\right\rangle_{AB}\bigg{|}_{\tau=\tau_0}\label{eq:init_L}\,,\\
&L'_{AB}|_{\tau=\tau_0}=\langle\tilde{\chi}'_A,\tilde{\chi}_B\rangle|_{\tau=\tau_0}=0\,,\label{eq:init_Lprime_chip_chi}\\
&\mathcal{Z}_{AB}|_{\tau=\tau_0}=\langle\tilde{\chi}'_A,\tilde{\chi}'_B\rangle|_{\tau=\tau_0}\nonumber\\
&=\frac{1}{2}\left(L^{-1}\right)_{AE} \left\langle\hat{\Sigma}_{\rm S}^{(p p)}\right\rangle_{EF}\bigg{|}_{\tau=\tau_0}  \left(L^{-1}\right)^{\mathrm{T}}_{FB}\,.\label{eq:init_z}
\end{align}
\end{subequations}
These initial conditions complete the specification of the initial state and, when 
combined with the evolution equations in Eqns.~\eqref{eq:chip_chi_mf}, 
\eqref{eq:z_evol}, and \eqref{eq:fin_eq_L}, provide a full definition of the dynamical 
coloring transformations. Moreover, we can apply the mode correlator redefinitions in 
Eqns.~(\ref{eq:gauge_vv_mf}--\ref{eq:gauge_F_mf}) to transform the results, and study 
alternative time-dependent linear combinations of field and momenta constituting the 
interaction operator $\hat{I}^{\rm S}_{\mathbf{k}}$ described in \subsecref{subsec:Lindblad_multi}.

With the complete evolution system in hand, the next subsection showcases only a small 
portion of the wide array of analyses enabled by our numerical implementation. Based on 
the dynamical coloring–transformation framework, this implementation lets us probe the 
first nontrivial corrections arising from the Lindblad equation in 
Eq.~\eqref{eq:LME_mf} for a multifield system interacting with an environment.

\trnglrschm

Although the implementation presented here is
restricted to trivial field-space metrics, the formalism can be 
extended to nontrivial field-space geometries without altering its 
essential structure. At the level of linear perturbations, we may 
introduce a local vielbein $(\Lambda_A^i)$ on field space and perform a variable 
redefinition that maps the system to an approximately flat field-space frame, following the standard treatment (see \eg \cite{Lee:2005bb}). Capital Latin indices label directions in the 
curved field space, while lowercase Latin indices correspond to the 
locally flat frame. Choosing the vielbein to satisfy 
parallel-transport equations along the background trajectory
\begin{align}
\mathcal{D}_t\Lambda^i_A\equiv\dot{\Lambda^i_A}+\Gamma^C_{AB}\dot{\phi}^B\Lambda^i_C=0\,,
\label{eq:par_tr}
\end{align}
allows for canonical quantization of the perturbation modes in this 
frame. This requires considering the vielbein as additional 
dynamical degrees of freedom. 
In this representation, the effects of field-space curvature are 
retained through corrections to the effective frequency matrix, 
while the kinetic structure remains canonical. Since the coloring 
transformation and tiling-based implementation of decoherence act 
directly on the mode evolution equations, their construction is not 
significantly modified in this frame. Sequences of decoherence 
events can therefore be implemented as in the previous sections, 
and the inverse vielbein may be used to reconstruct the 
perturbations in the original curved field-space basis.

\subsection{Results: evolution of the Wigner ellipse and generation of features 
in the power spectrum}
\label{subsec:results}

\sevolmf

In addition to the deformation of initial states to mimic the primordial  
spectrum of more complex (\ie with more fields, intricate potentials, nontrivial 
effects from 
field geometry or combinations of all of these) inflationary models, the inclusion of 
decoherence within the multifield framework raises a fundamental question: how do these 
field effects manifest and propagate across coupled degrees of 
freedom? Addressing this 
question requires characterizing the dynamics of each Wigner ellipse as it undergoes 
the simplest sequences of decoherence events. To this end, we adopt the same parameterization introduced in \secref{sec:OQSF} 
(Eqns.~\eqref{eq:tiling_source} and \eqref{eq:tiling_source_mod}) for the single-field 
case, which specifies the amplitude, location, duration, and wavelength band over which 
each event is effective.

In \Figref{fig:Wigner_mf}, we show the 
evolution of the Wigner ellipse as it 
undergoes decoherence and recoherence 
events of finite duration. As illustrated 
in the left panel of this figure, we 
considered three modes with different 
wavenumbers $k$ in order to  
compare the ellipse evolution in three 
cases: with a single decoherence event (in 
blue, passing through a brown rectangle), 
no events (in green) and with a 
single recoherence event (in red, crossing 
a yellow rectangle). Although some mild 
dephasing naturally occurs even in the 
absence of decoherence -- since modes with 
different $k$ cross the horizon at 
different times -- the effects of 
decoherence and recoherence events 
dominate the evolution. These events 
produce noticeable changes in the area of 
the Wigner ellipse, which in turn make the 
resulting squeezing and additional 
dephasing visually evident. Therefore, 
our approach of using different $k$-modes 
remains an efficient way to illustrate and 
contrast 
the deformations induced by such 
``accidents'' with those in the 
environment-free scenario. To setup the decoherence events, we consider the 
tile parameterization in Eq.~\eqref{eq:tiling_source}: the 
decoherence accident has $\alpha_\Gamma=0.35$ and $p=3$. It is centered at 
$(N,\ln\ell)=(10.0,9.0)$, has a height of $\Delta \ell=0.4$ and a width of 
$\Delta N=4$. The recoherence event has $\alpha_\Gamma=-0.35$, is located 
at $(N,\ln\ell)=(16.0,9.0)$ and has the same width, height and value of $p$ as the 
area-increasing accident. The shape of the horizon corresponds to the two-field 
nonlinear potential presented at the beginning of this section in 
Eq.~\eqref{eq:double_lp4_sfs}, producing almost $84$ e-folds of inflation.

\deccrrmf

In the right panel of \Figref{fig:Wigner_mf}, we show the projections of the Wigner 
ellipse at a fixed instant of time for the same modes displayed in the left panel. 
As a representative example, we consider the case where both fields contribute equally 
to the interaction operator 
\begin{align}
\hat{I}^{\rm S}_{\bf k}=\hat{v}^{\phi}_{\rm S}+\hat{v}^{\sigma}_{\rm S}\,,
\label{eq:int_mf_wigner_mf}
\end{align}
where this choice sets $\alpha_\phi=\alpha_\sigma=1$, $\beta_\phi=\beta_\sigma=0$, 
while the labels $\phi$ and $\sigma$ refer to the $\phi$ and $\sigma$ background 
field components of the nonlinear potential. Together 
with the tile parameters, this specifies all of the source 
term components $\mathcal{F}_{AB}$ entering the dynamical coloring transformations defined by 
Eqns.~\eqref{eq:chip_chi_mf}, \eqref{eq:z_evol}, and \eqref{eq:fin_eq_L}. 
Furthermore, from this panel, it is clear that the area of the ellipse is increased by 
the decoherence event, while the recoherence event reduces 
it. It is also evident that these events not only affect 
the area, but also the squeezing and the rotation 
phases, depicted by the orthogonal dashed lines in each 
projection. Variations in the area are further confirmed 
by the marginalized distributions $\rho_k(v_{\rm S}^A)$ and $\rho_k(\pi_{\rm S}^A)$, which broaden as the 
modes cross a decoherence tile with $\alpha_\Gamma>0$ and 
narrow when traversing a tile with $\alpha_\Gamma<0$. 
Although not shown here, as a consistency check, we 
compared the power spectrum produced 
by a separate code solving numerical mode evolution 
with our Cholesky decomposition transport scheme in the 
decoherence-free case for the nonlinear potential, 
obtaining the same results. We will 
present an alternative fast and slow scale separation 
approach based in the evolution of the mode equations in a future project. 

An animation accompanying our results in \Figref{fig:Wigner_mf}, provided as an 
ancillary file \cite{supplemental1}, illustrates the full evolution of the 
joint Wigner function as the modes traverse 
decoherence/recoherence events for the same tile 
parameters considered in this figure. 
In this animation, the evolution also includes the deformation of the ellipse during 
horizon crossing. A second animation shows the generation of 
squeezed states under a different sequence of tiles for the single-field version of this model (\ie $V(\phi)=\lambda\phi^4/4$). 

As in the single-field case, the evolution of the covariance matrix determinant $\mathcal{S}$ provides a diagnostic of the possible instabilities emerging during the dynamical coloring evolution. In 
\Figref{fig:s_evol_mf}, we show the evolution of 
$\mathcal{S}$ for the same three modes displayed in 
\Figref{fig:Wigner_mf}, corresponding respectively to 
cases where the area of the Wigner ellipse increases, 
decreases, and remains unchanged during the decoherence 
events. Similar to panel (d) in Figures 
\ref{fig:imp_rndm_acc_sf} and \ref{fig:cmpnstd_acc_sf} 
for the case of one field, variations in the determinant 
only occur while decoherence events are active. To test 
whether numerical errors accumulate over time, we computed 
the deviation $\Delta\mathcal{S}$ between the 
instantaneous value of the determinant and its mean value 
over the interval $N\in[20,83.5]$. The lower bound of this 
interval corresponds to an arbitrary instant of time 
sufficiently close to the onset of decoherence (or 
recoherence) events. 
The insets of \Figref{fig:s_evol_mf} show that the 
deviation remain at the level of double precision 
round-off errors, confirming the stability of 
the numerical evolution scheme. The code publicly available in this paper
uses an A-stable eighth-order symplectic Gauss-Legendre 
integrator \cite{10.2307/2003405} to numerically 
resolve the time evolution of the mode correlators. 

The interaction operator parameterization introduced in 
Eq.~\eqref{eq:gen_lin_par} suggests that the isentropic and the 
isocurvature components can be treated as distinct channels through which 
decoherence effects may manifest. For simplicity, the two-field realization 
considered in this section only requires interaction operators 
of the form
\begin{subequations}
\begin{align}
&\hat{I}^{\rm S}_{\parallel}=\alpha^\phi_{\parallel}(\tau)\hat{v}^{\phi}_{\parallel}+\alpha^\sigma_{\parallel}(\tau)\hat{v}^{\sigma}_{\parallel}\,,\label{eq:oper_par}\\
&\hat{I}^{\rm S}_{\perp}=\alpha^\phi_{\perp}(\tau)\hat{v}^{\phi}_{\perp}+\alpha^\sigma_{\perp}(\tau)\hat{v}^{\sigma}_{\perp}\,,
\label{eq:oper_perp}
\end{align}
\end{subequations}
which follow directly from the field 
definitions along the parallel and perpendicular directions in 
Eqns.~(\ref{eq:parallel}--\ref{eq:perp}), respectively. To treat separately the
decoherence effects associated with the parallel $\left\langle\{\hat{v}^A_{\parallel},\hat{v}^A_{\parallel}\}\right\rangle$ and perpendicular 
$\left\langle\{\hat{v}^A_{\perp},\hat{v}^A_{\perp}\}\right\rangle$ contributions of the 
power spectrum, it is sufficient to assign time-independent values to the coefficients $\alpha_\parallel$ and 
$\alpha_\perp$. Thus, for simplicity, we fix 
all the components of $\alpha_\parallel$ and $\alpha_\perp$ to one.
For compactness in the notation, we define the following normalized 
correlators with respect to these contributions:
\begin{subequations}
\begin{align}
&N_v\equiv \frac{k^3}{4\pi^2}\left[\frac{H}{a||\dot{\phi}||^2}\right]^2~~\,,~~\mathcal{P}_{\mathcal{RR}}\equiv N_v\left\langle\{\hat{v}^A_{\parallel},\hat{v}^A_{\parallel}\}\label{eq:par_par}\right\rangle\,,\\
&\mathcal{P}_{\mathcal{RS}}\equiv N_v\left\langle\{\hat{v}^A_{\parallel},\hat{v}^A_{\perp}\}\right\rangle~~\,,~~\mathcal{P}_{\mathcal{SS}}\equiv N_v\left\langle\{\hat{v}^A_{\perp},\hat{v}^A_{\perp}\}\right\rangle\label{eq:perp_perp}
\end{align}
\end{subequations}
where, apart from the normalization factor, this notation is consistent with previous 
work (\ie \cite{Gordon:2000hv,Byrnes:2006fr,Lalak:2007vi} and others). 

Considering the interaction operator definitions introduced 
above, our results in \Figref{fig:dec_crr_mf} display the 
power, cross-correlations and the dynamical effects from decoherence 
events in the evolution of the mode correlators defined in 
Eqns.~(\ref{eq:par_par}--\ref{eq:perp_perp}) for arbitrary values of $k$. 
These results are obtained by configuring a fixed sequence of 
decoherence events sourced separately by the parallel and perpendicular operators. 
Individual tiles are built following the parallelogram parameterization 
introduced in Eq.~\eqref{eq:tiling_source_mod}. 

\tablemf

In \Figref{fig:dec_crr_mf}, panel (a), we present the mode-injection scheme 
based on the same two-field nonlinear potential used throughout this section 
$V(\phi,\sigma)=\lambda\phi^4/4+g\phi^2\sigma^2/2$
with $\lambda=10^{-14}$ and $g=2\lambda$. The background initial conditions were chosen 
to yield approximately 83 e-folds of inflation. In this setup, both the parallel and 
perpendicular operators source the same distribution of decoherence events, arranged as 
a single sequence of tiles shown in panel (a). We report some of the main 
properties of each tile in Table \ref{tab:events}. From this scheme, it can be seen 
that the following features in the $(N,\ln\ell)$ space are common to all tiles: 
the position along the vertical axis, the height in the length scale and the width 
along the injection scale are fixed with $\ln\ell_{j}=7$,
$\Delta\ell_{ij}=2$ and $\sinh^{-1}\left(\Delta k_{ij}/2k_{i}\right)=5$.

In \Figref{fig:dec_crr_mf}, panels (b1) and (b2) show the evolution of the isocurvature 
$(\mathcal{P}_{\cal SS})$ and isentropic $(\mathcal{P}_{\cal RR})$ power spectra as the mode 
correlators pass through the same sequence of decoherence events depicted in panel (a). 
Panel (b1) corresponds to decoherence events driven by the environmental operator 
$\hat{I}^{\mathrm{S}}_{\parallel}$. At first glance, spectral deformations appear only 
in the isentropic spectrum $\mathcal{P}_{\cal RR}$. However, a more careful evaluation of 
the differences with respect to the 
accident-free scenario reveals faint 
imprints in the isocurvature spectrum 
$\mathcal{P}_{\cal SS}$. As highlighted in 
the inset at the lower-left corner, which 
displays the difference in adiabatic mode 
power $(\Delta \cal P_{SS})$ 
with respect to the case free of 
environmental effects, these residual 
structures follow the shape of the 
decoherence tiles 
but have almost negligible amplitudes. Panel (b2) presents the analogous effects 
induced by the perpendicular operator $\hat{I}^{\mathrm{S}}_{\perp}$. In contrast to 
panel (b1), the isocurvature power spectrum ($\mathcal{P}_{\cal SS}$) now 
exhibits visible intermittent deviations, while the isentropic spectrum remains  
approximately featureless. The results in both panels clearly indicate 
the presence of weak cross-correlations between entropy and curvature modes, 
allowing the propagation of decoherence events active in one direction to propagate 
to the other. Furthermore, the combined configuration of parallel and perpendicular 
environmental operators resembles the action of a beam-splitter, where the adiabatic 
and isocurvature modes play an role analogous to orthogonal polarization states in a 
quantum optics setup. The action of this approximate beam-splitter can be further 
refined by tuning the parameters $\alpha_\perp$ and $\alpha_\parallel$ to minimize 
cross-correlation effects, a possibility that will be explored in a future project.

In \Figref{fig:dec_crr_mf}, panel (c), we show the evolution of the constant-$k$
$\mathcal{P}_{\cal RR}$ mode correlator as it undergoes decoherence events driven by 
either the parallel or perpendicular environmental operators. The trajectory 
corresponds to the mode depicted by the solid purple line in panel (a) of the same 
figure.
To visualize the dynamical impact of each decoherence tile, we also display the
corresponding evolution in the absence of environmental coupling
($\hat{I}^{\mathrm{S}} = 0$). The evolution inside the horizon (before $N=40$) 
further illustrates 
the dynamical coloring transformations effectively implement a fast and slow scale 
separation scheme: as in the single-field case, none of the correlators computed 
involves algebraic phase cancellations through complex conjugation, since all 
quantities are real. Consequently, the plotted variables evolve slowly. 
This panel also confirms the expected behavior of the environmental 
operators: the parallel operator is more effective at deforming the
dynamics of the curvature mode correlators. Although not shown here, an analogous
result indicates that the perpendicular operator produces a similar effect on the
entropy mode correlator. The inset within this panel shows the evolution of the
differences with respect to the accident-free case, illustrating that -- even though the
curvature mode correlator affected by the perpendicular operator remains several
orders of magnitude smaller than in the case of the parallel decoherence event -- the
modes crossing the tiles sourced by $\hat{I}^{\mathrm{S}}_{\perp}$ still exhibit a
monotonic growth over time after crossing the horizon. This growth is consistent with 
the emergence of small structures observed in \Figref{fig:dec_crr_mf}, panel (b2) in 
the isentropic power spectrum. If the correlation modes are allowed to evolve for 
longer times outside the horizon, this growth may lead to features of larger 
amplitude. 

\landscapemf

The results obtained in panels (b1), (b2) and (c) consistently emphasize the relevance 
of cross-correlations. To explore this further, we plot in \Figref{fig:dec_crr_mf}, 
panel (d), the entropy-curvature mode cross-correlators $(\cal P_{RS})$ for 
the three configurations of tiles sourced by environmental operators. As in all 
of the previous cases, the cross-correlators retain the characteristic shape of the 
decoherence tiles that generate them, providing an additional check of the correct implementation of the dynamical system and a means to detect possible instabilities.  
In the high frequency regime, the tile patterns become convoluted with 
rapidly oscillating structures, which makes the underlying tile configuration harder to
discern solely from shape of the spectrum. As shown in the panel, cross-correlator 
modes are allowed to be negative as long as the full covariance matrix remains 
positive-definite. Moreover, the large blue spike demonstrates that decoherence 
events may be used to flip the signs of the correlations, thereby (in principle) altering the direction of power transfer between fields. 

From the form of the two-field potential in Eq.~\eqref{eq:double_lp4_sfs}, it follows that 
$g$ is one of the parameters tuning the coupling between the 
background. 
Taking this into account, the inset of panel (d) shows that the 
magnitude of the entropy-curvature cross-correlations increases with the ratio 
$g/\lambda$. For this comparison, we 
adjusted the background field parameters so 
that each case 
(\ie $g/\lambda = 2,$ $8$ and $14$) 
produces the same number of e-folds of 
inflation. 
In all three cases, the constant $k_{\rm phys}$ hypersurface is fixed, inducing small 
shifts in the position of the horizon. As $g$ increases, the horizon approaches the 
horizontal line denoting constant $k_{\rm phys}$, reducing the time of flight of the 
mode correlators inside the horizon and amplifying their magnitude. 
A closer inspection of the inset reveals that the characteristic features in the 
cross-correlations persist in every case, indicating that (a) the model parameters 
modulate the magnitude of the cross-correlators, and (b) features in the mode 
correlators can propagate between fields through their classical coupling. 
We observe that beyond mixing terms from the potential, the effective 
oscillation matrix $\boldsymbol{\Omega}^2$ in Eq.~\eqref{eq:eff_freq_mf} also couples mode 
perturbations through background field time derivatives. This observation motivates the 
construction of numerical routines capable of incorporating the effects of 
sharp trajectories and propagating decoherence across different degrees of freedom. 
These aspects will be explored in a forthcoming project. 

For sufficiently smooth evolution of the background trajectories, increasing the 
number of fields is not an obstacle to resolve a large amount of 
decoherence events generating state deformations during subhorizon evolution. We reuse  
the parallel and perpendicular operators defined in Eqns.~\eqref{eq:oper_par} and 
\eqref{eq:oper_perp} to construct an arrangement of $48\times48$ decoherence tiles in 
\Figref{fig:lndscp_crr_mf}. The left panel illustrates the mode injection scheme and 
displays a sequence of tiles that increase the area of the Wigner ellipse (\ie with 
$\alpha_\Gamma\geq0$), forming a landscape pattern in the $(N,\ln\ell)$ plane. Unlike 
the configuration used in \Figref{fig:dec_crr_mf}, panel (a), the tile parameterization 
here follows Eq.~\eqref{eq:tiling_source}, which assumes
rectangular rather than parallelogram tiles, as depicted in the small $5\times 4$ array 
of events in the upper corner of this panel. In the right panel, we illustrate the 
impact of two distinct configurations of events -- one sourced by the 
parallel environment operator $\hat{I}^{\rm S}_\parallel$ and the other by 
perpendicular ones $\hat{I}^{\rm S}_\perp$ -- both built using the same tile 
layout depicted in the left panel. In contrast with the results shown in 
\Figref{fig:dec_crr_mf}, panel (d), here we represent cross-correlations using the 
variable
\begin{align}
\cos(\Delta)\equiv \frac{\cal{P_{RS}}}{\sqrt{\cal{P_{RR}}\cal{P_{SS}}}}\,,
\label{eq:cos_delta}
\end{align}
which measures the relative strength of cross-correlations with respect to the 
amplitude of adiabatic and isocurvature spectra \cite{Gordon:2000hv}. 
Our results demonstrate that decoherence events not only produce localized 
features in the spectra of mode correlators across different wavelength ranges, 
but also modifies both the direction and the magnitude of power transfer between fields.

\section{Discussion}
\label{sec:conc}

Decoherence, viewed in the framework of open-quantum 
systems, is itself a physically motivated  
phenomenon capable of inducing nontrivial state 
deformations 
\cite{Martin:2015qta, Amin:2015ftc,Martin:2018lin,Martin:2018zbe,Martin:2021qkg,Martin:2021znx}. 
Within this context, decoherence events may be 
interpreted as accidents that alter the course of 
state evolution -- alterations that can be visualized 
as deformations of the Wigner ellipse representing the 
quantum state in phase space.
However, given the flexibility of this framework, 
we adopt an alternative viewpoint: sequences of 
decoherence events can form a laboratory for driving 
primordial states toward specific dynamical 
configurations or target properties. One of such 
objectives is the generation of targeted 
features in the primordial curvature power spectrum, 
which are of phenomenological 
interest in nonlinear structure formation 
\cite{Stahl:2025qru}. Although these features are 
usually associated with inflationary models that 
exhibit specific dynamical properties 
(\eg \cite{Achucarro:2014msa,Martin:2014kja,Bhattacharya:2022fze,Kallosh:2025rni} and others), our 
implementation offers a computationally efficient way to reproduce such effects by 
modifying initial states. Moreover, within the context of single-field models, a 
natural next step -- consistent with the framework of effective theory of inflation 
\cite{Salcedo:2024smn} -- would be to propagate these modifications into the background 
expansion history, thereby introducing wavelength-dependent corrections (similar to our 
results in \subsecref{subsec:FSCS}, Eq.~\eqref{eq:eff_zpp_z}), and opening an 
interesting avenue for future research. 

In this project, we present a numerical 
implementation that incorporates the leading 
non-unitary decoherence 
corrections into the dynamics of inflationary 
fluctuations at the level of two-point correlators. 
It is important to emphasize that the results presented here are intended to illustrate the diagnostic power of the framework, and should not be interpreted as predictions of any specific inflationary model.

These corrections enable controlled manipulations of 
the initial Gaussian states in a way analogous to a 
pumping protocol in quantum optics. 
The method accommodates arbitrary 
arrangements of decoherence events and 
accounts for nonlinear background field evolution 
without relying on the slow-roll approximations. 
Furthermore, this approach has been extended to 
evaluate systems with multiple fields that undergo 
coupling through the time-evolving background fields. 
The numerical implementation employs the Wigner 
representation of Gaussian states to perform 
dynamical coloring transformations -- which, as 
demonstrated in \cite{GalvezGhersi:2016wbu},
act as a fast-slow 
scale separation scheme that efficiently suppresses 
the highest oscillation scales of a problem. 

Part of the flexibility of this alternative viewpoint 
comes from the possibility of introducing 
events with negative decay rates, $\Gamma \tilde{C}_{E} < 0$, 
which we refer to as recoherence events \cite{Colas:2022kfu, Kranas:2025jgm}. While this may 
lead to controversial interpretations, including 
possible conflicts with the uncertainty principle, it 
also broadens the class of effective 
dynamical processes that can be explored and improves 
the precision with which the initial states can be 
prepared. We present an example of this versatility in \Figref{fig:cmpnstd_acc_sf}, 
where we built accident/anti-accident pairs. Here the effects 
generated by an event are compensated by a subsequent 
event with the opposite sign in its decay rate. This 
type of configuration induces intermittent state 
deformations that leave the primordial 
spectrum unaffected. Nevertheless, at the 
nonperturbative level, these intermittent decoherence 
events together with nonlinear potential deformations 
can source higher-order statistical moments in the 
distribution, including primordial non-Gaussianities.

As discussed in \cite{Martin:2021znx}, the open-quantum 
system formulation applied to inflationary cosmology 
does not impose restrictions in the number
of fields involved. Therefore, it is also 
reasonable to consider initial-state deformations 
arising from the coupling 
of a multifield system to an environment. 
In \subsecref{subsec:Lindblad_multi}, we derived the gauge transformations in 
Eqns.~(\ref{eq:gauge_vv_mf}--\ref{eq:gauge_F_mf}), which enable different forms of 
couplings through environment operators constructed from either fields or their 
conjugate momenta. In particular, an alternative coupling emerges from the projection 
into isocurvature and adiabatic modes, which suggests the 
existence of more than one channel through which decoherence may manifest. Our findings 
in \subsecref{subsec:results} exhibit a regime in which the effects of decoherence 
can be approximately decoupled: while the adiabatic spectrum is only slightly 
affected, the isocurvature sector undergoes stronger deformations.

As shown in \Appref{app:Gaussian}, the output of this coloring 
scheme can be used to produce three-dimensional field realizations suitable as 
initial conditions for nonlinear simulations of reheating, such as those in 
\cite{Frolov:2008hy}. This makes it computationally efficient to generate large
ensembles of field realizations, each corresponding to a distinct sequence of
decoherence events. For instance, when considering a potential that supports oscillon
formation -- such as the monodromy potential proposed in \cite{Amin:2011hj} -- 
the framework can be used to identify families of decoherence tiles that optimize 
oscillon production. This idea will be explored in forthcoming work.

\acknowledgements

The authors would like to thank Jonathan Braden for many fruitful conversations which 
motivated this project. We would also like to thank Andrei Frolov, Stefano Gonzales, 
Lorena Luján, Jorge Medina, Diego Suárez, Sashwat Tanay and Vincent Vennin for their 
technical and logistic support, feedback and many valuable discussions resulting in the 
first versions of this paper. 
Part of the computations were performed on the cloud computational resources provided 
by Oracle. The work of JP and JG was partially funded by Fondo Semilla 2023, granted by Universidad de Ingeniería y Tecnología. The work of GQ was 
funded in part by NSERC Discovery Grant ``Testing fundamental physics with B-modes on 
Cosmic Microwave Background anisotropy''.
This project was funded by CONCYTEC through the PROCIENCIA program in the context of 
the research opportunity ``E041-2024-03. Proyectos de Investigación Básica'', 
contract number PE501087705-2024-PROCIENCIA.

\begin{appendix}
\section{Other factorization schemes for dynamical coloring transformations}
\label{app:sym_sep}

As discussed in \subsecref{subsec:Cholesky_lind}, the two-point correlation function $\langle \hat{v}_A , \hat{v}_B \rangle$ defines a symmetric and positive-definite matrix and therefore admits a factorization of the form
\begin{align}
    \langle \hat{v}^{\rm S}_A , \hat{v}^{\rm S}_B \rangle = L_{AC} L^{\rm T}_{CB}\,,
\end{align}
where $\mathbf{L}$ is a real matrix. In constructing dynamical coloring transformations based on Cholesky decomposition, we fixed the shape of the 
factors to be triangular. However, as noted
previously, this choice is not unique. In this appendix, we examine an alternative 
scheme in which $\boldsymbol{L}$ is taken to be symmetric rather than triangular, 
noting that both decompositions involve the same number of degrees of freedom. 

\boxes

To this end, we revisit the derivation of the coloring transformations, 
beginning with the redefinition of the purely real (or imaginary) field multiplet and its conjugate 
momenta
\begin{align}
&\hat{v}^A_{\rm S}\rightarrow L_{AB}\tilde{\chi}_B\,,\label{eq:v_mf_dec_app}\\
&\hat{\pi}^A_{\rm S}\rightarrow L'_{AB}\tilde{\chi}_B+L_{AB}\tilde{\chi}'_{B}\,,
\end{align}
here the matrix $\mathbf{L}$ encodes the amplitude correlations among field components, 
and $\tilde{\chi}$ denotes a vector of Gaussian random variables spanning the relevant 
region of phase space. These random variables are normalized such that their two-point 
correlator satisfies
\begin{align}
    \langle \tilde{\chi}_A , \tilde{\chi}_B \rangle = \delta_{AB} \,.
    \label{eq:gaussian-vars-cf}
\end{align}
Differentiating Eq.~\eqref{eq:gaussian-vars-cf} twice with respect to conformal time yields the following consistency relations:
\begin{equation}
\label{eq:cons_rel}
    \begin{gathered}
        \mathrm{Sym}^{CD}_{AB} \left[ \langle \tilde{\chi}_C^{\prime} , \tilde{\chi}_D \rangle \right] = 0 \,, \\
        \mathrm{Sym}^{CD}_{AB} \left[ \langle \tilde{\chi}_C^{\prime\prime} , \tilde{\chi}_D \rangle + \langle \tilde{\chi}_C^{\prime} , \tilde{\chi}_D^{\prime} \rangle \right] = 0 \,,
    \end{gathered}
\end{equation}
where the symmetrization operator is defined as 
$\mathrm{Sym}\,\mathbf{X}\equiv \mathbf{X} + \mathbf{X}^{\rm T}$. Substituting these 
relations in the covariance matrix definitions in Eqns.~(\ref{eq:cov_val_vv_mf}--
\ref{eq:cov_val_pp_mf}), and then into the transport equations in 
Eqns.~(\ref{eq:mf_trans_vv}--\ref{eq:mf_trans_pp}), we obtain the following condition 
for the coloring factors 
\begin{align}
        &\text{Sym}^{CD}_{AB}\bigg[\left(L^{-1}\right)_{CE}{L}_{ED}^{\prime \prime} -\mathcal{Z}_{CD}+\left(L^{-1}\right)_{CE}\Omega^2_{EF}L_{FD}\nonumber\\
        &+2(L^{-1})_{CE}{L}_{EF}^{\prime}\langle{\tilde{\chi}}^{\prime}_F, \tilde{\chi}_{D}\rangle+\langle{\tilde{\chi}}^{\prime}_C,  {\tilde{\chi}}_E\rangle\langle{\tilde{\chi}}^{\prime}_E,  {\tilde{\chi}}_D\rangle\bigg]=0\,,
\end{align}
where $\boldsymbol{\mathcal{Z}}\equiv \langle\tilde{\boldsymbol{\chi}}',\tilde{\boldsymbol{\chi}}'\rangle+\langle\tilde{\boldsymbol{\chi}}',\tilde{\boldsymbol{\chi}}\rangle\langle\tilde{\boldsymbol{\chi}}',\tilde{\boldsymbol{\chi}}\rangle$ is the only other variable, aside from the 
decomposition matrices, that appears in the determinant of the covariance matrix. 
The overall symmetrization allows the inclusion of an antisymmetric matrix 
$\boldsymbol{\mathcal{A}}$ that fixes the shape of the $\boldsymbol{L}$, leading to the equations 
of motion 
\begin{widetext}
\begin{equation}\label{eq:L-EOM}
        L_{AB}^{\prime\prime} + \Omega^2_{AC} L_{CB} + 2 L_{AC}^{\prime} \langle \tilde{\chi}_C^{\prime} , \tilde{\chi}_B \rangle - L_{AC}\left[\mathcal{Z}_{CB}-\langle \tilde{\chi}_C^{\prime} , \tilde{\chi}_D \rangle\langle \tilde{\chi}_D^{\prime} , \tilde{\chi}_B \rangle\right]  + L_{AC} \mathcal{A}_{CB} = 0\,,
\end{equation}
\end{widetext}
which is the only instance where the form of $\boldsymbol{L}$ can change.
As in the Cholesky-based decomposition scheme, fixing 
the shape of the matrix requires to identify the shapes of each term in the equations 
of motion. For convenience, we define
\begin{align}
    P_{AB} \equiv ~&\Omega^2_{AC} L_{CB} + 2 L^{\prime}_{AC} \langle \tilde{\chi}_C^{\prime} , \tilde{\chi}_B \rangle - \nonumber\\
    &L_{AC}\left[\mathcal{Z}_{CB}-\langle \tilde{\chi}_C^{\prime} , \tilde{\chi}_D \rangle\langle \tilde{\chi}_D^{\prime} , \tilde{\chi}_B \rangle\right]\label{eq:Pdef}\,.
\end{align}
From Eq.~\eqref{eq:L-EOM}, $\boldsymbol{L}$ remains symmetric at all times if 
and only if the combination $\boldsymbol{P} + \boldsymbol{L}\boldsymbol{\mathcal{A}}$ is 
symmetric, which implies
\begin{equation}
    \mathbf{P} + \mathbf{L} \boldsymbol{\mathcal{A}} = \mathbf{P}^{\rm T} - \boldsymbol{\mathcal{A}} \mathbf{L}\,,
\end{equation}
rearranging terms in this expression, we find that $\boldsymbol{\mathcal{A}}$ satisfies the Sylvester equation
\begin{equation}
    \left\{\mathbf{L}\,,\boldsymbol{\mathcal{A}}\right\} = \mathbf{Q}\,,
\end{equation}
where $\{\cdot\,,\cdot\}$ is the anticommutator and
\begin{equation}
    \mathbf{Q} \equiv \mathbf{P}^{\rm T} - \mathbf{P} = - \mathrm{Antisym} \big[\mathbf{P}\big]\,. 
\end{equation}
Since $\boldsymbol{L}$ is symmetric, it can be diagonalized by an orthonormal matrix $\boldsymbol{O}$, such that
\begin{equation}
    \mathbf{O}^{\rm T} \mathbf{L} \mathbf{O} = \bf{L}^{\rm D}\,,
\end{equation}
where $\boldsymbol{L}^{\rm D}$ is diagonal. In this diagonal basis of $\boldsymbol{L}$, the Sylvester equation becomes
\begin{equation}
    \bf{L}^{\rm D} \boldsymbol{\mathcal{A}}^{\rm D} +  \boldsymbol{\mathcal{A}}^{\rm D} \bf{L}^{\rm D} = \mathbf{Q}^{\rm D}\,,
\end{equation}
with $\boldsymbol{\mathcal{A}}^{\rm D} = \boldsymbol{O}^{\rm T} \boldsymbol{A} \boldsymbol{O}$. In component form (no summation over repeated indices), this reads
\begin{align}
    &\sum_{J=1}^{\mathcal{N}} \bigg( L^{\rm D}_{A} \delta_{AJ} \mathcal{A}^{\rm D}_{JB} + \mathcal{A}^{\rm D}_{AJ} L^{\rm D}_{J} \delta_{JB} \bigg)\nonumber\\
    &= L^{\rm D}_{A} \mathcal{A}^{\rm D}_{AB} + \mathcal{A}^{\rm D}_{AB} L^{\rm D}_{B} = Q^{\rm D}_{AB}\,,
\end{align}
which implies that the elements of $\boldsymbol{\mathcal{A}}^{\rm D}$ are given by
\begin{equation}
    \mathcal{A}^{\rm D}_{AB} = \frac{Q^{\rm D}_{AB}}{L^{\rm D}_{A} + L^{\rm D}_{B}}\,.
\end{equation}
Finally, $\boldsymbol{\mathcal{A}}$ can be reconstructed as
\begin{equation}
    \boldsymbol{\mathcal{A}} = \mathbf{O} \boldsymbol{\mathcal{A}}^{\rm D} \mathbf{O}^{\rm T}\,.
\end{equation}
Hence, with this construction of the antisymmetric matrix $\boldsymbol{\mathcal{A}}$, it is possible to 
ensure that the factor $\boldsymbol{L}$, which evolves 
according to Eq.~\eqref{eq:L-EOM}, will be symmetric at 
all times. The correlators involving derivatives of the 
random variables close the dynamical system, and their 
equations of motion -- given in  
Eqns.~\eqref{eq:chip_chi_mf} and \eqref{eq:z_evol} -- 
remain unaltered. This example 
shows that it is possible to implement dynamical coloring transformations using other 
decomposition schemes. However, we are still not able to confirm that this 
procedure works effectively as a fast and slow separation scheme. A more detailed 
analysis of alternative decomposition schemes will be explored in a future project.

\section{Building three-dimensional Gaussian field realizations}
\label{app:Gaussian}

Apart from separating the fast and slow oscillation scales -- which already provides a 
computational advantage -- the dynamical coloring scheme also produces output 
in a format suitable directly compatible with nonlinear codes used to evolve fields 
during reheating (\eg \cite{Frolov:2008hy}). In these evolution routines, the first step is 
to generate spatial realizations of the fields and their derivatives that serve as 
initial conditions. To construct them, we must compute the mode correlators of the 
original field-perturbation variables in cosmic time
\begin{widetext}
\begin{align}
&\langle\Phi^A_{\bf k},\Phi^B_{\bf k}\rangle=\frac{1}{a^2}L^{AC}(L^{\rm T})^{CB}\,,\label{eq:fld_fld}\\
&\langle\Phi^A_{\bf k},\dot{\Phi}^B_{\bf k}\rangle=\frac{1}{a^3}L'^{AC}(L^{\rm T})^{CB}-\frac{H}{a^2}L^{AC}(L^{\rm T})^{CB}+\frac{1}{a^3}L^{AC}\langle\tilde{\chi}'^{C},\tilde{\chi}^E\rangle(L^{\rm T})^{EB}\,,\label{eq:fld_phid}\\
&\langle\dot{\Phi}^A_{\bf k},\dot{\Phi}^{B}_{\bf k}\rangle=\frac{1}{a^4}\bigg\{L'^{AC}(L'^{\rm T})^{CB}-L^{AC}\langle\tilde{\chi}'^{C},\tilde{\chi}^D\rangle\langle\tilde{\chi}'^{D},\tilde{\chi}^E\rangle (L'^{\rm T})^{EB}+L^{AC}\mathcal{Z}^{CD}(L'^{\rm T})^{DB}\nonumber\\
&-a^2H^2L^{AC}(L^{\rm T})^{CB}-\mathrm{Sym}^{AB}_{CD}\left[L'^{CE}\langle\tilde{\chi}'^E,\tilde{\chi}^F\rangle(L^{\rm T})^{FD}+aHL'^{CE}(L^{\rm T})^{ED}\right]\bigg\}\label{eq:phid_phid}\,,
\end{align}
\end{widetext}
which are written with respect to the dynamical variables of the coloring scheme in 
conformal time at initial time $t=t_0$. Here we only considered the real part of the 
field operators $(\rm S=R)$, since for our choice of interaction operators -- at the 
end of inflation -- the imaginary part does not introduce more information.
Together, the correlators in Eqns.~(\ref{eq:fld_fld}--\ref{eq:phid_phid}) 
form a $2\mathcal{N}\times2\mathcal{N}$ symmetric matrix 
\begin{align}
\boldsymbol{\Psi}_{\bf k}(t_0)=\left[\begin{matrix}
       \langle\Phi^A_{\bf k},\Phi^B_{\bf k}\rangle  & \langle\dot{\Phi}^A_{\bf k},\Phi^B_{\bf k}\rangle\vspace{.5em}\\
       \langle\dot{\Phi}^A_{\bf k},\Phi^B_{\bf k}\rangle  & \langle\dot{\Phi}^A_{\bf k},\dot{\Phi}^B_{\bf k}\rangle\\
    \end{matrix}\right]\,,
    \label{eq:full_matrix}
\end{align}
which follows the reality condition $\boldsymbol{\Psi}^*_{\bf k}=\boldsymbol{\Psi}_{-\bf k}$. For adequate spectral 
resolution, we first interpolate the correlation matrix on a logarithmic scale and then transform  
it to real space using an inverse discrete Fourier transform 
$\boldsymbol{\Psi}_{r}(t_0)\equiv \mathrm{DFT}^{-1}[\boldsymbol{\Psi}_{\bf k}(t_0)]$. 

Spatial realizations of the field and its derivative are constructed by computing the Cholesky factors of the real-space correlation matrix
\begin{align}
\boldsymbol{\Psi}_r(t_0)=\mathbf{\Lambda}(t_0){\mathbf{\Lambda}}^{\rm T}(t_0)\,,
\label{eq:ch_real_corr}
\end{align}
and multiplying the resulting factor by a $2\mathcal{N}$-component vector of unit-variance random 
variables 
\begin{align}
V^m(t_0)= \Lambda^{mn}(t_0)\tilde{\chi}^n\,.
\end{align}
Here $(m,n)\in [1,2\mathcal{N}]$, with the first $\mathcal{N}$ components of $V(t_0)$ corresponding 
to the field and the remaining $\mathcal{N}$ to its derivatives. In our implementation, each 
component of $\tilde{\chi}$ represents a vector of $128^3$ spatial samples. 
These provide a complete set of initial conditions for the $\mathcal{N}$ fields 
that can be evolved in a reheating routine. We follow this approach to illustrate in 
\Figref{fig:3d_real} the field, 
time-derivative, pressure and energy density fluctuations in three-dimensional space arising 
from our dynamical coloring transformation scheme at the end of inflation. We compute 
fluctuations in the energy density from 
\begin{align}
\rho(\mathbf{x})=\frac{1}{2}\left[\dot{\phi}^A\dot{\phi}_A+\frac{1}{a^2}(\nabla^i\phi^A)(\nabla_i\phi_A)\right]+V(\phi,\sigma)\,,
\label{eq:rho}
\end{align}
considering the same two-field nonlinear potential (in Eq.~\eqref{eq:double_lp4_sfs}) used 
throughout the multifield section of this paper. We employ periodic boundary conditions to construct the gradient operator in 
Fourier domain and compute their energy contributions. Fluctuations are computed with respect 
to the average values over the entire box. In the left panel, we illustrate spatial fluctuations in 
the energy density in a $128^3$ box, which are consistent with a normal distribution. 

As shown in the right panel of \Figref{fig:3d_real}, the histograms of the field, momentum, 
energy-density, and pressure fluctuations remain consistent with Gaussian statistics, as 
anticipated from the premise established earlier in this appendix. To perform a preliminary test of 
decoherence, we compare two cases: one where the mode correlators traverse a layout of decoherence 
tiles sourced by the perpendicular operator $\hat{I}^{\rm S}_\perp$, and another without 
decoherence. The resulting distributions show only minor differences, in line with our results in
\subsecref{subsec:results}, \Figref{fig:dec_crr_mf}, panel (b2), indicating regimes in which the perpendicular environmental operator 
exerts limited influence. This suggests that decoherence through the isocurvature channel can be 
included without necessarily introducing significant distortions in the primordial curvature 
spectrum.

\end{appendix}
\bibliography{Bibnotes.bib}

%======================
\end{document}